\documentclass[aps,prx,twocolumn,superscriptaddress,nofootinbib, preprintnumbers,longbibliography]{revtex4-2}

\usepackage{amsmath}
\usepackage{amssymb}
\usepackage{mathtools}
\usepackage{hyperref}
\usepackage{graphicx}
\usepackage{color}
\usepackage{comment}
\usepackage{url}
\usepackage{dcolumn}
\usepackage[caption=false]{subfig}
\usepackage{siunitx}
\usepackage[dvipsnames]{xcolor}
\usepackage{bm}
\usepackage{booktabs}
\usepackage{tikz}
\usepackage[normalem]{ulem}
\usepackage{multirow}
\usetikzlibrary{arrows.meta,calc,decorations.pathreplacing,decorations.pathmorphing}

\newcommand{\HIPetc}{\affiliation{
Department of Physics and Helsinki Institute of Physics,
PL 64, 
FI-00014 University of Helsinki,
Finland
}}

\newcommand{\bfx}{\bm{x}}
\newcommand{\bfr}{\bm{r}}
\newcommand{\bfk}{\bm{k}}

\newcommand{\GN}{G}

\newcommand{\Ho}{\mathcal{H}_0}

\newcommand{\LC}{\mathcal L}
\newcommand{\TL}{\mathcal T}

\newcommand{\Hell}{\mathcal H}

\newcommand{\OL}{\Omega_\Lambda}
\newcommand{\Om}{\Omega_\text{m}}

\begin{document}

\preprint{HIP-2026-23/TH}

\title{Filling the interior of the light cone}

\author{T\"ore Boybeyi}%
 \email{boybe001@umn.edu}
\affiliation{School of Physics and Astronomy, University of Minnesota, 55455 MN, USA}

\author{Lorenzo Giombi}
\email{lorenzo.giombi@helsinki.fi}
\HIPetc

\author{Niko Jokela}
\email{niko.jokela@helsinki.fi}
\HIPetc

\author{K. Kajantie}
\email{keijo.kajantie@helsinki.fi}
\HIPetc


\begin{abstract}

\noindent
The retarded Green's function of gravitational waves in a spatially flat Friedmann--Lema\^itre--Robertson--Walker universe contains, besides the light-cone term, a tail supported inside the light cone. We formulate the tail, for homogeneous and isotropic backgrounds, as the solution of a ``\textit{Goursat problem},'' {\emph{i.e.}}, a hyperbolic partial differential equation with boundary conditions given on two intersecting characteristic curves. In this sense, we fill the interior of the light cone. We solve this problem directly in position space for any expansion history, and compare exact solutions in $\Lambda$CDM independently with a convergent double series about the coincidence point, and with an approximate analytic interaction model. We discuss the interpretation of the tail as the sum of the waves scattered off the background curvature, as suggested by its expansion in powers of the curvature potential. Our novel solutions allow us to follow the tail throughout the entire expansion history, from the electroweak epoch to today. We find that the tail of an early source records the trace of the cosmic fluid along the light cone, ordered in retarded time: single scattering accounts for its features in the radiation era, while multiple scattering is needed in the matter era. 
This work establishes the accuracy of our calculation of the tail and lays the groundwork for its applications. From the position-space data we obtain the large-$k$ expansion of the propagator. Its first correction, the effect of the tail, depends on the curvature potential integrated along the light cone. The applications of the tail, the power spectrum of a background of bursts, and the memory waveform of a single burst, both of which require convolving the Green's function with its source, are presented in follow-up papers.
\end{abstract}

\maketitle


\section{Introduction}

\noindent
In the early twentieth century, Hadamard's work on the Cauchy problem for hyperbolic equations showed that radiation in a curved spacetime, electromagnetic or gravitational, has two components~\cite{hadamard2014lectures}: besides the usual propagation on the light cone, a \emph{tail} propagates inside it, thus reaching the observer after a time delay.  
The Friedmann--Lema\^itre--Robertson--Walker (FLRW) metric is a prime example of a curved spacetime, and cosmological modes therefore carry a tail, although it is usually small in practice. Tails have been studied in curved spacetimes~\cite{DeWitt:1960fc,Hobbs:1968zza,Friedlander:2010eqa,Poisson:2011nh,Harte:2013dba,Copi:2022ire}, around black holes~\cite{Price:1971fb,Gundlach:1993tp,DeAmicis:2024eoy}, in the post-Newtonian waveform of compact binaries~\cite{Blanchet:1993ec,Wiseman:1993aj,Trestini:2023wwg}, and in cosmology~\cite{Faraoni:1991xe,Sonego:1993fw,Caldwell:1993xw,deVega:1998ia,Chu:2015yua,Chu:2016ngc,Chu:2016qxp,Kehagias:2016zry,Chu:2021apx,Jokela:2022rhk,Jokela:2023suz,Jokela:2023ytv,Compere:2026xwy}.

The purpose of this paper is to show that a systematic and numerically effective treatment of the tail is obtained by relating it to the solution of the so-called ``\emph{Goursat problem}''~\cite{goursat2013course}: a hyperbolic partial differential equation with boundary conditions given on two intersecting characteristic curves. The key here is the boundary condition setup. While the data of a wave equation are usually given on a spacelike surface of constant time, here they are given instead on the light cone itself. The solution thus proceeds inward, from the light cone to the worldline of the source at its center, filling the interior of the light cone. The method applies to any homogeneous and isotropic background. We solve it for $\Lambda$CDM and for the whole expansion history with the Standard Model plasma.

The strength of the tail is set by the curvature of the background, through its Ricci scalar $R$. Boundary data on the light cone are given by the integral of the interaction $m^2(\eta)=a''/a=a^2R/6$ along the ray, from the source to the field point, and set the scale of the full solution inside the light cone. 
The linearized Einstein equations establish that a nonvanishing $R$ causes the Green's function to acquire a component connecting timelike-separated points. To interpret this component, we adopt a picture representing the tail as scatterings of the wave off the background curvature~\cite{Price:1971fb,Gundlach:1993tp,Faraoni:1991xe,Faraoni:1999us}, and make it quantitative by expanding the solution in powers of $R$. Each order is pictured as the wave traveling along null segments joined at events where it scatters off the background curvature. The light-cone value is the scattering along the ray itself, and the interior of the light cone collects the paths with any number of scatterings. We use this multiple-scattering picture to understand the features of the tail and suggest approximations in special regimes.

In the next section we introduce the general properties of the tail.
The reader interested in the method and its numerical solution can go directly to Section~\ref{sec:goursat}.

\section{General properties of the tail}
\noindent
In flat spacetime, electromagnetic and gravitational radiation propagate strictly on the light cone, so that a signal emitted at $x'$ is received at points $x$ null-separated from $x'$. In a curved spacetime, the Green's function also acquires a component supported inside the light cone, implying that the signal also reaches points after the light-cone front.
This delayed signal, the \emph{tail}, contains crucial information for high-precision waveform matching of compact binaries~\cite{Blanchet:1993ec}, for the late-time ringdown of merger remnants~\cite{Price:1971fb,Gundlach:1993tp,DeAmicis:2024eoy}, and for the memory effect~\cite{Jokela:2022rhk,Trestini:2023wwg}.

Tensor perturbations $h_{ij}$ of the FLRW metric obey a massless wave equation, and the strain is the convolution of its retarded Green's function with the source. The tail is the part of this single field that reaches the observer from inside the light cone. For $\chi_{ij}=ah_{ij}$ the equation takes the Klein--Gordon form with the potential $m^2=a^2R/6$, where the Ricci scalar $R$ of the background acts as the coupling to the curvature. That the Green's function has support inside the light cone is a mathematical fact, but the physical mechanism explaining it is still a matter of interpretation. The most compelling one is scattering off the Ricci potential~\cite{Faraoni:1991xe,Faraoni:1999us,Chu:2011ip,Chu:2015yua}. In this picture, part of the wave emitted away from the observer is scattered back toward it from behind the source, so that it reaches the observer after the light-cone front, inside the light cone. After the light-cone front, an observer thus measures the sum over all possible scattering paths from the source, with each scattering weighted by the Ricci potential $m^2$. When $m^2=0$, as in radiation domination, the equation for $\chi_{ij}$ is the flat one and there is no tail~\cite{Kulczycki:2016jpf}.

On a spatially flat FLRW background, $h_{ij}$ is gauge invariant at linear order, and the split of its retarded Green's function into light-cone and tail terms is purely geometric~\cite{DeWitt:1960fc,Poisson:2011nh}. The split is clear only in position space, while in Fourier space the tail is already contained in the exact mode functions. Whether the tail carries energy and momentum is less clear. For isolated sources in asymptotically flat spacetime, the tails generated by the curvature of the source mass contribute to the energy flux at infinity, and energy balance requires a tail-induced radiation reaction on the source~\cite{Blanchet:2013haa}. In cosmology, energy and momentum are encoded in Isaacson's effective stress-energy~\cite{Isaacson:1968zza}, which requires averaging over wavelengths much shorter than the curvature scale, so that for the long-wavelength modes, which carry most of the tail, the local energy density is ambiguous~\cite{Unruh:1998ic}. We do not address this problem here and focus on the tail alone.

Before General Relativity, Lorentz, Abraham, and Poincar\'e realized that accelerated charges in flat spacetime interact with their own electromagnetic field, an idea formalized by Dirac~\cite{Dirac:1938nz} (see~\cite{Rohrlich:1990zz} for a historical review). 
DeWitt and Brehme~\cite{DeWitt:1960fc}, later corrected by Hobbs~\cite{Hobbs:1968zza}, generalized Dirac's result to curved spacetime: the retarded field of the charge is not confined to the light cone, and the self-force acquires a tail sourced by the curvature along the past history of the charge. This failure of Huygens' principle is generic for hyperbolic equations with variable coefficients, and occurs even in flat spacetime with an even number of spatial dimensions, while sharp propagation on the light cone holds only in odd dimensions $d\geq3$~\cite{hadamard2014lectures,courant1963methods}.
In cosmology, gravitational waves behave as massless, minimally coupled scalar fields~\cite{Ford:1977dj}, and the propagator has support inside the light cone whenever $a''/a\neq0$. Ref.~\cite{Iliopoulos:1998wq} also studied the propagator on flat FLRW backgrounds for graviton loop corrections to Newton's law.
The gravitational self-force on a small body orbiting a black hole was derived in Refs.~\cite{Mino:1996nk,Quinn:1996am}, and the scalar-charge problem in Ref.~\cite{Quinn:2000wa}. Extreme-mass-ratio inspirals, loud and long-lived sources for space-based detectors, have since motivated efficient and accurate evaluations of the tail.

Tail Green's functions have been constructed for FLRW and de Sitter backgrounds~\cite{Caldwell:1993xw,deVega:1998ia}, leading to several physical implications. 
Because of the tail, the field at $x$ depends on the whole past history of the source inside the past light cone of $x$. Such hereditary contributions enter the post-Newtonian waveform of compact binaries~\cite{Wiseman:1993aj}, the self-force~\cite{Poisson:2011nh}, and the scattering of waves off plane-wave backgrounds~\cite{Harte:2013dba}. Tails have also been connected to the asymptotic symmetries of de~Sitter and cosmological spacetimes~\cite{Kehagias:2016zry}, and to a proposed signature from the scattering of gravitational-wave tails off stars and compact objects~\cite{Copi:2022ire}.

The tail is not the only hereditary contribution to the waveform. Gravitational radiation also exhibits memory, the non-oscillatory component of every burst, sourced by the net change in the quadrupole of unbound masses and by the energy of the emitted radiation. In flat spacetime it is a step, arriving with the light-cone front and leaving a permanent displacement between freely falling test masses. Memory is sourced by a net change in the quadrupole moment of unbound masses (linear memory)~\cite{Zeldovich:1974gvh,Braginsky:1986ia} and, due to the nonlinear nature of General Relativity, by the stress-energy of the emitted gravitational waves (nonlinear memory)~\cite{Christodoulou:1991cr,Thorne:1992sdb,Wiseman:1991ss}. It is an unavoidable component of the waveform of every compact binary~\cite{Favata:2009ii,Favata:2010zu,Bieri:2013ada}. In an expanding universe, tail and memory combine;\footnote{The tail should not be confused with memory: the former is a property of the propagation, the latter of the signal. In post-Newtonian theory both appear as hereditary contributions to the waveform.} the light-cone memory step is not permanent, and the tail adds a memory that grows long after the front~\cite{Jokela:2022rhk,Jokela:2023suz,Sarkkinen:2023egx}. In expanding backgrounds, the memory of idealized sources was first computed in spatially flat FLRW~\cite{Tolish:2016ggo}, and then in de Sitter and $\Lambda$CDM~\cite{Bieri:2015jwa,Bieri:2017vni}. Position-space Green's functions, separating the light-cone and tail contributions, were constructed for spatially flat FLRW with a constant equation of state~\cite{Chu:2015yua,Chu:2016ngc,Chu:2016qxp} and asymptotically de Sitter spacetimes~\cite{Chu:2021apx} to compute the memory.
The nonlinear memory of a merger grows over cosmological times and can contribute to the stochastic gravitational-wave background~\cite{Jokela:2022rhk}. The tail part of the propagator also sources the inflationary tensor spectrum~\cite{Jokela:2023ytv}. 

Memory, including its spin component~\cite{Nichols:2017rqr}, is an observational target for current and future detectors~\cite{Favata:2010zu,Inchauspe:2024ibs}. A statistical signal from a population of binary black-hole mergers may be detected by LIGO-Virgo-KAGRA~\cite{Lasky:2016knh,Boersma:2020gxx,Grant:2022bla}, while single loud events could be probed by third-generation observatories~\cite{Islam:2021old,Grant:2022bla}.
LISA is instead expected to detect the memory of individual massive black-hole binary mergers~\cite{Inchauspe:2024ibs,Cogez:2026frh}, and possibly that of stellar- and intermediate-mass binaries using triggers from ground-based detectors~\cite{Ghosh:2023rbe}.
Supermassive black-hole binaries are targets of pulsar timing arrays and of joint PTA and LISA inference~\cite{Middleton:2015oda,Middleton:2017nbg,NANOGrav:2023gor,Steinle:2023vxs}. The memory burst of a single supermassive merger could be detected by pulsar timing~\cite{seto2009search,vanHaasteren:2009fy,Cordes:2012zz}, while the whole population of mergers sources a stochastic memory background~\cite{Zhao:2021zlr,Buda:2026yxr,Unal:2025jnq,Boybeyi:2024aax}. Population estimates model the single-event signal using either an idealized step~\cite{Phinney:2001dii,Pollney:2010hs,Zhao:2021zlr,Boybeyi:2024aax,Zhang:2025xqh} or numerical-relativity memory waveforms~\cite{Buda:2026yxr}, and propagate it to the observer through redshift and luminosity distance alone, {\emph{i.e.}}, through the light-cone term of the Green's function only. A cosmologically consistent prediction requires the full retarded Green's function, with the tail computed for the actual expansion history throughout the interior of the light cone. Here we compute the tail $B$ in a homogeneous expanding universe and establish the accuracy of several approximations used in the literature~\cite{Jokela:2022rhk,Compere:2026xwy}. The effect of the tail on the memory waveform of a single merger, after the convolution with the source, as well as its imprints on the stochastic background of a population of mergers and on the temperature anisotropy on the cosmic microwave background, will be presented in a companion paper.
There, we will also clarify how to select the physical components of the tensor $h_{ij}$. Although only the transverse-traceless (TT) part of $h_{ij}$ is physical (gravitational waves), the Green's function is a scalar and is therefore independent of the projection, which enters only after the convolution with the source. The TT projection is exact and algebraic in momentum space, but non-local in position space, where it is usually approximated by its radiation-like form, transverse to the direction of propagation. We will discuss this in more detail in the companion paper, focusing here on the retarded Green's function alone.

In the Hadamard construction on globally hyperbolic spacetimes, the retarded Green's function splits into a light-cone term and a tail supported inside the light cone, quantified by the two-point function $B(x,x')$~\cite{hadamard2014lectures,Poisson:2011nh}. Refs.~\cite{Jokela:2022rhk,Jokela:2023suz} stated the computation of $B$ in FLRW as a Cauchy problem with data on a spacelike slice, and found analytic light-cone values for single-fluid cosmologies such as matter domination. No closed form exists inside the light cone, nor anywhere for multi-fluid cosmologies such as $\Lambda$CDM. 
Ref.~\cite{Jokela:2022rhk} computed the $\Lambda$CDM tail from the equations of its Fourier modes. Since $B$ jumps to zero across the light cone, its Fourier modes decay slowly, and the inverse transform, truncated at a finite wavenumber $K$, leaves ripples of period $2\pi/K$ in position space and other numerical artifacts. The position-space construction of the tail in spatially flat FLRW goes back to Ref.~\cite{Chu:2015yua}, and tails have recently been included in the linearized field of localized sources~\cite{Compere:2026xoe,Compere:2026xwy}. Here we solve for $B$ directly in position space, under full numerical control and throughout the light cone. We do so both for the exact $\Lambda$CDM interaction and for the full expansion history including the Standard Model plasma. 
Our formulation naturally lends itself to the interpretation of the tail as scattering off the background curvature~\cite{Price:1971fb,Gundlach:1993tp,Faraoni:1991xe,Faraoni:1999us}, which we quantify by expanding it in the number of scatterings off the Ricci potential.

We state the problem for $g=rB$ as a ``\emph{Goursat problem}''~\cite{goursat2013course}: $B$ is fixed on the characteristic future light cone of the source, where light-cone and tail terms together must satisfy the linearized Einstein equation. By reflection symmetry $r\to-r$ for isotropic sources, data on the second characteristic are traded for regularity on the worldline, $rB\to0$ as $r\to0$. This gives $B$ at high precision without numerical differentiation. Double-null methods for the tail have been developed in black-hole perturbation theory and self-force calculations, where the retarded Green's function of the Regge--Wheeler and Zerilli equations is built from light-cone data and propagated on a grid adapted to null 
coordinates~\cite{Mark:2017dnq,Jonsson:2020npo,OToole:2021crw}. 
We apply the ``diamond-cell'' marching scheme of that context, most directly of Ref.~\cite{OToole:2021crw}, to the cosmological tail. Independently, we derive a convergent double Taylor series about the coincidence point from a Neumann (Born) expansion of the Riemann function of the Goursat problem. The two constructions share no numerical ingredients beyond the light-cone data, and cross-check each other. Compared to other formalisms~\cite{Jokela:2022rhk, Compere:2026xwy}, our Goursat solutions are numerically faster and more stable, can be extended easily to any expansion history, and make transparent how the curvature potential weights the solution as it is marched inward from the light cone. 

Homogeneity and isotropy reduce the $(3+1)$-dimensional problem to a $(1+1)$-dimensional characteristic problem. Small inhomogeneities spoil this reduction: the light-cone data must then be supplied by ray tracing through the inhomogeneous past light cone of each field point, making the tail a probe of primordial inhomogeneities, absent from the light-cone signal. We leave this to future work.

Section~\ref{Methods} recalls the Hadamard decomposition of the Green's function, and Section~\ref{sec:goursat} recasts the tail as a Goursat problem, solved by a corner series and by double-null marching. Results are shown in Section~\ref{Results}, where the two methods are cross-validated for two interaction potentials. Section~\ref{sec:scattering}  quantifies the tail in the multiple scattering picture, Section~\ref{sec:full_history} follows it through the entire expansion history of the universe, and Section~\ref{sec:fourier} compares it with the light cone across scales. Section~\ref{sec:discussion} presents the conclusions.

We use $c=1$, $\Lambda$CDM with $\OL=0.7,\,\Om=0.3$, and conformal time $\eta$. The source event is $x'=(\eta',\bfx')$ and the field point $x=(\eta,\bfx)$, at time ${t}=\eta-\eta'$ after the source and comoving distance $r=|\bfx-\bfx'|$ from it.

\section{\texorpdfstring{Gravitational waves in $\Lambda$CDM}{Gravitational waves in Lambda CDM}}
\label{Methods}

\noindent
In a spatially flat FLRW background with conformal time $\eta$ and scale factor $a(\eta)$, the transverse-traceless perturbations of the metric $h_{ij}\equiv h_{ij}(\eta,\bfx)$ sourced by a transverse-traceless stress tensor $t_{ij}^{TT}$ obey the linearized Einstein equation
\begin{equation}
    h_{ij}'' + 2\mathcal{H}(\eta)\, h_{ij}'
    - \nabla^2 h_{ij} = 16\pi \GN\, t_{ij}^{TT}(\eta,\bfx)\ ,
    \label{eq:hij_wave}
\end{equation}
where $\mathcal{H}\equiv a'/a$ and a prime denotes a partial derivative with respect to conformal time, that is ${}^\prime = \partial_\eta$. The field redefinition $\chi_{ij}\equiv a(\eta)h_{ij}$ removes the friction term and casts \eqref{eq:hij_wave} as a Klein--Gordon-type equation with a time-dependent effective mass 
\begin{equation}
    \chi_{ij}'' - \nabla^2\chi_{ij} - m^2(\eta)\chi_{ij} = 4\pi a(\eta) S_{ij}(\eta,\bfx)\ ,
    \label{eq:chi_wave}
\end{equation}
where $S_{ij}\equiv 4\GN t_{ij}^{TT}$ and $m^2(\eta)\equiv a''(\eta)/a(\eta)$. A formal solution to the wave equation of tensor perturbations can be written as the convolution of the source $S_{ij}(\eta,\bfx)$ with the retarded propagator $G(\eta,\bfx;\eta',\bfx')$ of gravitational waves 
\begin{equation}
    h_{ij}(\eta,\bfx) = \frac{1}{a(\eta)}\int d^4x'\, a(\eta')\, G(x, x^\prime)\, S_{ij}(x')\ ,
\end{equation}
where for simplicity we defined the spacetime events $x = (\eta,\bfx)$ and $x^\prime =(\eta',\bfx')$. 
The retarded propagator $G(\eta,\bfx;\eta',\bfx')$ is the Green's function of the linear differential operator in equation~\eqref{eq:chi_wave}, that is
\begin{equation}
    \left[\partial_\eta^2 - \nabla_{\bfx}^2 - m^2(\eta)\right] G(x,x') = 4\pi\,\delta(\eta-\eta')\,\delta^{(3)}(\bfx-\bfx') \ ,
    \label{eq:green_eq}
\end{equation}
with the causality condition $G(\eta,\bfx;\eta',\bfx')=0$ for $\eta<\eta'$. 

Because \eqref{eq:green_eq} is a wave equation on a globally hyperbolic background spacetime, its retarded Green's function has the universal Hadamard structure~\cite{hadamard2014lectures,Friedlander:2010eqa,DeWitt:1960fc, Poisson:2011nh}
\begin{equation}\label{eq:G_decomp}
    G(x,x') = \frac{\delta(u)}{r} + B(x,x')\,\theta(u)\ ,
\end{equation}
with 
\begin{equation}
     r\equiv|\bfx-\bfx'|,\quad u\equiv(\eta-\eta')-r \ .
\end{equation}
The first term is supported strictly on the past light cone $u=0$, and corresponds to the flat-spacetime $(m(\eta)=0)$ propagator of the d’Alembert operator. The second term, $B(\eta,\bfx;\eta',\bfx')\,\theta(u)$, arises instead from the effect of the curved spacetime on the propagation of gravitational waves and is supported everywhere inside the light cone ($u>0$). This signal, commonly called the \textit{tail}, travels over timelike separation and thus affects points $x$ that lie inside the future light cone of $x^\prime$. In Minkowski spacetime $m^2=0$, and one checks directly by Fourier transforming equation~\eqref{eq:green_eq} in $\bfx-\bfx'$ and solving the resulting radial equation, that $G(x,x')=\delta(u)/r$, exactly, with no tail. The curvature of the background reintroduces a tail through $m^2(\eta)=a''/a=a^2R/6$, which in a spatially flat FLRW universe vanishes only when the stress-energy tensor has zero trace, as for an exactly conformal fluid~\footnote{This condition, as well as the relation $m^2(\eta)=a''/a=a^2R/6$, is only correct for spatially flat geometries. For example, in the Milne universe $a\propto t$ and the Green's function has no tail~\cite{milne1935relativity}. The Milne universe is the interior of the future light cone of one event in Minkowski space, and is identical to an FLRW metric with open spatial sections, where the tensor equation contains spatial-curvature terms besides $a''/a$ which make the tail term vanish.} with $a\propto\eta$.

The two-point function $B(x,x^\prime)$ quantifies the strength of the tail signal. 
Exact analytic solutions for $B(x, x^\prime)$ can be found in single-fluid cosmologies~\cite{Jokela:2022rhk,Jokela:2023suz}. The late expansion of the universe is, however, better described by the concordance $\Lambda$CDM model, for which the tail should be studied numerically. 
The exact $\Lambda$CDM interaction is given by
\begin{equation}
    m^2_{\Lambda{\rm CDM}}(\eta) \equiv \frac{a''(\eta)}{a(\eta)} = \Ho^2\left(2\OL a^2(\eta) + \frac{\Om}{2a(\eta)}\right)
    \label{eq:m2_LCDM} \ ,
\end{equation}
with $\Ho$ the Hubble parameter at the present time $\eta_0$ and the scale factor $a(\eta)$ determined implicitly by
\begin{equation}\label{eq:H0eta}
    \Ho\eta(a)=2\int_0^{\sqrt a}dx\,(\Om+\OL x^6)^{-1/2} \ .
\end{equation}
Here, $\Omega_{\rm m}$ and $\Omega_\Lambda$ are the dimensionless density parameters of matter and the cosmological constant in the universe, while the contributions from relativistic particles and spatial curvature are neglected. Throughout, unless otherwise specified, we consider the values $\Om=0.3$, $\OL=0.7$~\cite{Planck:2018vyg} and normalize the scale factor to unity today, $a(\eta_0)=1$, fixing the present time to $\Ho\eta_0=3.305$ through equation~\eqref{eq:H0eta}. With this normalization $\Ho=H_0$, and we measure times and distances in units of $\Ho^{-1}$. Redshift and conformal time are then related through the scale factor alone,
\begin{equation}
    1+z(\eta)=\frac{a(\eta_0)}{a(\eta)}=\frac{1}{a(\eta)}\ ,
    \label{eq:z_of_eta}
\end{equation}
and the conformal time of a given redshift follows from equation~\eqref{eq:H0eta} with $a=1/(1+z)$. 
 
\begin{figure}[t]
    \centering
    \includegraphics[width=\linewidth]{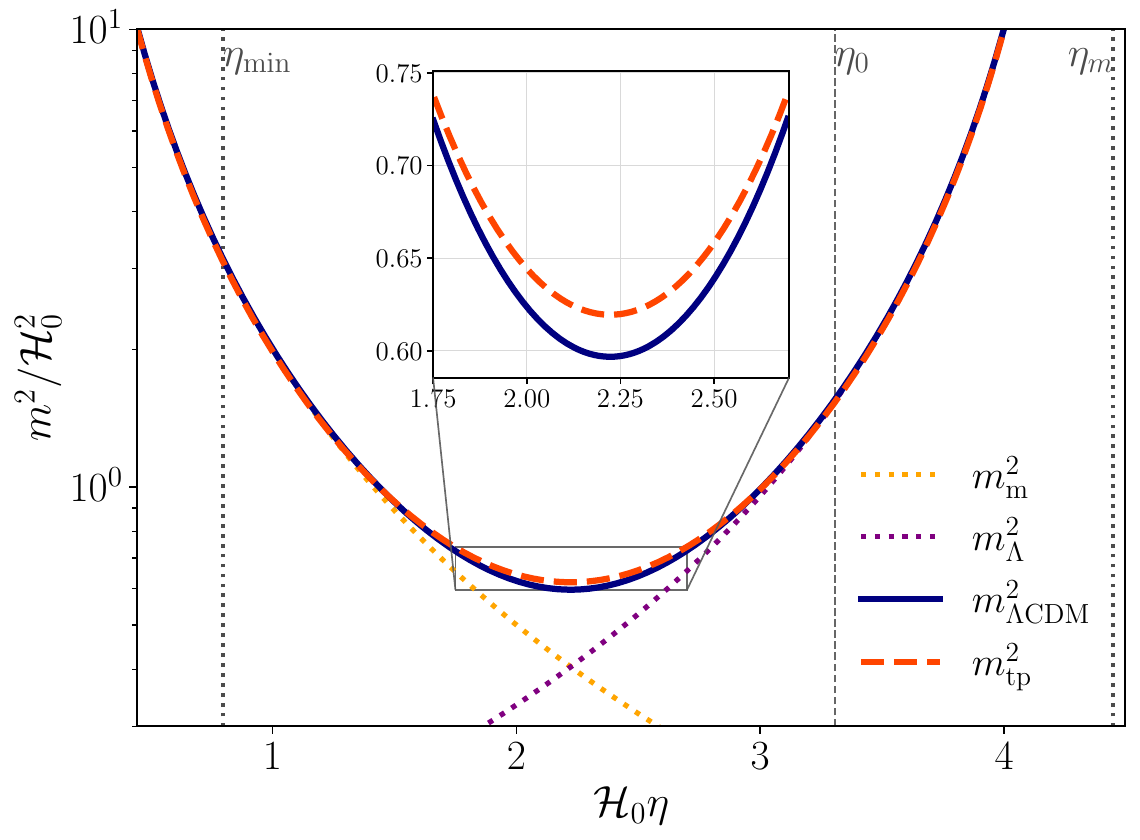}
    \caption{Ricci curvature interaction $m^2 = a^2R/6$ for pure matter $a=\Om(\Ho\eta)^2/4$ (dotted orange), pure $\Lambda$ $a=[\sqrt{\OL}\,\Ho(\eta_m-\eta)]^{-1}$ (dotted blue), $\Lambda$CDM (solid blue), and two-pole model (dashed orange). In all cases $\Omega_{\rm m} = 0.3$ and $\Omega_\Lambda =0.7$.
     Vertical lines mark the source time $\eta_{\min}$ ($z=20$, $\Ho\eta_{\min}\simeq0.8$) of the convergence test of Section~\ref{sec:numerics},  the present time $\eta_0$, and the conformal future infinity $\eta_m$.}
    \label{fig:scale_factor_comparison}
\end{figure}

A useful approximation of the concordance model is the analytic \emph{two-pole} interaction
\begin{equation}
    m^2_{\rm tp}(\eta) = 2 \left[ \frac{1}{\eta^2}+\frac{1}{(\eta_m-\eta)^2} - 0.095 H_0^2\right] \ ,
    \label{eq:m2_twopoles}
\end{equation}
built from the two limits of $\Lambda$CDM, as shown in Figure~\ref{fig:scale_factor_comparison}. The constant shift $- 0.095 H_0^2$ was introduced in Ref.~\cite{Jokela:2022rhk} from a fit to the exact $\Lambda$CDM interaction. At early times the universe is matter dominated, $a\simeq\Om(\Ho\eta)^2/4$, and $a''/a=2/\eta^2$. At late times the expansion approaches de Sitter, $a\simeq[\sqrt{\OL}\,\Ho(\eta_m-\eta)]^{-1}$, and $a''/a=2/(\eta_m-\eta)^2$. The scale factor diverges at the finite conformal time $\Ho\eta_m=4.4457$, as one obtains from equation~\eqref{eq:H0eta} setting $a\to\infty$, which is the conformal future infinity of $\Lambda$CDM. Up to a constant shift, the two-pole interaction is the sum of these two limits, which are mapped into each other by $\eta\to\eta_m-\eta$. 

In the limits $\eta\to0$ and $\eta\to\eta_m$, the two-pole interaction reproduces exactly both poles of $\Lambda$CDM. In between, the two-pole interaction underestimates the $\Lambda$CDM one for $\Ho\eta<1.46$ and for $2.98<\Ho\eta<\Ho\eta_m$, and overestimates it for $1.46<\Ho\eta<2.98$. For later reference, we report the values
\begin{center}
\begin{tabular}{c@{\hspace{1.5em}}c}
$\Ho\eta$ & $m^2_{\rm tp}/m^2_{\Lambda{\rm CDM}}-1$ \\[0.3ex]
\hline
$0.95$ & $-1.46\%$ \\
$2.22$ & $+3.80\%$ \\
$3.305$ & $-1.28\%$ \\
$3.50$ & $-1.47\%$ \\
\end{tabular}
\end{center}
and the two interactions coincide at $H_0\eta\simeq 1.46$ and $H_0\eta\simeq 2.98$.
The two-pole interaction offers good analytic control over the tail solution, and we use it to validate our solutions in exact $\Lambda$CDM.

In the next sections we reformulate the problem for the retarded propagator~\eqref{eq:green_eq} as a characteristic (Goursat) initial-value problem, giving a new perspective on the causal structure of the tail contribution. We obtain the solution in two ways: by double-null numerical marching and through a convergent double power series about the coincidence point.

\section{Goursat formulation for the tail term}
\label{sec:goursat}

\subsection{The tail as a characteristic initial-value problem}
\noindent
Inserting the decomposition \eqref{eq:G_decomp} into \eqref{eq:green_eq}, the light-cone piece alone reproduces the source term of a flat-space wave equation plus a leftover piece proportional to $m^2$,
\begin{equation}
    \left[\partial_\eta^2-\nabla_{\bfx}^2-m^2(\eta)\right]\frac{\delta(u)}{r} = 4\pi\delta^{(4)}(x-x') - m^2(\eta)\frac{\delta(u)}{r},
\end{equation}
which must be canceled by the tail term $B(x,x^\prime) \theta(u)$. Since $u$ is a characteristic (null) coordinate of the wave operator, {\emph{i.e.}}, $(\partial_\eta u)^2-|\nabla_{\bfx} u|^2=0$, the term proportional to $\delta'(u)$ vanishes. Demanding that the terms proportional to $\theta(u)$ and $\delta(u)$ cancel separately~\footnote{To be precise, one does not need to impose that the terms vanish identically everywhere. It is sufficient to demand that the term proportional to $\theta(u)$ vanishes for $u>0$ (leading to the bulk equation~\eqref{eq:bulk_eq}), and that the term proportional to $\delta(u)$ vanishes at $u=0$ (leading to the transport equation~\eqref{eq:transport_eq}).} against~\eqref{eq:green_eq} then gives two equations for $B(x,x^\prime)$, one homogeneous bulk equation inside the light cone, and one transport equation confined to the light cone:
\begin{align}
    \textbf{Bulk }(u>0): &\;\, \left[\partial_\eta^2-\nabla^2-m^2(\eta)\right]B=0\ \label{eq:bulk_eq}\\
    \textbf{Transport }(u=0): &\;\, 2\left(\partial_\eta+\partial_r+\frac1r\right)B = \frac{m^2(\eta)}{r} \ .\label{eq:transport_eq}
\end{align}
Along the outgoing ray $u=0$ ({\emph{i.e.}}, $\eta=\eta'+r$), the directional derivative $D/Dr\equiv\partial_\eta+\partial_r$ turns \eqref{eq:transport_eq} into a total derivative, $d(rB)/dr|_{u=0} = m^2(\eta)/2$. This fixes $rB$ along the ray up to its value at the vertex of the light cone, $r=0$, {\emph{i.e.}}, at the coincidence point $x=x'$, which is not determined by the equations~\eqref{eq:bulk_eq} and~\eqref{eq:transport_eq}, and requires an additional condition. We impose regularity of the tail at coincidence. Regularity of $B$ at $\mathbf x=\mathbf x'$, together with spherical symmetry, implies that $g\equiv rB$ is odd in $r$, so that on the worldline
\begin{equation}
    [rB]_{\vert r= 0} = 0 \ .
    \label{eq:regularity}
\end{equation}
If $B(x,x^\prime)$ diverged as $r\to 0$, this would yield an additional singular contribution to the Green's function on the light cone at coincidence, on top of the term $\delta(u)/r$, which already accounts for the full source. Integrating from the vertex, the tail on the light cone is
\begin{equation}
    B(\eta,\bfx;\eta',\bfx')\Big|_{u=0} = \frac{1}{2r}\int_{\eta'}^{\eta}m^2(\eta'')\,d\eta'', \qquad r=\eta-\eta' \ .
    \label{eq:B_lightcone}
\end{equation}
This is a well-posed \emph{characteristic} initial-value problem. Equation~\eqref{eq:B_lightcone} sets the boundary data for $B$ on the light cone by a single integral of $m^2$ between source and field point, while equation~\eqref{eq:regularity} imposes the regularity on the worldline. Together they define the boundary conditions on a wedge of the future or past light cone of the source $x'$. The function $B$ is then propagated from the light cone into the interior $u>0$ by the homogeneous wave equation~\eqref{eq:bulk_eq}.
When $m^2=0$, the light-cone data vanish identically and $B\equiv0$, recovering strict light cone propagation, while the interior of the light cone is filled by a non-vanishing tail for any FLRW background with $m^2\neq0$.

This is the Hadamard construction specialized to a homogeneous and isotropic background. Since $m^2(\eta)$ does not depend on $\bfx$, the problem with a point source retains full rotational symmetry about $\bfx'$ and translation invariance in $\bfx-\bfx'$, so that $B$ depends only on $(\eta,\eta',r)$. We exploit this residual symmetry to reduce the $(3+1)$-dimensional system \eqref{eq:bulk_eq}~and~\eqref{eq:B_lightcone} to a $(1+1)$-dimensional characteristic problem. Defining $g(\eta,r)\equiv rB(\eta,r)$ and using $\nabla^2B = r^{-1}\partial_r^2(rB)$ for a function of $r=|\bfx-\bfx'|$ alone, the bulk equation~\eqref{eq:bulk_eq} simplifies to the following wave equation
\begin{equation}
    \left[\partial_\eta^2-\partial_r^2-m^2(\eta)\right]g=0 \ .
    \label{eq:bulketar}
\end{equation}
Since the boundary data \eqref{eq:B_lightcone} live on the null surface $r=\eta-\eta'$, it is natural to move to null coordinates
\begin{equation}
    u=t-r,\qquad v=t+r,\qquad t\equiv\eta-\eta' \ ,
\end{equation}
in terms of which the bulk equation is
\begin{equation}
    \boxed{4g_{,uv}=m^2(\eta)\,g}\ , \qquad \eta=\eta'+\frac{u+v}{2}\ .
    \label{eq:bulk_uv}
\end{equation}
The physical domain is the wedge $u\geq0,\ v\geq u$ ({\emph{i.e.}}, $r\geq0$), bounded by the light cone $u=0$, on which $g$ is fixed by \eqref{eq:B_lightcone},
\begin{equation}
    \boxed{g(0,v) = \frac{v}{2}\,B\Big|_{u=0} = \frac12\int_{0}^{v/2}m^2(\eta'+t')\,dt'} \ ,
    \label{eq:g0v}
\end{equation}
and the source worldline $u=v$ ($r=0$), on which regularity~\eqref{eq:regularity} requires
\begin{equation}
    \boxed{g(u,u)=0\qquad\forall u } \ .
    \label{eq:worldline}
\end{equation}
This way, the original Green's function equation~\eqref{eq:green_eq} with Cauchy data on a spacelike slice~\cite{Friedlander:2010eqa,Jokela:2022rhk,Jokela:2023suz} is formulated for the tail as a \textit{Goursat problem}~\cite{goursat2013course}: a hyperbolic partial differential equation with data prescribed on two intersecting characteristic curves. Because $B(t,r)=B(t,-r)$ by spherical symmetry and $g=rB$, the exchange $r\to-r$ ({\emph{i.e.}},\ $u\leftrightarrow v$) implies the exact antisymmetry
\begin{equation}
    g(u,v) = -g(v,u) \ ,
    \label{eq:antisym}
\end{equation}
of which \eqref{eq:worldline} is the diagonal restriction. To be precise, the worldline $u=v$, {\emph{i.e.}}, $r=0$, is not a characteristic of the wave equation~\eqref{eq:bulk_uv}, whose characteristics are the lines $u={\rm const}$ and $v={\rm const}$.  However, since equation~\eqref{eq:bulk_uv} is symmetric under $u\leftrightarrow v$, the solution in the wedge is the restriction of the solution in the quadrant $u,v\geq0$ with data on its two characteristics, $g(0,v)$ of equation~\eqref{eq:g0v} and $g(u,0)=-g(0,u)$, which is antisymmetric and therefore satisfies equation~\eqref{eq:worldline}. 

The term $m^2g$ contains no derivatives and leaves the characteristic null lines $u,v={\rm const}$ of equation~\eqref{eq:bulk_uv} unchanged and identical for any expansion history. 
The light-cone front $\delta(u)/r$ and the jump of the tail to its light-cone value $B|_{u=0}$ therefore travel together, at the speed of light, along $u=0$, while the rest of the tail, $u>0$, fills the interior of the light cone and reaches a given point only after the front.

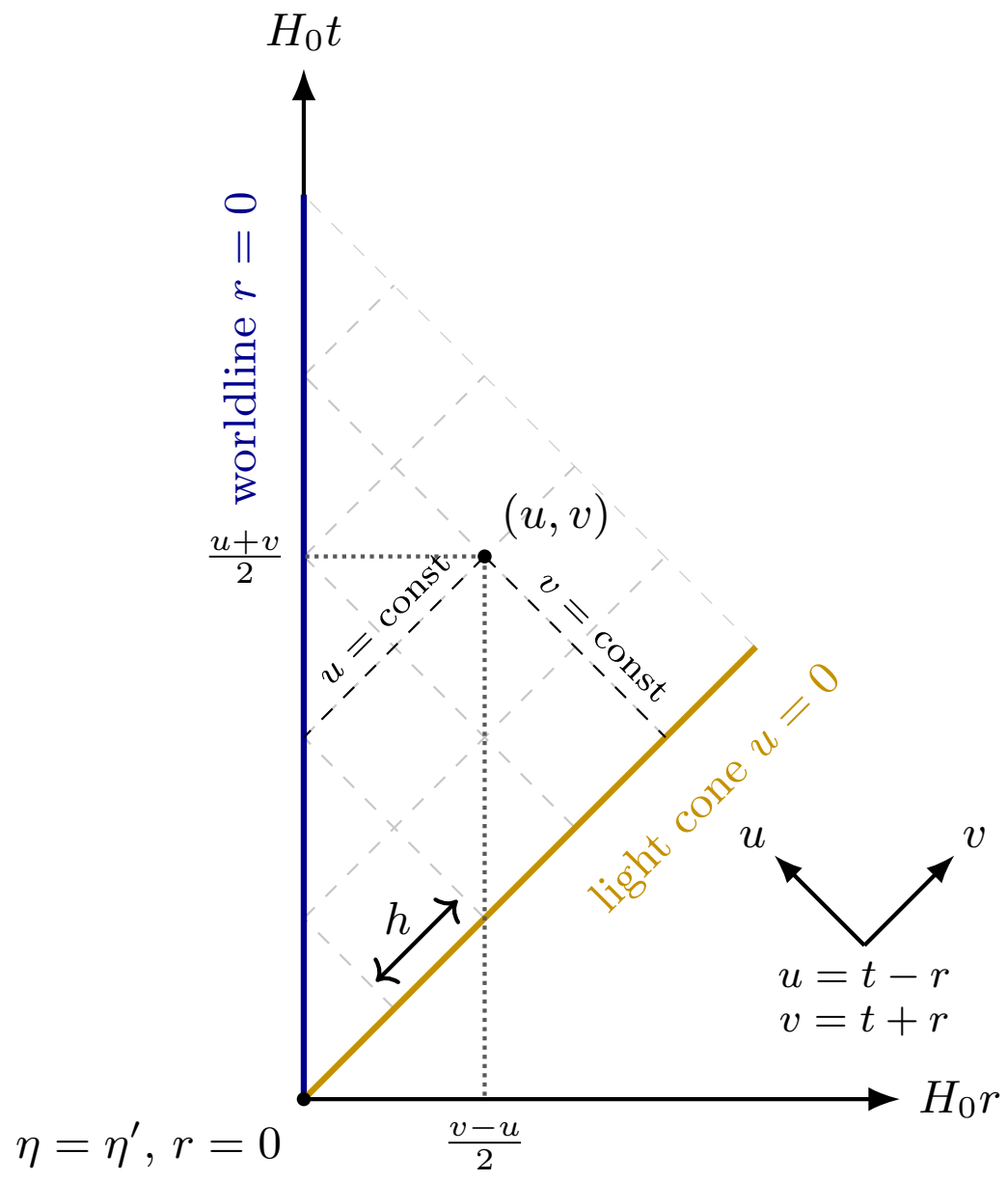
\begin{figure}[t]
\centering
\begin{tikzpicture}[scale=0.62]
  \definecolor{lcCol}{RGB}{196,146,0}
  \def\N{5}
  \foreach \i in {0,...,\N} {
    \foreach \j in {0,...,\N} {
      \pgfmathsetmacro{\X}{\j-\i}
      \pgfmathsetmacro{\Y}{\i+\j}
      \coordinate (P\i-\j) at (\X,\Y);
    }
  }
  \draw[-{Latex[length=2.2mm]}, thick] (P0-0) -- (6.6,0) node[right] {$H_0r$};
  \draw[-{Latex[length=2.2mm]}, thick] (P0-0) -- (0,11.4) node[above] {$H_0{t}$};
  \begin{scope}
    \clip (P0-0) -- (P0-5) -- (P5-5) -- cycle;
    \foreach \i in {0,...,\N} {\draw[gray!45,dashed] (P\i-\i) -- (P\i-5);}
    \foreach \j in {0,...,\N} {\draw[gray!45,dashed] (P0-\j) -- (P\j-\j);}
  \end{scope}
  \draw[very thick, blue!55!black] (P0-0) -- (P5-5);
  \node[blue!55!black, rotate=90, anchor=south] at ($(P4-4)+(-0.35,0.3)$) {worldline $r=0$};
  \draw[very thick, lcCol] (P0-0) -- (P0-5);
  \node[lcCol, rotate=45, anchor=north] at ($(P0-4)+(0.28,-0.28)$) {light cone $u=0$};

  \fill (P0-0) circle (2.2pt);
  \node[below left=1pt] at (P0-0) {$\eta=\eta',\,r=0$};

  \coordinate (F) at (P2-4);
  \draw[densely dotted, thick, gray!70!black] (F) -- (F |- P0-0);
  \draw[densely dotted, thick, gray!70!black] (F) -- (P0-0 |- F);
  \node[below, font=\footnotesize] at (F |- P0-0) {$\tfrac{v-u}{2}$};
  \node[left, font=\footnotesize] at (P0-0 |- F) {$\tfrac{u+v}{2}$};
  \draw[dashed] (F) -- (P2-2) node[pos=0.45, sloped, above, font=\scriptsize] {$u=$ const};
  \draw[dashed] (F) -- (P0-4) node[pos=0.55, sloped, above, font=\scriptsize] {$v=$ const};
  \fill (F) circle (2.2pt);
  \node[above right=0pt] at (F) {$(u,v)$};

  \coordinate (C) at (6.2,1.7);
  \draw[-{Latex[length=2mm]}, thick] (C) -- ++(1.0,1.0) node[above right=-2pt] {$v$};
  \draw[-{Latex[length=2mm]}, thick] (C) -- ++(-1.0,1.0) node[above left=-2pt] {$u$};
  \node[below, font=\footnotesize, align=center] at (C) {$u={t}-r$\\[-1pt]$v={t}+r$};

  \draw[<->, thick] ($(P1-1)+(.8,-.7)$) -- ($(P1-1)+(1.7,0.2)$);
  \node[right] at ($(P0-0)!0.5!(P1-1)+(0.7,1.)$) {$h$};
\end{tikzpicture}
\caption{The Goursat wedge $u\geq0$, $v\geq u$ in which the tail term $g(u,v)=rB$ is determined. Data are given on the two boundaries of the wedge, the light cone $u=0$~\eqref{eq:g0v}, and the source worldline $r=0$ ($u=v$)~\eqref{eq:worldline}. The dashed grid marks the uniform spacing $h$ of the double-null grid used for the marching scheme of Section~\ref{sec:numerics}, on which $g$ is obtained at every point $(u,v)$ inside the wedge; a generic field point $(u,v)$ is shown for reference. The axes are the comoving distance $r$ and the time ${t}=\eta-\eta'$ after the source, in arbitrary units, with the worldline along the ${t}$ axis. The null coordinates $u={t}-r$ and $v={t}+r$ increase along the two diagonals (arrows), so that the field point lies at the intersection of its characteristics $u=$ const and $v=$ const (dashed), at $r=(v-u)/2$ and ${t}=(u+v)/2$.}
\label{fig:goursat_wedge}
\end{figure}

In Section~\ref{sec:scattering}, we discuss a model that interprets the tail as a wave scattered off the background curvature~\cite{Price:1971fb,Gundlach:1993tp,Faraoni:1991xe,Faraoni:1999us}. In this picture, the scattered waves travel along broken null-paths weighted by the curvature interaction $m^2$, thus delaying their propagation unless the scattering occurs along the ray itself, between source and field point. 
The tail on the light cone~\eqref{eq:B_lightcone} is fixed by the integral of $m^2$ along the ray, {\emph{i.e.}}, by the curvature crossed by the front, and is the value of the tail immediately behind it. In the scattering picture it is the wake of the front: only the waves scattered along the ray itself reach the field point together with the front.

The worldline condition~\eqref{eq:worldline} states that there is no source at $r=0$ other than the delta function of the light-cone term, {\emph{i.e.}}, no singular contribution to $B$ as $r\to 0$. In the $(1+1)$-dimensional formulation, the worldline acts as a reflecting boundary on which $g$ vanishes, like the fixed end of a string: a wave incoming along $v$ is reflected into an outgoing one along $u$ with opposite sign, which is the antisymmetry~\eqref{eq:antisym}. 

Integrating equation~\eqref{eq:bulk_uv} over the characteristic rectangle $0\leq u'\leq u$, $0\leq v'\leq v$, with equation~\eqref{eq:antisym} and $g(0,0)=0$, gives the exact integral form
\begin{align}
g(u,v)=&g(0,v)-g(0,u) \nonumber \\
& +\frac14\int_0^u\!\!\int_0^v m^2(\eta'' )\,g(u',v')\,du'dv' \ ,
\label{eq:g_integral}
\end{align}
with $\eta'' =\eta'+\frac{u'+v'}{2}$.
The first two terms carry the light-cone data along the null lines, as for $m^2=0$. The integral accumulates the contribution of $m^2$ over the whole rectangle in the past of $(u,v)$. In Section~\ref{sec:scattering} we will discuss a model that identifies the tail as the collection of all broken null-paths along which a signal can travel from the source to the observer, each vertex representing a scattering off the background curvature potential, weighted by $m^2$. In this scattering picture, the first two terms of equation~\eqref{eq:g_integral} are the waves scattered once, while the integral collects the waves scattered at least twice.

Although the Goursat formulation efficiently poses the problem in terms of $g = rB$, the map $g\mapsto B=g/r$ appears singular as $r\to 0$ $(u=v)$. This apparent singularity is removed by condition~\eqref{eq:worldline}. This is just a coordinate singularity which we can remove at any time after the source. Since $g$ vanishes as $r\to0$ and is antisymmetric, on the worldline $\partial_ug=-\partial_vg$, and $g/r\to\partial_rg=2\partial_vg$. Differentiating the integral form~\eqref{eq:g_integral} in $v$, we then find
\begin{align}\label{eq:B_worldline}
B(u,u) =& \frac12m^2\Big(\eta'+\frac u2\Big) \nonumber\\
&+\frac12\int_0^u m^2\Big(\eta'+\frac{u'+u}{2}\Big)\,g(u',u)\,du' \ ,
\end{align}
finite for any bounded curvature potential, and equal to $m^2(\eta')/2$ at the corner $u=0$. In the scattering picture, the first term arises from single scattering off the curvature, whose window of scattering times collapses onto the worldline at the intermediate time $\eta'+u/2=(\eta+\eta')/2$, while the integral collects the multiple scatterings.

The system~\eqref{eq:bulk_uv}--\eqref{eq:worldline}, or equivalently equation~\eqref{eq:g_integral}, is exact, and we solve it in three ways. First we derive two approximations. (i) The recursive Neumann (Born) series, Section~\ref{sec:neumann}, expands $g$ in powers of the coupling $m^2$. In the multiple-scattering picture, each order adds one scattering off the curvature, weighted by $m^2$. These are functions on the whole wedge, and the series converges everywhere inside the light cone. (ii) The double series, Section~\ref{sec:series}, instead expands $g$ in powers of the distance $(u,v)$ from the coincidence point $u=v=0$. The series coefficients follow analytically from the Taylor expansion of $m^2$, but the series converges only within a finite radius around the corner. Truncated at a finite order, both are approximations to the exact solution. (iii) Finally, we discuss exact numerical solutions in Section~\ref{sec:numerics}. We use a double-null marching scheme to solve for $g$ numerically on a discrete grid. Its update integrates equation~\eqref{eq:bulk_uv} over one cell, adding $\frac14m^2g\,\Delta u\Delta v$, and summing it over the cells below a node reproduces equation~\eqref{eq:g_integral}, with the integral replaced by a sum over cells.

\subsection{Recursive Neumann (Born) series}
\label{sec:neumann}
\noindent
Equation~\eqref{eq:g_integral} is a linear Volterra integral equation of the second kind for $g$, where the integration only samples points $(u',v')$ in the past of $(u,v)$. It can therefore be solved by iteration, in powers of the coupling $m^2$. Writing $g=\sum_{n\geq1}g_n$, the first order collects the light-cone data, and each further order is sourced by the previous one,
\begin{align}
g_1(u,v)&=g(0,v)-g(0,u)\label{eq:neumann_1}\\
g_{n+1}(u,v)&=\frac14\int_0^u\!\!\int_0^v m^2(\eta'')\,g_n(u',v')\,du'dv'\ ,\label{eq:neumann_n}
\end{align}
with $\eta''=\eta'+\frac{u'+v'}{2}$ as in equation~\eqref{eq:g_integral}. Since the light-cone data~\eqref{eq:g0v} are themselves linear in $m^2$, the $n$-th term is of order $(m^2)^n$. Equivalently, $g_1$ solves the bulk equation~\eqref{eq:bulk_uv} with $m^2=0$ and carries all the light-cone data, while $g_{n+1}$ solves $4\partial_u\partial_vg_{n+1}=m^2g_n$ with vanishing data on the light cone. Every order also satisfies the worldline condition~\eqref{eq:worldline}.

For a source at any $\eta'>0$, with $m^2$ bounded on the domain, the series converges throughout the light cone.
If $|m^2|\leq M$ and $|g_1|\leq A$ on the rectangle $[0,u]\times[0,v]$, one can verify by induction using equation~\eqref{eq:neumann_n} that
\begin{equation}
|g_n(u,v)|\leq A\,\frac{(Muv/4)^{n-1}}{[(n-1)!]^2}\ ,
\label{eq:neumann_bound}
\end{equation}
so that the whole series converges on any bounded part of the wedge. This is the Born series of the Green's function. In Section~\ref{sec:scattering}, we describe its interpretation as a sum over multiple broken null-paths, each vertex representing a scattering off the background curvature, so that $g_n$ collects the paths with $n$ scatterings. 

The recursive Neumann series has been used in Ref.~\cite{Compere:2026xwy} as a valid method to solve the tail. We stress that, besides being an approximation, the Neumann series is computationally more expensive than solving the exact problem~\eqref{eq:bulk_uv}--\eqref{eq:worldline}, since each term is computed with the same double-null marching as the full solution, but with the source $m^2g_n$ in place of $m^2g$ (see Section~\ref{sec:numerics}). A quantitative comparison of the accuracy and of the computational cost of the two methods is given in Appendix~\ref{app:literature} (Figure~\ref{fig:CPU_time}). The relevance of the Neumann series in this article is limited to the interpretative picture it offers for the tail in terms of multiple scatterings, while the direct numerical integration of the system~\eqref{eq:bulk_uv}--\eqref{eq:worldline} is both numerically cheaper and more accurate.

\subsection{Double-series expansion near the corner}
\label{sec:series}
\noindent
If the interaction potential $m^2$ is analytic at $\eta'$, it is possible to expand $g$ in powers of the null coordinates $u$ and $v$ about the coincidence point $u=v=0$.
Since $\eta'\in(0,\eta_m)$ lies away from any singularity, $m^2(\eta'+\tfrac{u+v}{2})$ near the corner $(0,0)$ admits a Taylor expansion in the time difference $t=\eta-\eta'=\tfrac{u+v}{2}$,
\begin{equation}
    m^2(\eta'+t) = \sum_{p\geq0}c_p\,t^p \ , \quad \quad c_p=\frac{1}{p!}\,\frac{d^p m^2}{d\eta^p}\Big|_{\eta'} \ .
    \label{eq:cp_def}
\end{equation}
We then posit the double Taylor Ansatz
\begin{equation}
    g(u,v) = \sum_{m,n\geq0}a_{mn}\,u^mv^n \ .
    \label{eq:Ansatz}
\end{equation}
We call $N=m+n$ the total degree of the term $a_{mn}u^mv^n$, and denote by $g_{[N]}$ the series truncated at total degree $N$, {\emph{i.e.}}, the sum~\eqref{eq:Ansatz} restricted to $m+n\leq N$.
The Ansatz~\eqref{eq:Ansatz} is the Born series of Section~\ref{sec:neumann}, with each order expanded in powers of $u$ and $v$ and regrouped by total degree. Since each iteration of equation~\eqref{eq:neumann_n} integrates over both $u'$ and $v'$, the order $g_\ell$ starts at $u^{\ell-1}v^{\ell-1}$: the first order, $g_1=g(0,v)-g(0,u)$, gives only the light-cone coefficients $a_{0n}$ and $a_{n0}$, and a coefficient $a_{mn}$ receives contributions only from the orders $\ell\leq\min(m,n)+1$. Each coefficient is therefore a finite sum over scattering orders, while each scattering order contributes to infinitely many coefficients. In Appendix~\ref{app:riemann} we make this correspondence rigorous, deriving the recursion below from both the Born series~\eqref{eq:neumann_n} and the Riemann function of the Goursat problem.

Substituting the expansion~\eqref{eq:cp_def} into the light-cone data~\eqref{eq:g0v} and integrating term by term, we obtain
\begin{equation}
    g(0,v) = \frac12\int_0^{v/2}\sum_pc_p\,t'^p\,dt' = \sum_{j\geq1}a_{0,j}\,v^j \ , 
    \label{eq:a0j}
\end{equation}
with
\begin{equation}\label{a_lc}
    a_{0,j} = \frac{c_{j-1}}{j\,2^{j+1}}, \qquad  a_{0,0} = 0 \ .
\end{equation}
This holds for any smooth and homogeneous interaction $m^2(\eta)$, with the potential entering only through its own Taylor coefficients $c_p$.

Expanding the binomial in the interaction~\eqref{eq:cp_def}, we can write $m^2(\eta'+\tfrac{u+v}{2})=\sum_{k,l\geq 0}c_{k,l}u^kv^l$ with $c_{k,l}=c_{k+l}2^{-(k+l)}\binom{k+l}{k}$. 
Multiplying by the Ansatz~\eqref{eq:Ansatz} via the Cauchy product and matching term by term in \eqref{eq:bulk_uv} gives the recurrence
\begin{widetext}
\begin{equation}
       a_{\alpha+1,\beta+1} = \frac{1}{4(\alpha+1)(\beta+1)}\sum_{m=0}^\alpha\sum_{n=0}^\beta a_{mn}\,\frac{c_{\alpha+\beta-m-n}}{2^{\alpha+\beta-m-n}}\binom{\alpha+\beta-m-n}{\alpha-m} \ .
    \label{eq:interior}
\end{equation}
\end{widetext}

Equations \eqref{eq:a0j} and \eqref{eq:interior} determine $a_{mn}$ for all $m,n\geq1$ and the row $a_{0,j}$, but leave the column $a_{j,0}$ undetermined. These are fixed by the worldline condition~\eqref{eq:worldline}, $g(u,u)=\sum_{N}u^N\sum_{m+n=N}a_{mn}=0$, which gives the closure condition
\begin{equation}
    a_{N,0} = -\sum_{k=0}^{N-1}a_{k,N-k}\ .
    \label{eq:closure}
\end{equation}
We apply equations~\eqref{a_lc}--\eqref{eq:closure} order by order in the total degree $N$, so that the closure at degree $N$ feeds directly into the interior recursion at degree $N+1$. The resulting coefficients are antisymmetric, $a_{mn}=-a_{nm}$, as required by~\eqref{eq:antisym}.

On the worldline $r\to0$, the map $g\mapsto B=g/r$ has a coordinate singularity, and $B=\lim_{r\to0}g(u,v)/r$ is given by equation~\eqref{eq:B_worldline}. The series then reads
\begin{equation}
    B(u,u) = 2\,\partial_vg(u,v)\big|_{v=u} = 2\sum_{m\geq0,\,n\geq1}n\,a_{m,n}\,u^{m+n-1}\ ,
    \label{eq:coincidence}
\end{equation}
a finite quantity given directly by the series coefficients.

In the double series expansion, all information about the curvature potential of the background is inside the coefficients $c_p$ defined in equation~\eqref{eq:cp_def}. In particular, in a homogeneous universe, these encode the evolution of the scale factor $a(\eta)$, as discussed below.  

\subsubsection{\texorpdfstring{Expansion coefficients in $\Lambda$CDM}{Expansion coefficients in $\Lambda$CDM}}
\noindent
For the two-pole approximation, the coefficients $c_p$ of the series~\eqref{eq:cp_def} follow from closed-form differentiation of~\eqref{eq:m2_twopoles}, resulting in
\begin{equation}\label{eq:cp_tp}
        c_p^{\rm tp} = 2\left[ (-1)^p\frac{p+1}{(\eta^\prime)^{p+2}} + \frac{p+1}{(\eta_m -\eta^\prime)^{p+2}} - 0.095H_0^2\,\delta_{p0} \right]\ .
\end{equation}
In the concordance model~\eqref{eq:m2_LCDM} instead, the interaction depends on time only through the scale factor, $m^2=H_0^2\big(2\OL\,a^2+\tfrac12\Om/a\big)$. Writing the Taylor series of the scale factor about the source time as
\begin{equation}
    a(\eta'+t)=\sum_{k\geq0}\alpha_k t^k \ ,
    \label{eq:alpha_def}
\end{equation}
the Taylor coefficients of the interaction follow analytically from those of $a$,
\begin{equation}
    c_p^{\Lambda\rm CDM } = H_0^2\left(2\OL\,(a^2)_p + \tfrac12\Om\,(1/a)_p\right) \ ,
    \label{eq:cp_exact}
\end{equation}
where $(a^2)_p=\sum_{i=0}^p\alpha_i\alpha_{p-i}$ is the Cauchy product, and $(1/a)_0=1/\alpha_0$ and $(1/a)_p=-\alpha_0^{-1}\sum_{i=1}^p\alpha_i\,(1/a)_{p-i}$ for $p\geq1$ are the coefficients of the reciprocal series. The problem is thus reduced to the coefficients $\alpha_k$. The scale factor $a(\eta)$ is known only numerically, by inverting equation~\eqref{eq:H0eta}, and computing the coefficients $\alpha_k$ by finite differences is unstable, since the numerical error grows with the order of the derivative. Instead we exploit the fact that $a(\eta)$ solves the first-order Friedmann equation
\begin{equation}
    \frac{da}{d\eta} = H_0\sqrt{\Om\,a+\OL\,a^4},
    \label{eq:friedmann_ode}
\end{equation}
whose right-hand side is again a function of $a$ alone. The Taylor coefficients of the argument of the square root, $\Om\,a+\OL\,a^4=\sum_kA_kt^k$, are
\begin{equation}
    A_k = \Omega_m  \alpha_k + \Omega_\Lambda  \sum_{j=0}^k \sum_{i=0}^j \sum_{\ell = 0}^{k-j} \alpha_i \alpha_{j-i}  \alpha_{\ell}\alpha_{k-j-\ell} \ ,
    \label{eq:Ak}
\end{equation}
and those of the square root, $\sqrt{\Om\,a+\OL\,a^4}=\sum_kw_kt^k$, follow from $w^2=A$ order by order,
\begin{equation}
    w_0 = \sqrt{A_0}, \quad 
    w_k = \frac{A_k - \sum_{i=1}^{k-1} w_i w_{k-i}}{2 w_0} \quad (k\ge 1) \ .
    \label{eq:sqrt_series}
\end{equation}
Matching the coefficients of $t^k$ on the two sides of equation~\eqref{eq:friedmann_ode} gives the forward recursion
\begin{equation}
    \alpha_{k+1} = \frac{H_0\,w_k}{k+1} \ ,\qquad k=0,1,2,\dots \ .
    \label{eq:alpha_recursion}
\end{equation}
This generates the Taylor series of $a(\eta)$ about $\eta'$ to arbitrary order from the single value $\alpha_0 \equiv a(\eta^\prime)$, computed from equation~\eqref{eq:H0eta}, and, through equation~\eqref{eq:cp_exact}, the coefficients $c_p^{\Lambda\rm CDM}$ of the interaction.
With $c_p$ in hand for either background, the expressions~\eqref{eq:a0j}--\eqref{eq:closure} can be applied identically for the two-pole or $\Lambda$CDM interactions by supplying coefficients~\eqref{eq:cp_tp} or~\eqref{eq:cp_exact}, respectively.

\subsubsection{Estimate of the truncation error}
\noindent
The truncation error of the series is controlled analytically.
The Taylor series~\eqref{eq:cp_def} of $m^2_{\Lambda{\rm CDM}}$ about $\eta'$, with coefficients~\eqref{eq:cp_exact}, converges up to the nearest singularity of $m^2_{\Lambda{\rm CDM}}$. For the purpose of this section, we will assume that the nearest singularity is the pole at $\eta=0$, at distance $\eta'$, while the pole at $\eta_m$ is at distance $\eta_m-\eta'$. Near this pole, the $\Lambda{\rm CDM}$ interaction is given by the matter-era one, $m^2_{\rm md}=2/\eta^2$, up to a correction that is analytic at $\eta=0$.
For $m^2_{\rm md}$, one can show that the tail is $B_{\rm md}=1/(\eta\eta')$ everywhere inside the light cone, and its truncated series can be summed explicitly. With $\eta=\eta'+(u+v)/2$ and $x\equiv{t}/\eta'=(u+v)/(2\eta')$,
\begin{equation}
    g_{\rm md}=r B_{\rm md}=\frac{v-u}{2\eta'^2}\,\frac{1}{1+x}=\frac{v-u}{2\eta'^2}\sum_{k=0}^{\infty}(-x)^k\ ,
\end{equation}
where the $k$-th term is a homogeneous polynomial of degree $k+1$ in $(u,v)$. Truncating at total degree $N$ leaves the remainder $g_{\rm md}-g_{{\rm md},N}=(-x)^N g_{\rm md}$,
so that the relative error of the truncated matter-era series is $({t}/\eta')^N$, independent of $r$. Assuming that the pole dominates the truncation error, $|\Delta B_{[N]}|\simeq({t}/\eta')^N B_{\rm md}$, with $\Delta B_{[N]}\equiv B-g_{[N]}/r$, and relative to the $\Lambda$CDM tail
\begin{equation}\label{eq:series_error}
    \left|\frac{\Delta B_{[N]}}{B}\right|\simeq\left(\frac{{t}}{\eta'}\right)^N\frac{1}{\eta\eta'B}\ .
\end{equation}
The error decreases with $N$ for ${t}<\eta'$ and grows for ${t}>\eta'$. At ${t}=\eta'$ it is independent of $N$ and of order one.
Outside the radius of convergence the truncated series diverges from the marched solution, and only the numerical solution can be trusted.

\vspace{0.5cm}

Within its radius of convergence, the series coefficients~\eqref{eq:interior} follow analytically from the Taylor coefficients~\eqref{eq:cp_exact}, without any integration or grid. This makes the double series expansion an independent check of the exact numerical integration, with which it shares only the light-cone data. Both the double series and the recursive Neumann series are used independently to validate the numerical integration of the exact Goursat problem~\eqref{eq:bulk_uv}--\eqref{eq:worldline}, which is one of the main results of this article. 

\subsection{Numerical strategy}
\label{sec:numerics}

\begin{figure}[t]
\centering
\begin{tikzpicture}[scale=1.0, every node/.style={font=\small}]
  \definecolor{lcCol}{RGB}{196,146,0}
  \definecolor{wlCol}{RGB}{33,102,172}
  \definecolor{ord1}{RGB}{199,233,192}
  \definecolor{ord2}{RGB}{116,196,118}
  \definecolor{ord3}{RGB}{35,139,69}

  \def\N{3}
  \foreach \i in {0,...,\N} {
    \foreach \j in {0,...,\N} {
      \pgfmathsetmacro{\X}{\j-\i}
      \pgfmathsetmacro{\Y}{\i+\j}
      \coordinate (P\i-\j) at (\X,\Y);
    }
  }
  \fill[blue!8] (P0-2) -- (P0-3) -- (P1-3) -- (P1-2) -- cycle;
  \begin{scope}
    \clip (P0-0) -- (P0-3) -- (P3-3) -- cycle;
    \foreach \i in {0,...,\N} {\draw[gray!35,thin] (P\i-\i) -- (P\i-3);}
    \foreach \j in {0,...,\N} {\draw[gray!35,thin] (P0-\j) -- (P\j-\j);}
  \end{scope}
  \draw[thick] (P0-2) -- (P0-3) -- (P1-3) -- (P1-2) -- cycle;
  \draw[thick, wlCol] (P0-0) -- (P3-3);
  \draw[thick, lcCol] (P0-0) -- (P0-3);

  \draw[green!35!black, thick, dashed] (P0-0) -- (P1-2) -- (P1-3);
  \draw[-{Latex[length=2mm]}, green!35!black, thick, dashed] (P1-3) -- (P2-3);

  \foreach \j in {0,...,\N} {\fill[lcCol] (P0-\j) circle (2.6pt);}
  \foreach \i in {1,...,\N} {\fill[wlCol] (P\i-\i) circle (2.6pt);}
  \fill[ord1] (P1-2) circle (2.6pt);
  \fill[ord2] (P1-3) circle (2.6pt);
  \fill[ord3] (P2-3) circle (2.6pt);

  \node[font=\scriptsize] at ($(P1-2)+(0.28,0.05)$) {\textbf{1}};
  \node[font=\scriptsize] at ($(P1-3)+(0.22,0.17)$) {\textbf{2}};
  \node[font=\scriptsize] at ($(P2-3)+(0.28,0.05)$) {\textbf{3}};

  \node[below left=1pt] at (P0-0) {$g_{0,0}$};
  \node[left=3pt] at (P2-2) {$g_{2,2}$};

  \node[wlCol, rotate=90, anchor=south] at ($(P1-1)+(-0.32,0)$) {worldline $g_{i,i}=0$};
  \node[lcCol, rotate=45, anchor=north] at ($(P0-1)+(0.22,-0.22)$) {light cone $g_{0,j}$};

  \begin{scope}[shift={(3.55,1.35)}, every node/.style={font=\scriptsize, anchor=west}]
    \fill[lcCol] (0,0.7) circle (2.4pt); \node at (0.2,0.7) {light-cone data $g_{0,j}$};
    \fill[wlCol] (0,0.35) circle (2.4pt); \node at (0.2,0.35) {worldline data $g_{i,i}=0$};
    \fill[ord1] (0,0) circle (2.4pt); \node at (0.2,0) {sweep $1$: $g_{1,2}$};
    \fill[ord2] (0,-0.35) circle (2.4pt); \node at (0.2,-0.35) {sweep $2$: $g_{1,3}$};
    \fill[ord3] (0,-0.7) circle (2.4pt); \node at (0.2,-0.7) {sweep $3$: $g_{2,3}$};
  \end{scope}

  \begin{scope}[shift={(4.85,3.85)}, scale=1.25]
    \coordinate (ZB) at (0,0);
    \coordinate (ZL) at (-1,1);
    \coordinate (ZR) at (1,1);
    \coordinate (ZT) at (0,2);
  \end{scope}
  \draw[gray!55] (P0-2) -- (ZB);
  \draw[gray!55] (P1-3) -- (ZT);
  \fill[blue!8] (ZB) -- (ZR) -- (ZT) -- (ZL) -- cycle;
  \draw[thick] (ZB) -- (ZR) -- (ZT) -- (ZL) -- cycle;
  \draw[-{Latex[length=2mm]}, gray!60] (ZB) -- (ZT);
  \draw[-{Latex[length=2mm]}, gray!60] (ZL) -- (ZT);
  \draw[-{Latex[length=2mm]}, gray!60] (ZR) -- (ZT);
  \fill (ZB) circle (2.4pt);
  \fill (ZL) circle (2.4pt);
  \fill (ZR) circle (2.4pt);
  \fill[red!75!black] (ZT) circle (2.8pt);
  \node[below] at (ZB) {$g_{i,j}$};
  \node[left] at (ZL) {$g_{i+1,j}$};
  \node[right] at (ZR) {$g_{i,j+1}$};
  \node[above] at (ZT) {$g_{i+1,j+1}$};
  \draw[-{Latex[length=1.6mm]}] ($(ZB)+(0.75,0.05)$) -- ++(0.5,0.5) node[right] {$v$};
  \draw[-{Latex[length=1.6mm]}] ($(ZB)+(-0.75,0.05)$) -- ++(-0.5,0.5) node[left] {$u$};
\end{tikzpicture}
\caption{Graphical representation of the double-null marching of equation~\eqref{eq:marching} on a small block of the grid of Figure~\ref{fig:goursat_wedge}. Light-cone data $g_{0,j}$ (yellow) and worldline data $g_{i,i}=0$ (blue) seed the sweep, and each interior point is filled from the three already-known corners of its cell. The numbers and the green dashed path trace the three sweeps needed to advance from the corner $g_{0,0}$ through row $i=1$ ($g_{1,2}$, then $g_{1,3}$) into row $i=2$ ($g_{2,3}$). Top right: the shaded cell, here with $i=0$ and $j=2$, enlarged. The three known values $g_{i,j}$, $g_{i+1,j}$, and $g_{i,j+1}$ on its lower edges determine the new value $g_{i+1,j+1}$ at its upper vertex.}
\label{fig:march_sweep}
\end{figure}
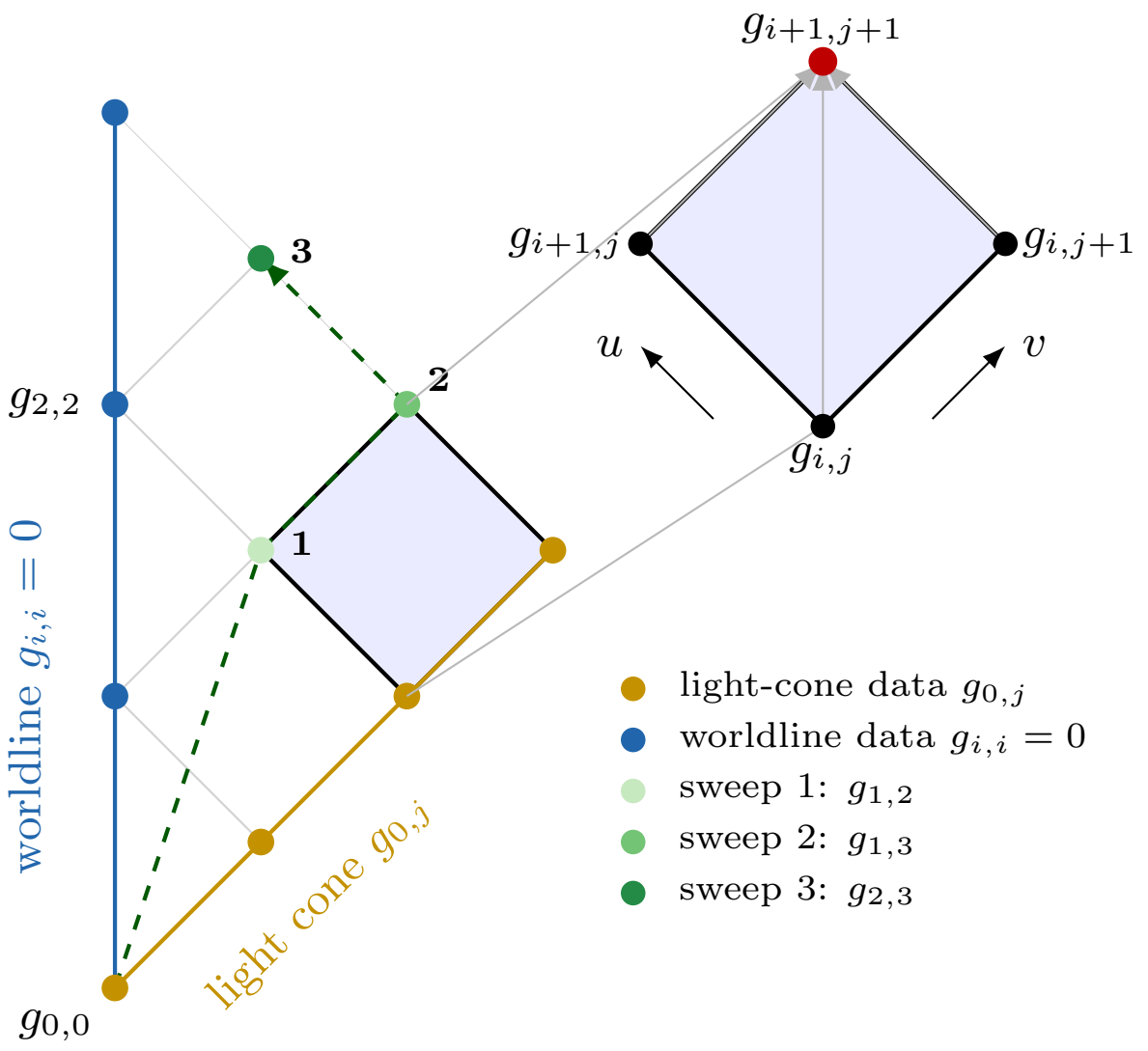

\noindent
In this section we collect the numerical methods used to perform the direct integration of the Goursat problem~\eqref{eq:bulk_uv}--\eqref{eq:worldline}.
We define a uniform grid in the null coordinates, $u_i=ih$ and $v_j=jh$, restricted by spherical symmetry to the wedge $j\geq i$ ($r\geq0$) of Figure~\ref{fig:goursat_wedge}, and to $i+j\leq N_{\rm grid}$. The time axis is bounded by ${t}\leq N_{\rm grid}h/2$ after the source. The geometry of the discretized grid is made clearer in Figure~\ref{fig:march_sweep}.  We write $g_{i,j}\equiv g(u_i,v_j)$. The row $i=0$ lies on the light cone, and the diagonal $i=j$ on the worldline. Four neighboring nodes, $(i,j)$, $(i+1,j)$, $(i,j+1)$, and $(i+1,j+1)$, bound a cell of the grid, a square of side $h$ in the $(u,v)$ plane, which appears as a diamond in the $(r,{t})$ plane, with its earliest vertex at $(i,j)$ and its latest at $(i+1,j+1)$.

\subsubsection{Double-null marching.} 
\noindent
We solve the Goursat problem~\eqref{eq:bulk_uv}--\eqref{eq:worldline} by the standard second-order double-null (``diamond-cell'') scheme~\cite{Gundlach:1993tp} on the grid defined above. Higher-order variants of the same construction are used, for example, in black-hole perturbation theory~\cite{OToole:2021crw}. Integrating \eqref{eq:bulk_uv} over an elementary diamond cell and using the midpoint rule for the right-hand side gives the explicit update
\begin{align}
    g_{i+1,j+1} =& g_{i,j+1}+g_{i+1,j}-g_{i,j} \nonumber\\
    &+ \frac{h^2}{8}m^2(\eta_{\rm mid})\big(g_{i,j+1}+g_{i+1,j}\big) \ ,
    \label{eq:marching}
\end{align}
with local truncation error $\mathcal{O}(h^4)$, where $\eta_{\rm mid}=\eta'+\tfrac h2(1+i+j)$ is the conformal time at the cell center. The integration starts near the corner from the light cone with the discretized boundary data
\begin{equation}\label{g_0j}
    g_{0,j} = \frac12\int_0^{jh/2}m^2(\eta'+t')\,dt', \qquad g_{i,i}=0 \ ,
\end{equation}
and proceeds inward toward the worldline and outward toward increasing time. 
In $\Lambda$CDM, the interaction $m^2$ of equation~\eqref{eq:m2_LCDM} is an explicit function of the scale factor, while $a(\eta)$ is known only implicitly. We therefore change the integration variable from $t'$ to $a$, with $dt'=da/a'$ and $a'$ given by the Friedmann equation~\eqref{eq:friedmann_ode}, leading to
\begin{equation}
g_{0,j}=\frac{H_0}{2}\int_{a(\eta')}^{a(\eta'+jh/2)}\frac{2\OL\,a^2+\Om/(2a)}{\sqrt{\Om\,a+\OL\,a^4}}\,da \ .
\label{eq:g0j_a}
\end{equation}
The integrand is a smooth function of $a$, and only the endpoints require the scale factor explicitly at the light-cone nodes. We accumulate the integral node by node, evaluating each piece with Gauss-Legendre quadrature, and give the boundary values of the march on the light-cone row of the grid.

The marching formula~\eqref{eq:marching} is illustrated in Figure~\ref{fig:march_sweep}.
The enlarged cell at the top right shows the elementary update, and the grid shows how it propagates across a small block. The boundary data are the light-cone row $g_{0,j}$ and the worldline diagonal $g_{i,i}=0$~\eqref{eq:worldline}. Every other point is filled by repeated application of \eqref{eq:marching}, one unit cell at a time, sweeping row by row in $i$ and, within each row, left to right in $j$ starting from the diagonal. For example, reaching $g_{2,3}$ from the corner $g_{0,0}$ takes three such sweeps. 
The update~\eqref{eq:marching} requires $m^2$ at the cell centers, $\eta_{\rm mid}=\eta'+\tfrac h2(1+i+j)$, and the light-cone data~\eqref{eq:g0j_a} require values of the scale factor along the light cone up to the node $(0,j)$, at the conformal time $\eta'+jh/2$. Since $\eta=\eta'+(u+v)/2$, all these times lie on the lattice $\eta'+kh/2$, with $k=j$ at the light-cone nodes, and $k=i+j+1$ at the cell centers. 
We evaluate $\eta(a)$ through the quadrature of equation~\eqref{eq:H0eta} using Gauss--Legendre panels. Since the integrand is smooth, each value is exact to machine precision. We invert it at each point of the lattice, solving $\eta(a)=\eta'+kh/2$ for $a$ with a Newton iteration, safeguarded by bisection, with $d\eta/da=1/a'$ and $a'$ given by the Friedmann equation~\eqref{eq:friedmann_ode}. Each step of the Newton iteration evaluates the quadrature once.
From $a$ we obtain the curvature potential $m^2=a''/a$ through equation~\eqref{eq:m2_LCDM} analytically. 

Once $g_{i,j}$ is known, we reconstruct $B(u_i,v_j)=2g_{i,j}/(v_j-u_i)$ away from the diagonal. On the diagonal $i=j$ ($r=0$), we use the exact relation~\eqref{eq:B_worldline}, which requires no numerical derivative.  We evaluate its integral with the composite Simpson rule, running along the line $v=u_i$ of the grid, over the nodes $(i',i)$ with $i'=0,\dots,i$, from the light cone to the worldline. We verified that this agrees with the derivative $2\partial_vg$ computed on the grid with a sixth-order one-sided finite-difference stencil. 

Unless otherwise stated, the numerical integration is performed over the half wedge $0\leq u\leq v$, $u+v\leq N_{\rm grid}h$, {\emph{i.e.}}, $0\leq r\leq{t}\leq N_{\rm grid}h/2$, on a grid of nodes $u_i=ih$, $v_j=jh$, with $N_{\rm grid}=2500$ and step $h=0.002$ (in units where $H_0=1$). The grid thus reaches $H_0{t}=2.5$. For a source at time $H_0\eta' = 1$, this implies that the latest observation time on the grid is $H_0\eta=3.5$, slightly beyond the present time.

\begin{figure}[t]
    \centering
    \includegraphics[width=0.99\linewidth]{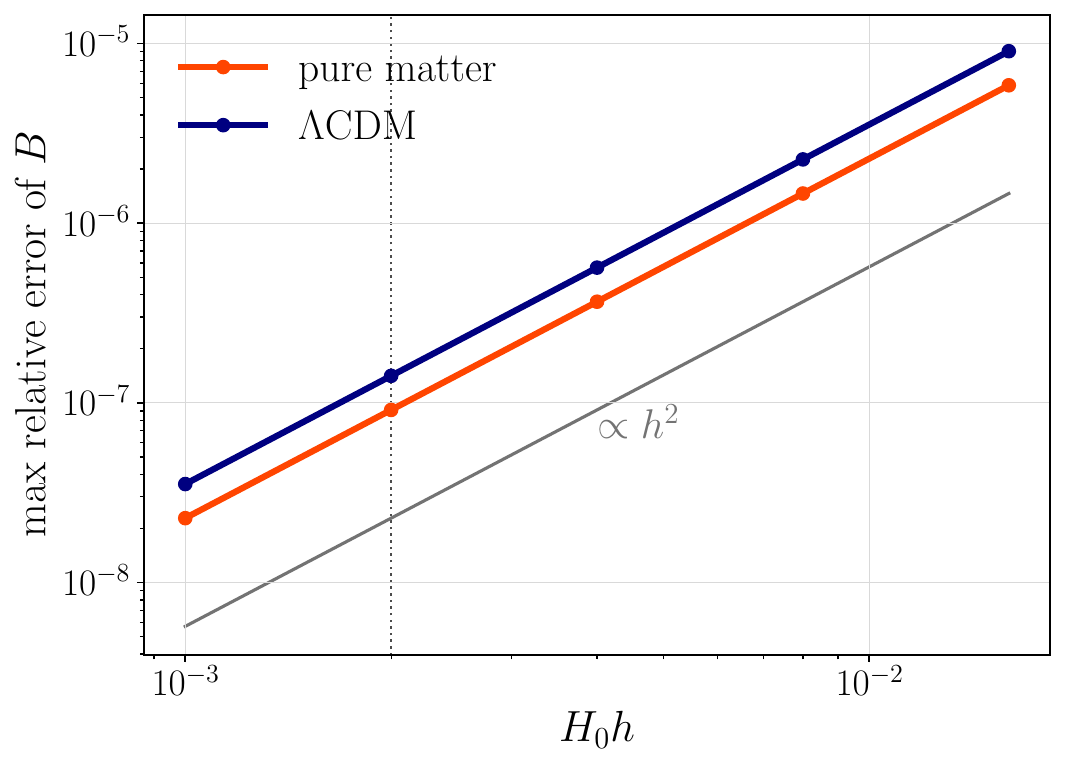}
    \caption{Convergence of the double-null marching~\eqref{eq:marching} with the step size $h$, for $\eta'=0.8\,\Ho^{-1}$ and $0\le\eta-\eta'\le2.4\,\Ho^{-1}$: maximum relative error of $B$ over the Goursat wedge. Orange: pure matter domination, compared with the exact tail $B=1/(\eta\eta')$. Blue: $\Lambda$CDM, compared with the Richardson extrapolation of the two finest runs. The dotted vertical line marks the step $\Ho h=0.002$ used throughout. The errors decrease as $h^2$.}
    \label{fig:convergence_goursat}
\end{figure}

We test the convergence of the double-null marching~\eqref{eq:marching} directly, solving the Goursat problem for $\eta'=0.8\,\Ho^{-1}$ over $0\le{t}\le2.4\,\Ho^{-1}$ with step sizes $\Ho h=0.016,\,0.008,\dots,0.001$. Halving the step keeps all the nodes of the coarser grids, so that the nodes of the coarsest one, of spacing $0.016\,\Ho^{-1}$ in $u$ and $v$, are common to all the runs. On these common nodes we compare each solution with a reference. For pure matter domination, the reference is the exact tail $B_{\rm ref}=1/(\eta\eta')$. For $\Lambda$CDM, the reference is the Richardson extrapolation of the two finest runs, $B_{\rm ref}=(4B_{0.001}-B_{0.002})/3$. In Figure~\ref{fig:convergence_goursat} we plot the maximum relative error of $B$ over the wedge against the spacing $h$.
This tests the accuracy of both the light-cone data~\eqref{eq:g0v} and the marching~\eqref{eq:marching} together. The case of pure matter domination also verifies the accuracy with which we solve the background evolution numerically, through equation~\eqref{eq:H0eta} with $\Omega_m=1$ and $\Omega_\Lambda =0$, against the exact analytic prediction. 
Both errors decrease as $h^2$, as expected from the local truncation error $\mathcal O(h^4)$ of equation~\eqref{eq:marching}. With the step size $\Ho h=0.002$, the maximum relative error over the integration wedge in the calculation of the tail $B$ with the exact $\Lambda$CDM interaction is approximately $1.4\times10^{-7}$.

\subsubsection{Derivatives of the tail}
\label{sec:numerics_deriv}
\noindent
For several spectral applications, such as the calculation of the stochastic gravitational-wave memory background, it is useful to compute the time derivative $\partial_\eta B$ at fixed $(\eta',r)$. Varying $\eta$ while holding $r$ and $\eta'$ fixed corresponds to moving diagonally in the $(u,v)$ plane, $u\to u+dt$, $v\to v+dt$. Thus $\partial_\eta|_{r,\eta'}=\partial_u+\partial_v\equiv\partial_t$. Since $g=rB$ with $r$ fixed along this direction, $\partial_\eta B=P(u,v)/r$ with $P\equiv g_u+g_v$. Applying $\partial_u+\partial_v$ to the bulk equation \eqref{eq:bulk_uv} shows that $P$ obeys the \emph{same} characteristic equation as $g$, but with a source term built from $g$ itself, that is
\begin{equation}
    4P_{,uv} = m^2(\eta)P + \frac{dm^2}{d\eta}(\eta)\,g(u,v) \ .
    \label{eq:P_bulk}
\end{equation}
Therefore, $P(u,v)$ can be marched with the same double-null scheme,
\begin{align}
    P_{i+1,j+1} = P_{i,j+1}+P_{i+1,j}-P_{i,j}  \qquad\qquad \nonumber\\
    + \frac{h^2}{8}m^2(\eta_{\rm mid})\big(P_{i,j+1}+P_{i+1,j}\big)  \nonumber\\
    +  \frac{h^2}{8}\frac{dm^2}{d\eta}(\eta_{\rm mid})\big(g_{i,j+1}+g_{i+1,j}\big)\ ,
    \label{eq:P_marching}
\end{align}
at the cost of one additional known source term per cell, evaluated with $dm^2/d\eta=H_0^2a'(\eta)\big(4\OL a-\Om/(2a^2)\big)$ and already available from the scale factor computed at each node. The characteristic data on the light cone, $P(0,v)=g_u(0,v)+g_v(0,v)$, needed to seed~\eqref{eq:P_marching}, are inherited from $g(u,v)$. The second term follows from the identity $g_v(0,v)=m^2(\eta)/4$, while the first term is computed by integrating the bulk equation along $u=0$ to obtain $g_u(0,v)$,
\begin{align}\label{eq:P_lightcone}
    P(0,v) =&  \frac{m^2(\eta)-m^2(\eta')}{4} \nonumber\\
    & + \frac14\int_0^vm^2\!\left(\eta'+\frac{v'}2\right)g(0,v')\,dv' \ ,
\end{align}
with $\eta=\eta'+\frac v2$. The integral in equation~\eqref{eq:P_lightcone} uses only the light-cone row $g(0,\cdot)$ and $m^2$ at the grid nodes, and is accumulated once, alongside the marching, by a running trapezoidal rule. The coincidence limit of $P/r$, needed on the diagonal, follows from the same steps that led to equation~\eqref{eq:B_worldline}. Since $P$ vanishes on the worldline and is antisymmetric, $\partial_\eta B=P/r\to2\partial_vP$. Integrating equation~\eqref{eq:P_bulk} over the characteristic rectangle below $(u,v)$, differentiating in $v$, and using $\partial_vP(0,v)=\frac18\frac{dm^2}{d\eta}(\eta)+\frac14m^2(\eta)\,g(0,v)$ from equation~\eqref{eq:P_lightcone}, we find
\begin{multline}
\partial_\eta B(u,u)=\frac14\frac{dm^2}{d\eta}\Big(\eta'+\frac u2\Big)+\frac12m^2\Big(\eta'+\frac u2\Big)\,g(0,u)\\
+\frac12\int_0^u\Big[m^2(\eta'')\,P(u',u)+\frac{dm^2}{d\eta}(\eta'')\,g(u',u)\Big]du' \ ,
\label{eq:dB_worldline}
\end{multline}
with $\eta''=\eta'+(u'+u)/2$. The integral runs along the line $v=u$ of the grid, uses only the marched $g$ and $P$, and is evaluated with the composite Simpson rule. We verified that it agrees with the direct numerical derivative $2\partial_vP$ computed on the grid with a sixth-order one-sided finite-difference stencil. Within the radius of convergence of the corner series, $P/r$ on the diagonal also follows from differentiating the series term by term, $P(u,v)=\sum_{m,n}a_{mn}\big(mu^{m-1}v^n+nu^mv^{n-1}\big)$.

\section{Results}\label{Results}

\noindent
We organize the results as follows. We first present the tail computed in the exact $\Lambda$CDM interaction~\eqref{eq:m2_LCDM} with the double-null marching scheme~\eqref{eq:marching} and validate it against the analytic two-pole interaction~\eqref{eq:m2_twopoles} and the independent corner series expansion of Section~\ref{sec:series}. We then discuss and quantify the tail in the multiple-scattering picture (Section~\ref{sec:scattering}) and finally follow the tail of a single source through the whole expansion history (Section~\ref{sec:full_history}). Unless stated otherwise, all results use the parameters $\Om=0.3$, $\OL=0.7$, $H_0\eta'=1$, and (for the two-pole potential) $H_0\eta_m=4.445744$.

\subsection{The tail term in \texorpdfstring{$\Lambda$}{Lambda}CDM}
\label{sec:res_lcdm}
\begin{figure}[t]
    \centering
    \includegraphics[width=0.99\linewidth]{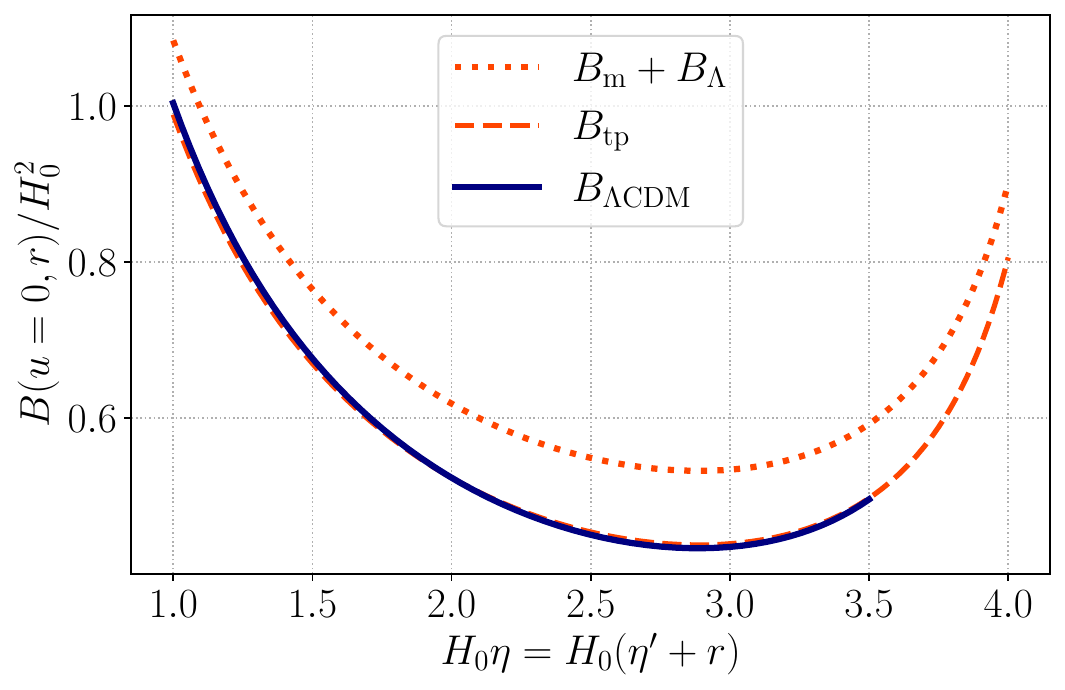}
    \caption{Light-cone data $B(\eta,\bfx,\eta',\bfx')|_{u=0}$, in units of $H_0^2$, for a source at $H_0\eta^\prime = 1$, as a function of $H_0\eta=H_0(\eta'+r)$. Solid: exact $\Lambda$CDM value~\eqref{eq:B_lightcone} with the interaction~\eqref{eq:m2_LCDM}. Dashed: two-pole fit~\eqref{eq:B_tp_fit}. Dotted: $B_{\rm m}+B_\Lambda$, the sum of the light-cone tails of pure matter, $B_{\rm m}=1/(\eta\eta')$, and of pure $\Lambda$ domination, $B_\Lambda=1/[(\eta_m-\eta)(\eta_m-\eta')]$.}
    \label{fig:B_LC}
\end{figure}

\noindent
Figure~\ref{fig:B_LC} shows the light-cone value of the tail~\eqref{eq:B_lightcone} for a source at $H_0\eta'=1$, as a function of the observation time $\eta=\eta'+r$. These are the boundary data of the Goursat problem, seeding the double-null marching on the light-cone row of the grid~\eqref{g_0j}. At the source the tail equals $m^2(\eta')/2$, as indicated by~\eqref{eq:B_worldline}. Then it decreases, due to the average of $m^2$ over a longer interval in the matter era, reaching a minimum at $H_0\eta\simeq 2.89$, and grows again as $\Lambda$ domination sets in. The figure also shows the two-pole formula, for which the data on the light cone are given analytically by equation~\eqref{eq:B_tp_fit}.

\begin{figure*}[t]
    \centering
    \includegraphics[width=\linewidth]{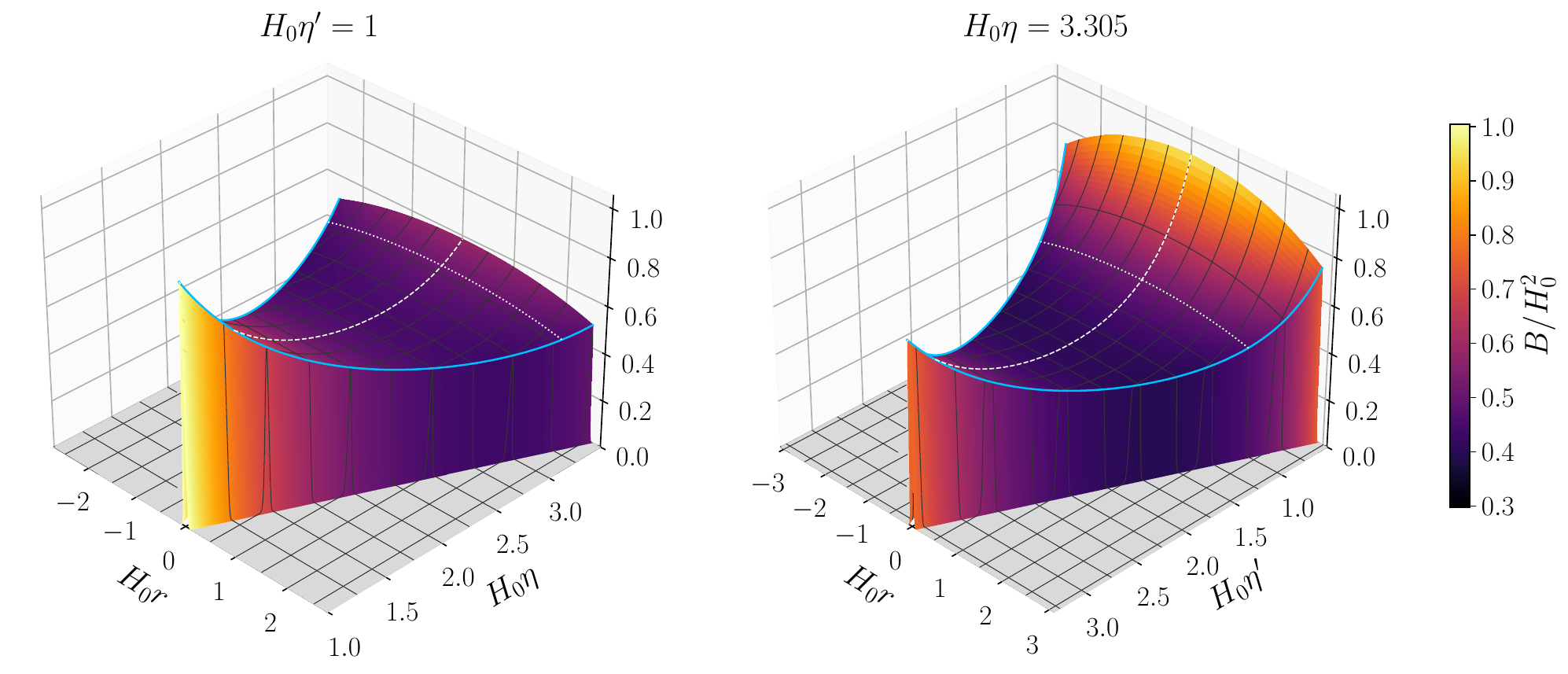}
    \caption{Tail $B(\eta,\eta',r)$ of the retarded Green's function for the exact $\Lambda$CDM potential with $\Om=0.3$, $\OL=0.7$. Left: $B$ in the future light cone from a source fixed at $H_0\eta'=1$ as a function of the observation time $\eta$ and of the distance $r$. Right: $B$ in the past light cone of an observer fixed at $H_0\eta=3.305$ as a function of the emission time $\eta'$ and of $r$. They have one line in common, $H_0\eta'=1$ and $H_0\eta=3.305$ (white dotted line). The tail is supported inside the light cone, $|r|\leq{t}$: the Green's function jumps across the light cone from zero to the light-cone value~\eqref{eq:B_lightcone}. To show the jump, the surfaces are continued outside the light cone by steep walls, $B|_{u=0}\,e^{-(u/w)^2}$ with $H_0 w=0.03$, down to a flat floor. The height of the walls is the light-cone value. Light-blue lines mark the light cone at $r=\pm{t}$, and the white dashed line the profile on the worldline, $r=0$.}
    \label{fig:B_3D}
\end{figure*}

Figure~\ref{fig:B_3D} shows the tail inside the light cone for the physical $\Lambda$CDM interaction~\eqref{eq:m2_LCDM}. The tail $B(\eta,\eta',r)$ depends on the observation time, the emission time, and the distance. The two panels show two sections, with each section mirrored about $r=0$: in the left panel the source time $\eta^\prime$ is fixed, and $B$ fills the future light cone of the source; in the right panel the observation time $\eta$ is fixed, and $B$ fills its past light cone.

The left panel of Figure~\ref{fig:B_3D} shows the double-null solution over the interior of the future light cone of a burst event at $H_0\eta^\prime = 1$. On the light cone the tail takes the values of the boundary data of Figure~\ref{fig:B_LC}, which set the height of the walls and the approximate scale of the tail in the interior of the light cone at each time $\eta$. Indeed, as pointed out above, the tail is largest near the corner $\eta\to\eta'$, where, recalling equation~\eqref{eq:B_worldline}, $B\to m^2(\eta')/2$. Along the worldline it then decreases to a shallow minimum at $H_0\eta\simeq2.72$, and grows again at later times, as $\Lambda$ domination sets in. Notice that the minimum along the worldline occurs earlier than the minimum along the light cone, because the two are fixed by different combinations of the interaction. On the light cone, equation~\eqref{eq:B_lightcone} is the average of $m^2/2$ over the whole interval $[\eta',\eta]$, which keeps decreasing even after the minimum of $m^2$ itself, at $H_0\eta\simeq2.22$. On the worldline instead, equation~\eqref{eq:B_worldline} is the sum of two terms: the local value $m^2/2$ at the intermediate time $\eta'+{t}/2$, and the integral of $m^2g$ along the line $v=u$, which is positive and grows with ${t}$.

The right panel of Figure~\ref{fig:B_3D} shows the same function from the point of view of an observer today, at $H_0\eta=3.305$, that is the interior of its past light cone collecting the tail of all sources at $H_0\eta'\geq0.5$. This is the section relevant for an observation, and the one computed with Fourier methods in Ref.~\cite{Jokela:2022rhk} (see Appendix~\ref{app:literature} for a comparison and Figure~\ref{fig:jks_extended_3D} for a look at the tail from a different angle). The two panels share the line $H_0\eta'=1$, $H_0\eta=3.305$. 
At the vertex, $\eta'\to\eta_0$, the present interaction is dominated by the cosmological constant, $m^2\simeq2\Omega_\Lambda a^2H_0^2$, imprinted in the peak around the corner. Moving the source to earlier times, the tail first decreases, tracing the transition from $\Lambda$ to matter domination, reaches a minimum and then grows again in the deep matter era as the term $\Omega_m/(2a)$ takes over. 

\begin{figure}[t]
\centering
    \centering
    \includegraphics[width=0.94\linewidth]{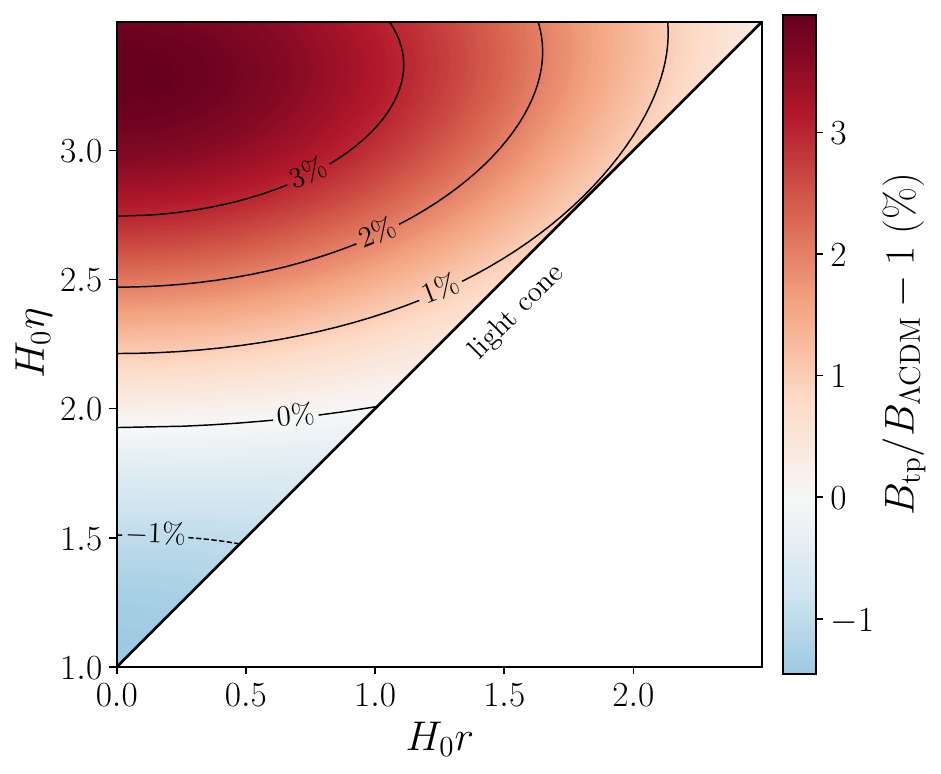}
    \caption{Relative difference between the two-pole and the exact $\Lambda$CDM tails, $B_{\rm tp}/B_{\Lambda{\rm CDM}}-1$ (in percent), over the Goursat wedge $0\leq r\leq{t}$ for a source at $H_0\eta'=1$, from the same double-null solutions as the left panel of Figure~\ref{fig:B_3D}. The diagonal is the light cone, $r={t}$, and the vertical axis the worldline, $r=0$. Black lines are contours of constant difference (dashed for negative values).} 
    \label{fig:B_ratio}
\end{figure}

\subsection{Validation: the two-pole model}
\label{sec:res_twopeak}
\noindent
In the two-pole model, the value of $B$ on the light cone can be obtained analytically by plugging the interaction~\eqref{eq:m2_twopoles} into equation~\eqref{eq:B_lightcone}, giving
\begin{equation}\label{eq:B_tp_fit}
    B_{\rm tp}(x,x^\prime)\Big\vert_{u=0} = \frac{1}{\eta\eta^\prime} + \frac{1}{(\eta_m-\eta)(\eta_m - \eta^\prime)} - 0.095H_0^2 \ .
\end{equation}
This is the light-cone tail fitted to $\Lambda$CDM in Refs.~\cite{Jokela:2022rhk,Jokela:2023suz}. Figure~\ref{fig:B_LC} displays the tail contribution $B$ on the light cone. The two-pole formula~\eqref{eq:B_tp_fit} agrees well with the exact numerical integration described in Section~\ref{sec:goursat}.

In the two-pole model the bulk equation~(\ref{eq:bulketar}) can be separated (separation parameter $k$) into a harmonic equation for $r$ and an ordinary differential equation for $\eta$ with regular singular points at $\eta=0,\,\eta_m$, and an irregular singular point at $\eta\to\infty$. Its solutions, for nonzero $k$, are Confluent Heun functions of $\eta$~\cite{NIST:DLMF}, describing a flow in the direction $r$ in the strip $0<\eta<\eta_m$~\cite{Sasaki:2018oxw}. Relating this formulation to the Goursat solution demands specific boundary conditions.

The double-null solution of the tail for the two-pole model has qualitatively the same structure as the $\Lambda$CDM one in Figure~\ref{fig:B_3D}, since both interactions have a second-order pole at the conformal future infinity and reproduce a pure matter interaction in the limit $\eta\to 0$.
Figure~\ref{fig:B_ratio} quantifies the difference between the tail in the two models over the whole wedge. Both cases are computed with the double-null marching scheme of Section~\ref{sec:numerics}, the exact $\Lambda$CDM case using the interaction~\eqref{eq:m2_LCDM}, and the approximate two-pole model using~\eqref{eq:m2_twopoles}. On the light cone the two-pole tail agrees with the exact one to within $1.5\%$ at all times, owing to the constant $-0.095H_0^2$ fitted in Refs.~\cite{Jokela:2022rhk,Jokela:2023suz}. Inside the light cone, the tail computed with the two-pole model underestimates the exact tail in $\Lambda$CDM by up to $1.5\%$ for $H_0{t}\lesssim1$, while it overestimates it in the region $H_0t >1$, reaching $\simeq3.8\%$ for $H_0{t}\simeq2.2$--$2.5$.

This difference in the tail can be explained by the difference $m^2_{\rm tp}-m^2_{\Lambda{\rm CDM}}$ (Figure~\ref{fig:scale_factor_comparison}): the two-pole model overestimates the exact $\Lambda$CDM interaction in the region $1.46<H_0\eta<2.98$, and underestimates it for $H_0\eta<1.46$. The value of the interaction potential $m^2$ is imprinted in the tail. For simplicity, we can focus on the worldline, where equation~\eqref{eq:B_worldline} gives $B$ as the local value $m^2/2$ at the intermediate time $(\eta+\eta')/2$, plus the integral of $m^2g$ along the line $v=u$. The relative difference of the local term is negative as long as $H_0(\eta+\eta')/2<1.46$, explaining the blue region in Figure~\ref{fig:B_ratio}, and is largest when $(\eta+\eta')/2$ crosses the maximum excess of $m^2_{\rm tp}$, around $H_0\eta\simeq 3.4$. The full tail on the worldline follows the same pattern.

\subsection{Radial profile of the tail}
\noindent
Figure~\ref{fig:B_ratio} also shows that the difference of the tail in the two models increases toward the worldline.  As we will discuss in detail in Section~\ref{sec:scattering}, to a first approximation one can estimate the tail as the first two terms of equation~\eqref{eq:g_integral}, with light-cone values~\eqref{eq:g0v}, leading to 
\begin{equation}
	B(u,v) \simeq \frac{1}{2r}\int_{\eta'+u/2}^{\eta'+v/2}m^2(\eta'')\,d\eta'' \ .
\end{equation} 
That is, the tail is approximately half the average of $m^2$ over a window of width $r$ centered at $(\eta+\eta')/2$, which shrinks from the whole interval $[\eta',\eta]$ on the light cone to the single time $(\eta+\eta')/2$ on the worldline. On the light cone, the tail~\eqref{eq:B_lightcone} averages $m^2$ over the whole interval $[\eta',\eta]$, where $m^2_{\rm tp}$ is partly below and partly above $m^2_{\Lambda{\rm CDM}}$, and the difference largely cancels. On the worldline instead, the tail samples the local curvature at the intermediate time $(\eta+\eta^\prime)/2$, and the difference between the models does not cancel.  
Quantities controlled by $B$ close to the light cone, such as the tail memory of a burst shortly after its arrival, are therefore well approximated by the two-pole interaction, while those that depend on the interior of the light cone are more sensitive to the local curvature.
\begin{figure}[t]
  \centering
  \includegraphics[width=.95\linewidth]{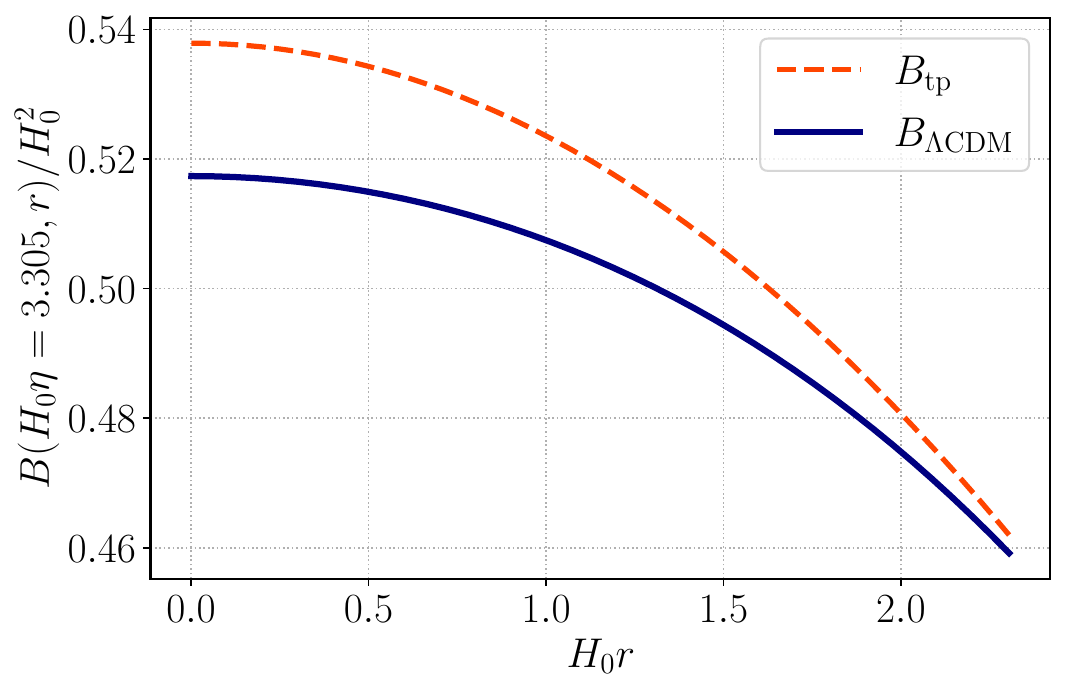}
\caption{Tail $B(\eta,\bfx;\eta',\bfx')$, in units of $H_0^2$, on the slice of fixed observation time today, $H_0\eta_0=3.305$, as a function of $r$, for a source at $H_0\eta'=1$: exact $\Lambda$CDM (solid) and two-pole interaction~\eqref{eq:m2_twopoles} (dashed).}
\label{fig:B_spatial}
\end{figure}

We show the radial profile of the tail on a fixed-$\eta$ slice for the two models in Figure~\ref{fig:B_spatial}. The two tails are very close on the light cone and grow toward the worldline, the two-pole model showing a steeper profile.   
The radial profile also depends strongly on the source time. We compute, at present time $\eta_0$, the excess of the tail on the worldline over its light-cone value,
\begin{center}
\begin{tabular}{c@{\hspace{1.5em}}c@{\hspace{1.5em}}c}
$\Ho\eta'$ & \multicolumn{2}{c}{$\vert 1-B(r=0)/B(r={t})\vert$} \\[0.3ex]
\hline
 & two-pole & $\Lambda$CDM \\[0.3ex]
$0.501$ & $0.318$ & $0.280$ \\
$0.801$ & $0.218$ & $0.179$ \\
$1.001$ & $0.164$ & $0.127$ \\
$1.501$ & $0.070$ & $0.044$ \\
$2.001$ & $0.022$ & $0.010$ \\
$2.501$ & $0.003$ & $0.001$ \\
\end{tabular}
\end{center}
\noindent
finding that the increment of $B$ on the worldline compared to its value on the light cone grows for earlier sources and is larger for the two-pole interaction at all source times.

In pure matter or pure $\Lambda$ domination, $B$ does not depend on $r$; it is exactly constant on each time slice~\cite{Jokela:2022rhk,Jokela:2023suz}. This follows from the original equation~\eqref{eq:bulketar}: for $g=rf(\eta)$, one finds that $B$ is independent of $r$ if and only if its light-cone value~\eqref{eq:B_lightcone}, as a function of $\eta$, solves $f''-m^2f=0$. This is the case for $B=1/(\eta\eta')$ in matter domination, where $a\propto\eta^2$, and for $B=1/[(\eta_m-\eta)(\eta_m-\eta')]\propto a$ in $\Lambda$ domination. Other single-fluid cosmologies,~\footnote{For a barotropic fluid with constant equation of state parameter $w$, the scale factor evolves as $a\propto \eta^{2/(1+3w)}$, and the tail on the light cone follows from~\eqref{eq:B_lightcone} as $B=\frac{1-3w}{(1+3w)^2}\,\frac{1}{\eta\eta'}$, which solves $B''=m^2B$ only for $w=0$ (matter) and $w=-1$ (de Sitter).} except pure radiation, for which $m^2 =0$, do not satisfy this relation, and thus lead to radial dependence in $B$.
The radial dependence in $\Lambda$CDM is thus produced by the transition between the two eras, which the tail of an early source samples over a longer time ${t}$. In Refs.~\cite{Jokela:2022rhk,Jokela:2023suz} the tail was approximated as independent of $r$ by its light-cone fit~\eqref{eq:B_tp_fit}. This is accurate to about $1\%$ for $H_0\eta'\geq2$, but underestimates the tail near the worldline by up to $13\%$ for sources at $H_0\eta^\prime \simeq 1.0$ and $28\%$ at $H_0\eta'\simeq 0.5$.

\subsubsection*{Radial derivative of the tail near the light cone}
\label{sec:res_deriv}
\noindent
\begin{figure}[t]
    \centering
    \includegraphics[width=0.9\linewidth]{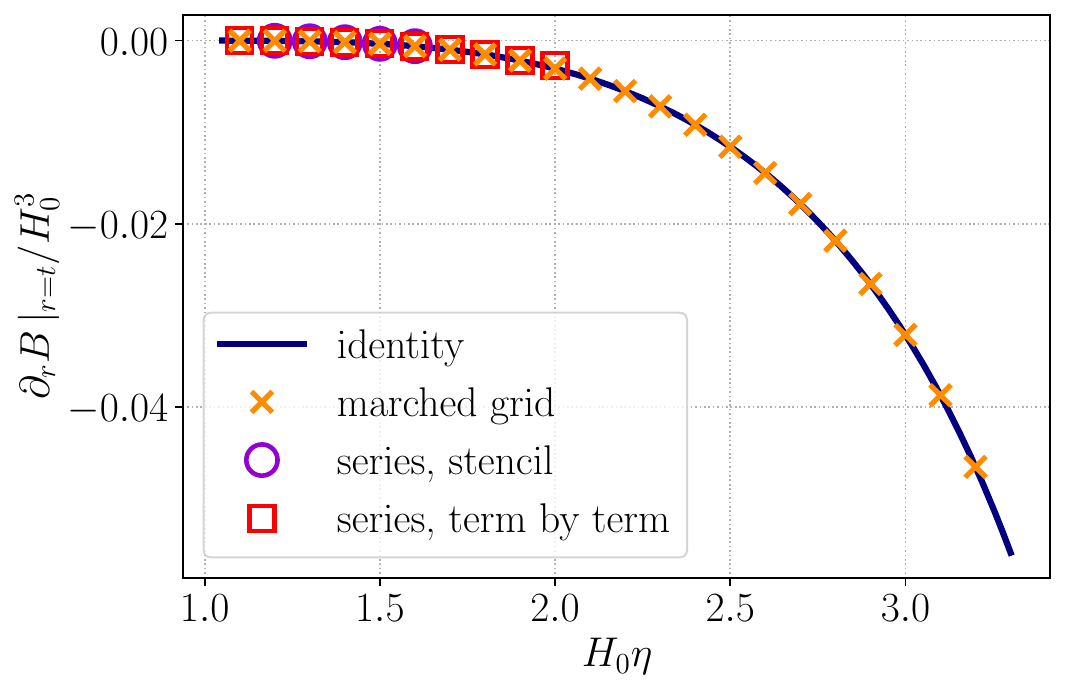}
    \caption{Radial derivative $\partial_rB$ of the $\Lambda$CDM tail term on the light cone ($u\to0$), in units of $H_0^3$, as a function of observation time $\eta$, for a source at $H_0\eta'=1$. Solid line: exact light-cone value, equation~\eqref{eq:dr_B_lc}. Crosses: sixth-order stencil applied to the double-null grid. Circles: same stencil on the corner series truncated at $N=30$ (Section~\ref{sec:series}). Squares: equation~\eqref{eq:dr_B_general} evaluated term by term with the series coefficients.}
    \label{fig:lcdm_radial_derivative}
\end{figure}
The radial profile of the tail starts at the light-cone front, where its slope is fixed in closed form by the light-cone data and the Goursat equation alone. We study the radial derivative $\partial_rB$ at fixed $\eta$ and $\eta'$ near the light cone ($u\to0$). With $\partial_r=\partial_v-\partial_u$ and $B=g/r$,
\begin{equation}
    \partial_rB\Big|_{u\to0} = \frac{1}{r^2}\Big[r\,(\partial_vg-\partial_ug)-g\Big]_{u=0} \ .
    \label{eq:dr_B_general}
\end{equation}
On the light cone, equation~\eqref{eq:g0v} gives $\partial_vg(0,v)=m^2(\eta)/4$, with $\eta=\eta'+v/2$, and equation~\eqref{eq:P_lightcone} gives $\partial_ug(0,v)=-m^2(\eta')/4+\frac14\int_0^vm^2g\,dv'$. The integral is known in closed form. Denoting the light-cone value of $g$ by
\begin{equation}
\kappa\equiv g(0,2{t})=\frac12\int_{\eta'}^{\eta}m^2(\eta'')\,d\eta'' \ ,
\label{eq:kstar}
\end{equation}
and noting that along the light cone $dv'=2\,d\eta''$ and $m^2\,d\eta''=2\,d\kappa$, we obtain $\int_0^vm^2g\,dv'=2\kappa^2$. The radial derivative at fixed $\eta$ and $\eta'$ therefore obeys the exact identity
\begin{equation}
\partial_r\big[rB(\eta,\eta',r)\big]_{r={t}}=\frac{m^2(\eta)+m^2(\eta')}{4}-\frac{\kappa^2}{2}\ ,
\label{eq:dr_g_lc}
\end{equation}
or, since $B|_{r={t}}=\kappa/{t}$,
\begin{equation}
\partial_rB\big|_{r={t}}=\frac{1}{{t}}\left[\frac{m^2(\eta)+m^2(\eta')}{4}-\frac{\kappa^2}{2}-\frac{\kappa}{{t}}\right] \ .
\label{eq:dr_B_lc}
\end{equation}
In pure matter or pure $\Lambda$ domination, the right-hand side of equation~\eqref{eq:dr_B_lc} vanishes identically, since $B$ does not depend on $r$. In $\Lambda$CDM, it is negative, as shown in Figures~\ref{fig:B_spatial}~and~\ref{fig:lcdm_radial_derivative}, confirming that the tail grows from the light cone inward, and its magnitude grows with $\eta$ for a source at $H_0\eta'=1$. 
The identity also fixes the next-to-leading order of the large-$k$ expansion of the propagator in Fourier space (Section~\ref{sec:fourier}).

Figure~\ref{fig:lcdm_radial_derivative} shows $\partial_rB$ evaluated in the limit $r\to t=\eta-\eta'$ ({\emph{i.e.}},\ $u\to0$) as a function of the observation time, comparing the identity~\eqref{eq:dr_B_lc} with three estimates. The first is purely numerical, a sixth-order forward stencil applied to a fixed-$\eta$ spatial slice of the data obtained from the marched grid. The other two use the corner series of Section~\ref{sec:series}, truncated at $N=30$. One uses the same stencil applied to the series-reconstructed data, and the other is analytic, from equation~\eqref{eq:dr_B_general} evaluated term by term as $\partial_vg(0,v)=\sum_n(n+1)a_{0,n+1}v^n$ and $\partial_ug(0,v)=\sum_na_{1,n}v^n$. The series methods agree well with the identity~\eqref{eq:dr_B_lc} within $\eta \leq 2$, the radius of convergence of the series, while the numerical stencil on the exact Goursat data agrees perfectly with the identity~\eqref{eq:dr_B_lc} across all times $\eta$.

\subsection{Series expansion versus double-null marching}
\label{sec:res_series}

\begin{figure}[t]
    \centering
    \includegraphics[width=0.95\linewidth]{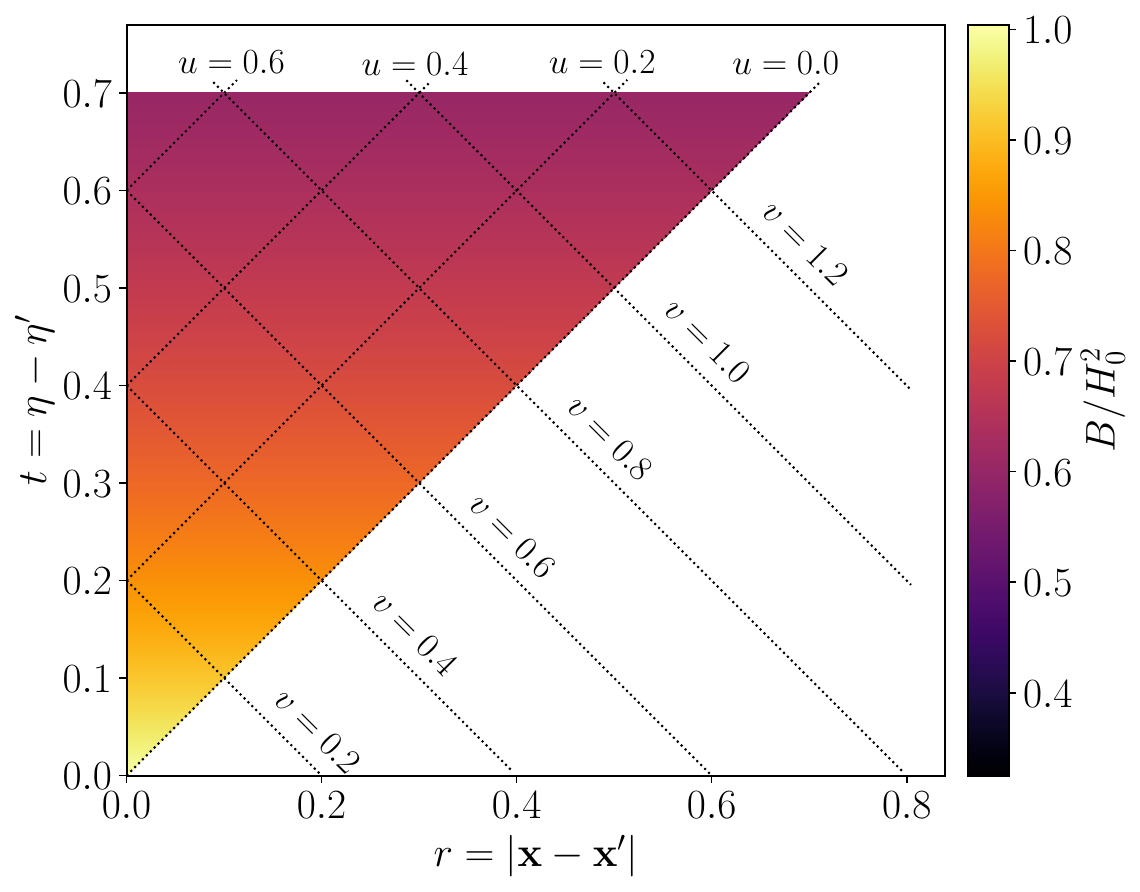}
    \caption{Tail term for the $\Lambda$CDM potential reconstructed from the corner series \eqref{eq:Ansatz}--\eqref{eq:cp_exact}, truncated at degree $N=30$.}
    \label{fig:lcdm_wedge_series}
\end{figure}

\begin{figure}[t]
    \centering
    \includegraphics[width=0.95\linewidth]{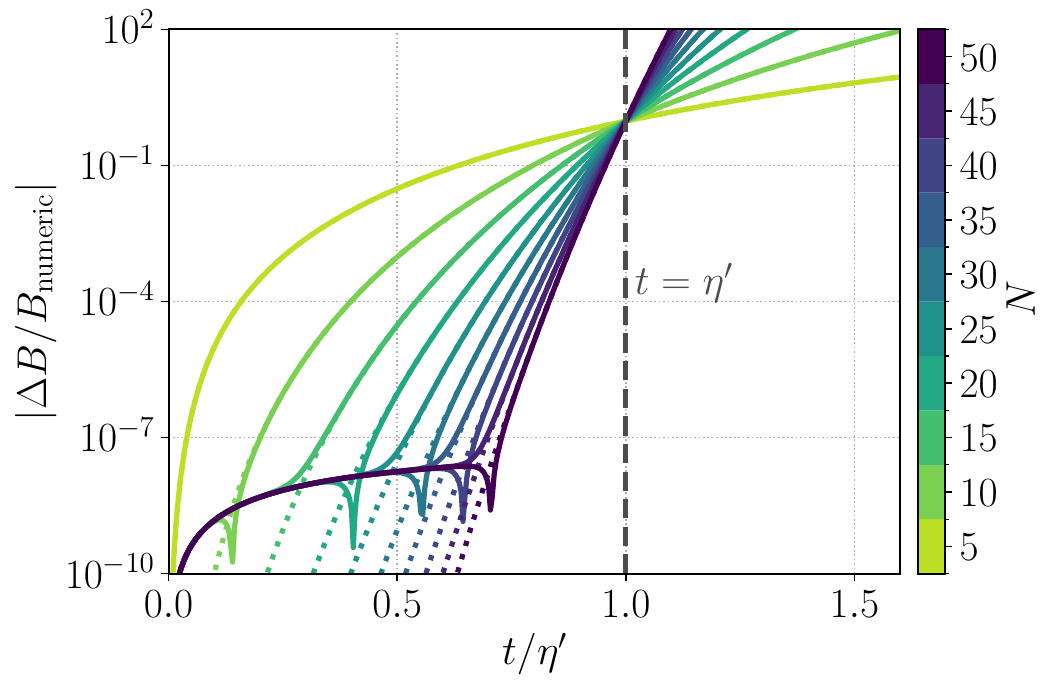}
    \caption{Relative difference $|\Delta B_{[N]}/B|=|B^{\rm series}_{[N]}-B|/B$ between the corner series~\eqref{eq:Ansatz}, truncated at total degree $N$, and the double-null solution, as a function of ${t}/\eta'$, for the $\Lambda$CDM interaction with $\Om=0.3$, $\OL=0.7$, and a source at $H_0\eta'=1$. The two solutions are compared on the same field points along the line $r={t}/2$, away from both the light cone and the worldline. Dotted lines show the analytic estimate~\eqref{eq:series_error} of the truncation error. Dashed line marks the radius of convergence of the series, ${t}=\eta'$.}
    \label{fig:series_convergence}
\end{figure}
\noindent
An independent validation of the numerical results is given by the double series expansion outlined in Section~\ref{sec:series}. We stress that the marching scheme and the series expansion share no numerical ingredients besides the common light-cone data~\eqref{eq:g0v}.
Figure~\ref{fig:lcdm_wedge_series} shows the tail term reconstructed from the corner series \eqref{eq:Ansatz}, truncated at total degree $N=30$ and built order by order from \eqref{eq:a0j}--\eqref{eq:closure}, for the $\Lambda$CDM interaction.
Qualitatively, Figure~\ref{fig:lcdm_wedge_series} shows the same behavior as the full tail solution in the left panel of Figure~\ref{fig:B_3D}, with a marked peak at the coincidence corner $u=v=0$, and the tail filling the interior of the light cone with decreasing value moving forward in time from the source (before the cosmological $\Lambda$ term takes over).

Figure~\ref{fig:series_convergence} compares the two solutions quantitatively, as a function of the distance ${t}$ from the corner and of the truncation order $N$. The truncation error of the series $\Delta B_{[N]}$ is controlled analytically as in equation~\eqref{eq:series_error}.
Inside the radius of convergence, ${t}<\eta'$, the truncation error decreases with $N$ as $({t}/\eta')^N$, and the series agrees with the numerical solution down to $\simeq2\times10^{-8}$, which is the floor of the discretization error of the double-null march (Section~\ref{sec:numerics}). This floor is reached for ${t}\lesssim0.55\,\eta'$ at $N=30$ and ${t}\lesssim0.7\,\eta'$ at $N=50$, while closer to the radius the error grows, and at ${t}=\eta'$ it is of order one for any $N$.

\begin{figure}[t]
\centering
\colorlet{fcCol}{red!70!black}
\colorlet{pcCol}{blue!70!black}
\colorlet{isCol}{violet!85!white!80!black}
\definecolor{lcCol}{RGB}{196,146,0}
\def\el{30}
\def\az{-118}
\pgfmathsetmacro{\exx}{-sin(\az)}\pgfmathsetmacro{\exy}{-sin(\el)*cos(\az)}
\pgfmathsetmacro{\eyx}{cos(\az)}\pgfmathsetmacro{\eyy}{-sin(\el)*sin(\az)}
\pgfmathsetmacro{\ezy}{cos(\el)}
\def\R{2.0}
\def\D{3.0}
\pgfmathsetmacro{\ea}{\D/2}                       
\pgfmathsetmacro{\ec}{\R/2}                       
\pgfmathsetmacro{\eb}{sqrt(\ea*\ea-\ec*\ec)}      
\def\Hf{2.6}                                     
\def\Hp{2.6}                                     
\def\zs{-2.4}                                     
\pgfmathsetmacro{\sf}{acos(tan(\el))}             
\pgfmathsetmacro{\sp}{acos(-tan(\el))}            
\def\ph{268}                                      
\begin{tikzpicture}[scale=0.95, >=Stealth, font=\small,
  x={(\exx cm,\exy cm)}, y={(\eyx cm,\eyy cm)}, z={(0cm,\ezy cm)},
  every node/.style={inner sep=1.5pt},
  conef/.style={fcCol, thin},
  conep/.style={pcCol, thin}]
\fill[violet!5] (-1.3,-1.6,\zs) -- (3.3,-1.6,\zs) -- (3.3,1.6,\zs) -- (-1.3,1.6,\zs) -- cycle;
\fill[violet!13] plot[domain=0:360, samples=90, smooth cycle] ({\ec+\ea*cos(\x)},{\eb*sin(\x)},\zs);
\draw[isCol, thick] plot[domain=0:360, samples=90, smooth cycle] ({\ec+\ea*cos(\x)},{\eb*sin(\x)},\zs);
\pgfmathsetmacro{\angx}{atan2(\exy,\exx)}
\node[isCol, rotate=\angx, anchor=south east, inner sep=1.5pt] at (3.2,-1.6,\zs) {\scriptsize space};
\fill[red!30, opacity=0.22] (0,0,0) -- ({\Hf*cos(\az-\sf)},{\Hf*sin(\az-\sf)},\Hf)
   -- plot[domain=\az-\sf:\az+\sf-360, samples=60] ({\Hf*cos(\x)},{\Hf*sin(\x)},\Hf) -- cycle;
\draw[conef] plot[domain=0:360, samples=90, smooth cycle] ({\Hf*cos(\x)},{\Hf*sin(\x)},\Hf);
\draw[conef] (0,0,0) -- ({\Hf*cos(\az-\sf)},{\Hf*sin(\az-\sf)},\Hf);
\draw[conef] (0,0,0) -- ({\Hf*cos(\az+\sf)},{\Hf*sin(\az+\sf)},\Hf);
\fill[blue!30, opacity=0.2] (\R,0,\D) -- ({\R+\Hp*cos(\az-\sp)},{\Hp*sin(\az-\sp)},{\D-\Hp})
   -- plot[domain=\az-\sp:\az+\sp, samples=60] ({\R+\Hp*cos(\x)},{\Hp*sin(\x)},{\D-\Hp}) -- cycle;
\draw[conep] plot[domain=0:360, samples=90, smooth cycle] ({\R+\Hp*cos(\x)},{\Hp*sin(\x)},{\D-\Hp});
\draw[conep] (\R,0,\D) -- ({\R+\Hp*cos(\az-\sp)},{\Hp*sin(\az-\sp)},{\D-\Hp});
\draw[conep] (\R,0,\D) -- ({\R+\Hp*cos(\az+\sp)},{\Hp*sin(\az+\sp)},{\D-\Hp});
\draw[densely dashed, gray] (0,0,\zs) -- (0,0,0);
\draw[densely dashed, gray] (\R,0,\zs) -- (\R,0,{\D+0.35});
\fill[violet!18, opacity=0.6] plot[domain=0:360, samples=120, smooth cycle]
   ({\ec+\ea*cos(\x)},{\eb*sin(\x)},{\ea+\ec*cos(\x)});
\draw[isCol, line width=1.4pt] plot[domain=0:360, samples=120, smooth cycle]
   ({\ec+\ea*cos(\x)},{\eb*sin(\x)},{\ea+\ec*cos(\x)});
\coordinate (P) at ({\ec+\ea*cos(\ph)},{\eb*sin(\ph)},{\ea+\ec*cos(\ph)});
\coordinate (P0) at ({\ec+\ea*cos(\ph)},{\eb*sin(\ph)},\zs);
\draw[densely dotted, isCol] (P) -- (P0);
\draw[fcCol, thin] (0,0,\zs) -- (P0);
\draw[pcCol, thin] (P0) -- (\R,0,\zs);
\draw[->, fcCol, thick] (0,0,0) -- ($(0,0,0)!0.97!(P)$);
\draw[->, pcCol, thick] (P) -- ($(P)!0.96!(\R,0,\D)$);
\fill[isCol] (P) circle (1.9pt);
\fill[isCol] (P0) circle (1.3pt);
\node[isCol, anchor=north west] at (P) {$x'', m^2(\eta^{\prime\prime})$};
\node[isCol, anchor=west, yshift=-2pt] at (P0) {\small $\bfx''$};
\fill[isCol] ({\ec-\ea},0,{\ea-\ec}) circle (1.6pt);
\fill[isCol] ({\ec+\ea},0,{\ea+\ec}) circle (1.6pt);
\node[isCol, anchor=east, xshift=-3pt] at ({\ec-\ea-0.01},0,{\ea-\ec-0.2}) {\small $\eta'+\frac u2$};
\node[isCol, anchor=west, xshift=7pt] at ({\ec+\ea+0.05},0,{\ea+\ec+0.1}) {\small $\eta'+\frac v2$};
\coordinate (Q1) at (0.155,0.580,0.600);  
\coordinate (Q2) at (1.482,0.897,1.964);  
\draw[lcCol, thick, densely dashed] (0,0,0) -- (Q1) -- (Q2) -- (\R,0,\D);
\fill[lcCol] (Q1) circle (1.5pt);
\fill[lcCol] (Q2) circle (1.5pt);
\draw[lcCol, densely dotted] (Q1) -- (0.155,0.580,\zs);
\draw[lcCol, densely dotted] (Q2) -- (1.482,0.897,\zs);
\draw[lcCol, thick, densely dashed] (0,0,\zs) -- (0.155,0.580,\zs) -- (1.482,0.897,\zs) -- (\R,0,\zs);
\fill[lcCol] (0.155,0.580,\zs) circle (1.2pt);
\fill[lcCol] (1.482,0.897,\zs) circle (1.2pt);
\fill[fcCol] (0,0,0) circle (2pt);
\fill[fcCol] (0,0,\zs) circle (1.6pt);
\fill[pcCol] (\R,0,\zs) circle (1.6pt);
\fill[pcCol] (\R,0,\D) circle (2.2pt);
\node[fcCol, anchor=north east, yshift=-3pt] at (0,0,0) {$x'$};
\node[fcCol, anchor=south west, inner sep=1pt] at (0,0,\zs) {$\bfx'$};
\node[pcCol, anchor=west] at (\R+0.05,0,\zs) {$\bfx$};
\node[pcCol, anchor=south west, yshift=5pt] at (\R,0,\D) {$x$};
\pgfmathsetmacro{\thtop}{atan2(\eyy,\exy)}   
\node[fcCol, anchor=south, inner sep=1.5pt] at ({\Hf*cos(\thtop)},{\Hf*sin(\thtop)},\Hf) {\scriptsize future light cone of $x'$};
\node[pcCol, anchor=north west, inner sep=2pt, xshift=3pt] at ({\R+\Hp*cos(\thtop+180)},{\Hp*sin(\thtop+180)},{\D-\Hp}) {\scriptsize past light cone of $x$};
\node[isCol, anchor=north] at ({\ec+\ea*cos(234)},{\eb*sin(234)-0.4},\zs) {\small $|\bfx''-\bfx'|+|\bfx-\bfx''|={t}$};
\draw[->] (-1.3,1.6,\zs) -- (-1.3,1.6,{\zs+1.0}) node[anchor=south] {$\eta$};
\end{tikzpicture}
\caption{Single scattering in spacetime. The wave emitted at the source event $x'=(\eta',\bfx')$ propagates on its future light cone (red), and the field point $x=(\eta,\bfx)$ receives on its past light cone (blue) the waves scattered by the background curvature. A wave scattered once at $x''=(\eta'',\bfx'')$ travels along two null legs (arrows), the first on the future cone of the source and the second on the past cone of the field point. The set of all single-scattering events $x''$ that lie at the intersection of the two cones forms an ellipse,
with the earliest event at $\eta''=\eta'+u/2$ behind the source, and the latest at $\eta''=\eta'+v/2$ behind the field point. 
Projected onto space (bottom, purple), the intersection is the ellipse with foci $\bfx'$ and $\bfx$, $|\bfx''-\bfx'|+|\bfx-\bfx''|={t}$.
The dashed yellow path is a double scattering: its first event lies on the future cone of the source, its second on the past cone of the field point, every leg being null.}
\label{fig:shell_geometry_3d}
\end{figure}
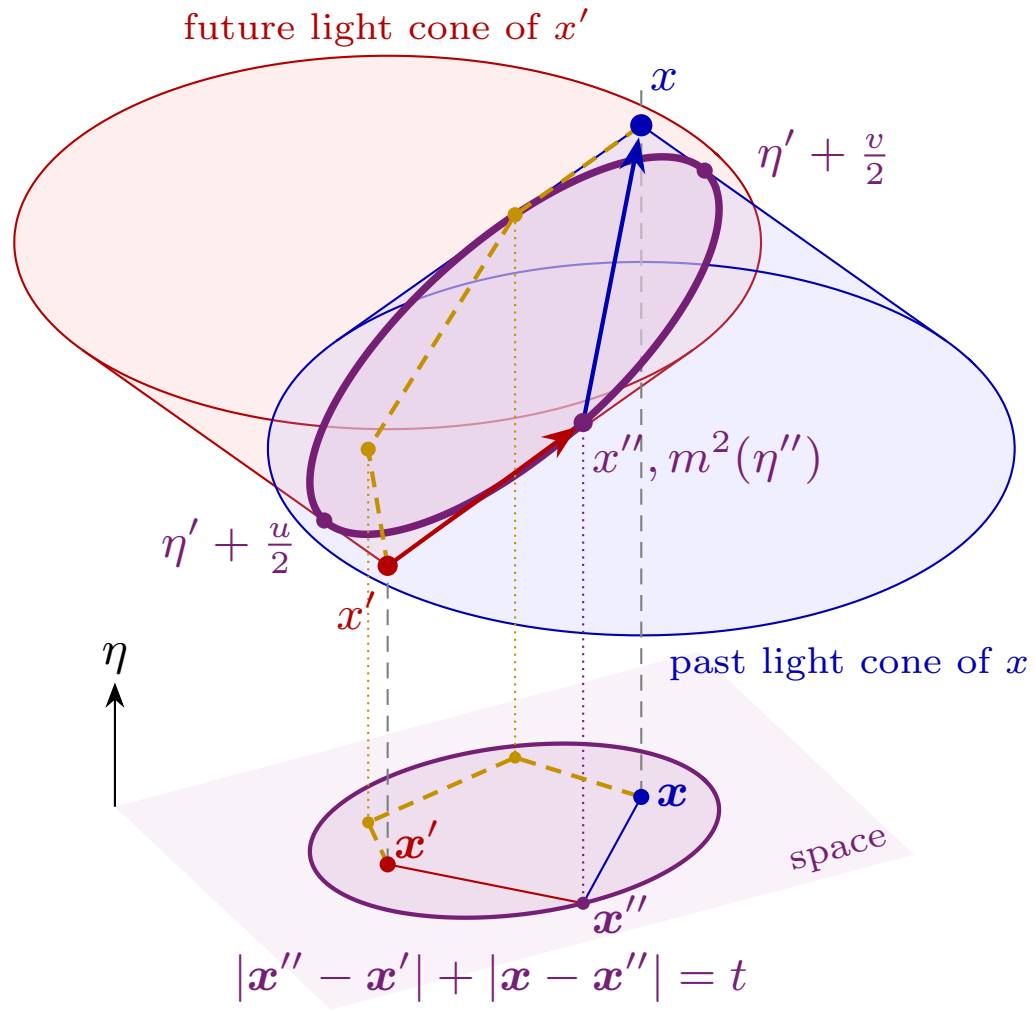

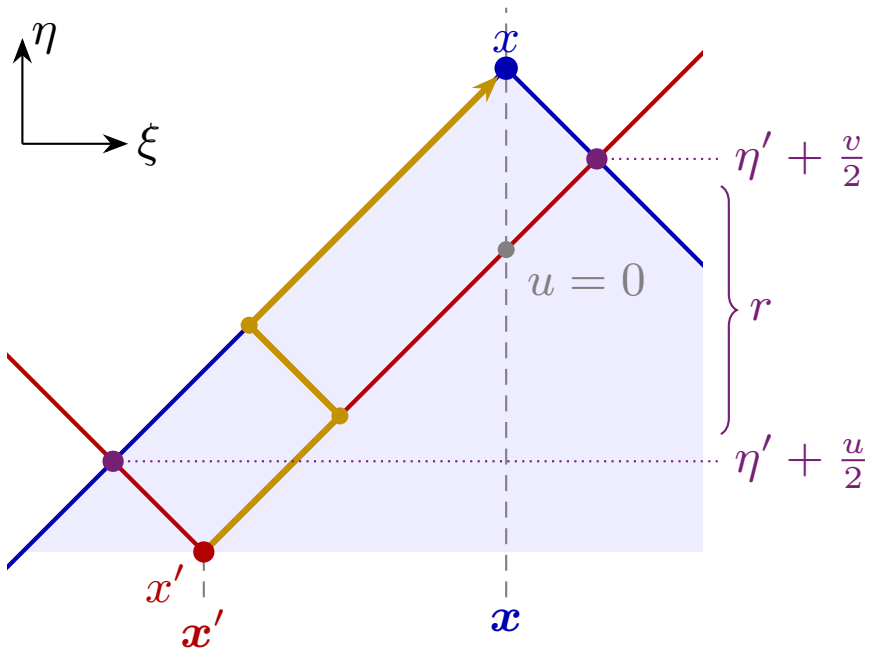
\begin{figure}[t]
\centering
\colorlet{fcCol}{red!70!black}
\colorlet{pcCol}{blue!70!black}
\colorlet{isCol}{violet!85!white!80!black}
\definecolor{lcCol}{RGB}{196,146,0}            
\begin{tikzpicture}[scale=1.0, >=Stealth, font=\small, every node/.style={inner sep=1.5pt}]
\def\R{2.0}
\def\tauv{1.2}                       
\pgfmathsetmacro{\Py}{\R+\tauv}      
\pgfmathsetmacro{\fx}{\R+\tauv/2}    
\pgfmathsetmacro{\bx}{\tauv/2}       
\begin{scope}
 \clip (-1.3,-0.3) rectangle (3.3,3.6);
 \fill[blue!7] (\R,\Py) -- (\R-\Py,0) -- (\R+\Py,0) -- cycle;       
 \draw[pcCol, thick] (\R,\Py) -- (\R-\Py-0.3,-0.3);
 \draw[pcCol, thick] (\R,\Py) -- (\R+\Py+0.3,-0.3);
 \draw[fcCol, thick] (0,0) -- (-1.5,1.5);                          
 \draw[fcCol, thick] (0,0) -- (3.6,3.6);
 \draw[densely dashed, gray] (\R,-0.3) -- (\R,3.6);                
 \draw[densely dashed, gray] (0,-0.3) -- (0,0);
 \draw[lcCol, line width=1.2pt, -{Stealth[length=1.8mm]}, shorten >=2pt] (0,0) -- (0.9,0.9) -- (0.3,1.5) -- (\R,\Py);
\end{scope}
\draw[isCol, densely dotted] (-\bx,\bx) -- (3.4,\bx);
\draw[isCol, densely dotted] (\fx,\fx) -- (3.4,\fx);
\draw[isCol, decorate, decoration={brace, amplitude=3pt}] (3.42,{\fx-0.18}) -- (3.42,{\bx+0.18});
\node[isCol, anchor=west] at (3.55,{(\bx+\fx)/2}) {$r$};
\node[isCol, anchor=west] at (3.45,\fx) {$\eta'+\frac v2$};
\node[isCol, anchor=west] at (3.45,\bx) {$\eta'+\frac u2$};
\fill[isCol] (\fx,\fx) circle (2pt);
\fill[isCol] (-\bx,\bx) circle (2pt);
\fill[lcCol] (0.9,0.9) circle (1.6pt);
\fill[lcCol] (0.3,1.5) circle (1.6pt);
\fill[fcCol] (0,0) circle (2pt);
\fill[pcCol] (\R,\Py) circle (2.2pt);
\fill[gray] (\R,\R) circle (1.6pt);
\node[fcCol, anchor=north east] at (-0.05,-0.02) {$x'$};
\node[fcCol, anchor=north] at (0,-0.32) {$\bfx'$};
\node[pcCol, anchor=south] at (\R,\Py+0.05) {$x$};
\node[gray, anchor=west] at (\R+0.08,\R-0.2) {$u=0$};
\node[pcCol, anchor=north] at (\R,-0.32) {$\bfx$};
\draw[->] (-1.2,2.7) -- (-1.2,3.4) node[anchor=west] {$\eta$};
\draw[->] (-1.2,2.7) -- (-0.5,2.7) node[anchor=west] {$\xi$};
\end{tikzpicture}
\caption{Scattering off the background curvature, in the section of Figure~\ref{fig:shell_geometry_3d} through the worldlines (dashed) of the source $x'=(\eta',\bfx')$ and of the field point $x=(\eta,\bfx)$, at distance $r=|\bfx-\bfx'|$ ($\xi$: comoving position). Future light cone of the source in red, past light cone of the field point and its interior in blue. Every leg of a scattering path is null, and each scattering event (dot) contributes a factor $m^2$ to the series~\eqref{eq:born_series}. A single scattering happens where the two cones cross (purple dots), between the earliest event at $\eta'+u/2$ and the latest at $\eta'+v/2$. These define the window of width $r$ over which equation~\eqref{eq:B_born} averages $m^2$. The yellow path is a double scattering. If we move the field point back in time down its worldline to the light cone (gray dot, $u=0$), the earliest and latest scattering events approach $x'$ and $x$, respectively; the two cones then touch along the entire ray, and only single scattering contributes. There, equation~\eqref{eq:B_born} is exact.}
\label{fig:shell_geometry}
\end{figure}

\section{The tail as multiple scattering}
\label{sec:scattering}

\noindent
In this section we discuss the interpretation of the tail as the wave scattered by the background curvature. We emphasize that this is a representation, not a physical mechanism. The linearized Einstein equations give the Green's function a component filling the interior of the light cone, without reference to any scattering event. In cosmology, this picture was put forward qualitatively in Refs.~\cite{Faraoni:1991xe,Faraoni:1999us}. Here we make the picture quantitative. 

Writing the operator of the wave equation~\eqref{eq:green_eq} as $\Box_0-m^2$, with $\Box_0=\partial_\eta^2-\nabla^2$ and $G_0=\delta(u)/r$ as its retarded Green's function, we can formally express the total Green's function as the integral equation 
\begin{equation}
G(x,x') =G_0(x,x')+\frac1{4\pi}\int d^4y\,G_0(x,y)\,m^2(\eta_y)\,G(y,x') \ .
\label{eq:G_integral}
\end{equation}
Iteration at any order $n$ gives the Born series
\begin{multline}
G(x,x')=G_0(x,x')+\sum_{n\geq1}\frac{1}{(4\pi)^n}\int d^4y_1\cdots d^4y_n
\\
\times G_0(x,y_n)\,m^2(\eta_n)\,G_0(y_n,y_{n-1})\cdots
\\
\cdots m^2(\eta_1)\,G_0(y_1,x')\ ,
\label{eq:born_series}
\end{multline}
with $\eta_i$ the time of the event $y_i$. Since $G_0$ is supported on the light cone only, the $n$-th term is a sum over paths made of $n+1$ null segments, joined at $n$ events where the wave is scattered with amplitude $m^2$. This expression is equivalent to the Neumann series~\eqref{eq:neumann_1}--\eqref{eq:neumann_n}. Each order of equation~\eqref{eq:born_series} is the retarded solution of $\Box_0G^{(n)}=m^2G^{(n-1)}$, starting from $G^{(0)}=G_0=\delta(u)/r$.  In analogy to~\eqref{eq:bulk_eq}--\eqref{eq:transport_eq}, writing $G^{(n)}=\theta(u)\,g_n/r$ for $n\geq1$, the equation for $G^{(n)}$ splits into two parts. Inside the light cone it gives the bulk equation $4\partial_u\partial_vg_n=m^2g_{n-1}$. On the light cone, the derivatives of the step $\theta(u)$ give a term $\delta(u)$ times the derivative of $g_n$ along the outgoing ray, which must match the part of the source $m^2G^{(n-1)}$ that is traveling exactly with the front. Only the unscattered wave $G^{(0)}=\delta(u)/r$ has such a part. For single scattering, $n=1$, the source $m^2G_0$ lies entirely on the light cone, and as the front advances along the ray, it crosses the curvature $m^2$ at each point. The waves scattered forward stay on the ray, so that $g_1$ grows along it by $m^2/2$ per unit time. Integrating from the source gives the light-cone data~\eqref{eq:g0v}, the curvature accumulated by the front. For $n\geq2$, the source $m^2\theta(u)g_{n-1}/r$ is a bounded function, with nothing concentrated on the light cone. Orders $n\geq2$ are sourced by waves already scattered inside the cone, which lag behind the front, and therefore vanish on the light cone.

\subsubsection*{Single scattering} 
\noindent
The single-scattering term ($n=1$) is computed in closed form. The integral
\begin{equation}
\int d^3x''\frac{\delta(\eta''-\eta'-|\bfx''-\bfx'|)\,\delta(\eta-\eta''-|\bfx-\bfx''|)}{|\bfx''-\bfx'|\,|\bfx-\bfx''|}
\end{equation}
equals $2\pi/r$ for $\eta'+u/2\leq\eta''\leq\eta'+v/2$ and vanishes otherwise. All scattering times within this interval have the same weight. Therefore,
\begin{equation}
B\simeq\frac{1}{2r}\int_{\eta'+u/2}^{\eta'+v/2}m^2(\eta'')\,d\eta''=\frac{g(0,v)-g(0,u)}{r}\ ,
\label{eq:B_born}
\end{equation}
that is, half the average of $m^2$ over a window of width $r$ centered at $(\eta+\eta')/2$. This is the first order $g_1/r$ of the Neumann series~\eqref{eq:neumann_1}, obtained here from the geometry of the scattering events rather than from the Goursat problem. The single-scattering term is exact on the light cone, where only the scattering along the ray itself contributes since all higher orders $g_{n\geq2}$ vanish, and it reduces to the boundary condition~\eqref{eq:B_lightcone}. The integration window $[\eta^\prime +u/2, \eta^\prime+v/2]$, of length $r=(v-u)/2$, collects all scattering events that can connect $x'$ to $x$ through a single scattering. The wave leaves the source event $(\eta',\bfx')$, is scattered at an event $(\eta'',\bfx'')$ and reaches the field point $(\eta,\bfx)$, along two null legs, $\eta''-\eta'=|\bfx''-\bfx'|$ and $\eta-\eta''=|\bfx-\bfx''|$. The scattering events therefore lie on the intersection of the future light cone of the source with the past light cone of the field point, which projects onto space as the prolate spheroid with foci $\bfx'$ and $\bfx$ (see Figure~\ref{fig:shell_geometry_3d}) on which $|\bfx''-\bfx'|+|\bfx-\bfx''|=t$. Its earliest point, at $\eta''=\eta'+u/2$, lies on the line through $\bfx$ and $\bfx'$ behind the source. Its latest point, at $\eta''=\eta'+v/2$, lies just behind the field point. The retarded time $u$ thus sets the earliest time at which the wave can be scattered. Figure~\ref{fig:shell_geometry} shows the same construction along the line through $\bfx'$ and $\bfx$.

\subsubsection*{Scattering order by order} 
\noindent
The paths with $n\geq2$ scatterings are the orders $g_n$ of the recursion~\eqref{eq:neumann_n}. They vanish on the light cone, and each is computed with the marching~\eqref{eq:marching}, with the source $m^2g_{n-1}$ in place of $m^2g$. The dimensionless expansion parameter is $\mu=\eta^2m^2$: single scattering is accurate where $\mu\ll1$, as in the radiation-dominated plasma (Section~\ref{sec:full_history}). In $\Lambda$CDM instead,  $\mu\geq2$ at all times, and multiple scatterings are expected to provide significant corrections to the tail.

\begin{figure}[t]
\centering
\includegraphics[width=\linewidth]{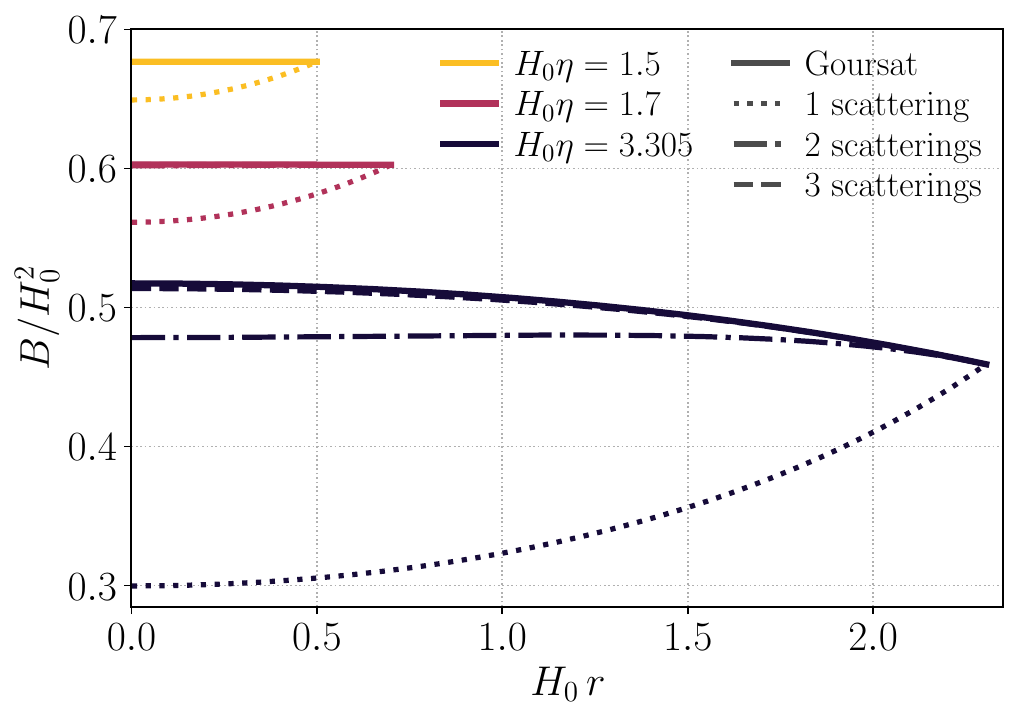}
\caption{Radial profile of the tail in $\Lambda$CDM ($\Om=0.3$, $\OL=0.7$) for a source at $H_0\eta'=1$, at different times after the source (colors). Each profile ends on the light cone, $r=t$. Solid: exact Goursat solution; dotted: single scattering~\eqref{eq:B_born}; dash-dotted and dashed: sum of the terms up to double and triple scattering~\eqref{eq:neumann_1}--\eqref{eq:neumann_n}.}
\label{fig:scattering_orders}
\end{figure}

Figure~\ref{fig:scattering_orders} compares the exact $\Lambda$CDM tail with the first three orders, for a source at $H_0\eta'=1$. All orders coincide on the light cone, where single scattering is exact. Inside the cone, single scattering underestimates the tail increasingly toward the worldline and with time, since the interaction potential $\mu$ increases in $\Lambda$ domination. Despite $\mu\geq2$, the series converges rapidly, with the three-scattering approximation reproducing the exact tail almost perfectly.
Near the light cone the tail is controlled by the amplitude of the single scattering, while deep inside the cone and at late times, it accumulates the contribution of multiple-scattering paths.

\section{The tail through the history of the universe}
\label{sec:full_history}
\noindent
The Goursat formulation of Section~\ref{sec:goursat} applies equally to sources emitted at any time, provided the potential $m^2=a''/a$ is known. We now consider the fate of a radiation burst of unspecified nature, emitted at some very early time. The source could be, for example, a cosmological first-order transition.

In exact radiation domination $a\propto\eta$, so that $a''=0$, the potential vanishes, and the propagator has no tail. However, the actual expansion history departs from this limit at all times. From the Friedmann equations in conformal time,
\begin{align}
m^2(\eta) & \equiv \frac{a''}{a}=\frac{4\pi G}{3}\,a^2\,(\rho-3p) \nonumber\\
\mu(\eta)&\equiv\eta^2m^2=\frac{(H_0\eta)^2a^2}{2\rho_{c}}\sum_X\Theta_X\ ,
\label{eq:mu_trace}
\end{align}
with $\rho_c$ the critical energy density today. 
The curvature interaction $m^2$ is then sourced by the trace $\Theta_X\equiv\rho_X-3p_X$ of every component. Massless particles have $\Theta=0$ and enter only through the expansion rate. A species of mass $m_i$, with $g_i$ internal states and momentum distribution $f_i(q)$, has
\begin{equation}
\Theta_i=\frac{g_i}{2\pi^2}\int_0^\infty dq\,\frac{q^2\,m_i^2}{E_i}\,f_i(q)\ ,\qquad E_i=\sqrt{q^2+m_i^2}\ ,
\label{eq:theta_species}
\end{equation}
which vanishes for $m_i\ll T$ and tends to $\rho_i$ once the species is non-relativistic. Cold matter contributes $\Theta_m=\rho_m$, and the cosmological constant, with $p_\Lambda=-\rho_\Lambda$, $\Theta_\Lambda=4\rho_\Lambda$. For the Standard Model plasma in thermal equilibrium at the photon temperature $T$, we use the effective numbers of degrees of freedom of Ref.~\cite{Saikawa:2018rcs}, defined by $\rho_{\rm pl}=\frac{\pi^2}{30}g_{*\rho}T^4$ and $s_{\rm pl}=\frac{2\pi^2}{45}g_{*s}T^3$, which include the QCD and electroweak crossovers. With vanishing chemical potentials $p_{\rm pl}=Ts_{\rm pl}-\rho_{\rm pl}$, so that
\begin{equation}
\Theta_{\rm pl}=4\rho_{\rm pl}-3Ts_{\rm pl}=\frac{2\pi^2}{15}\,(g_{*\rho}-g_{*s})\,T^4\ .
\end{equation}
This holds only while all species share the temperature $T$. After neutrino decoupling, $T_\nu\neq T$, and below $T=20 \,\mathrm{MeV}$ we compute the trace $\Theta_{\rm pl}$ from equation~\eqref{eq:theta_species}, summing $e^\pm$, $\mu^\pm$, and light hadrons in equilibrium, $f_i=1/(e^{E_i/T}\pm1)$, with their measured masses~\cite{ParticleDataGroup:2024cfk}. Muons and pions dominate the trace at $T\simeq10 \,\mathrm{MeV}$, while $e^\pm$ dominate below a few $\mathrm{MeV}$. We include one massive neutrino with $m_\nu=0.06$~eV, whose relic distribution keeps the relativistic form $f=1/(e^{q/T_\nu}+1)$, with $T_\nu=(4/11)^{1/3}T_0/a$. It contributes to the trace~\eqref{eq:theta_species} with $g=2$. The scale factor follows from entropy conservation, $g_{*s}a^3T^3={\rm const}$. In this section, we use the Planck 2018 parameters~\cite{Planck:2018vyg} ($h=0.6736$, $\Om=0.3138$), which give $H_0\eta_0=3.180$.

\begin{figure}[t]
\centering
\includegraphics[width=\linewidth]{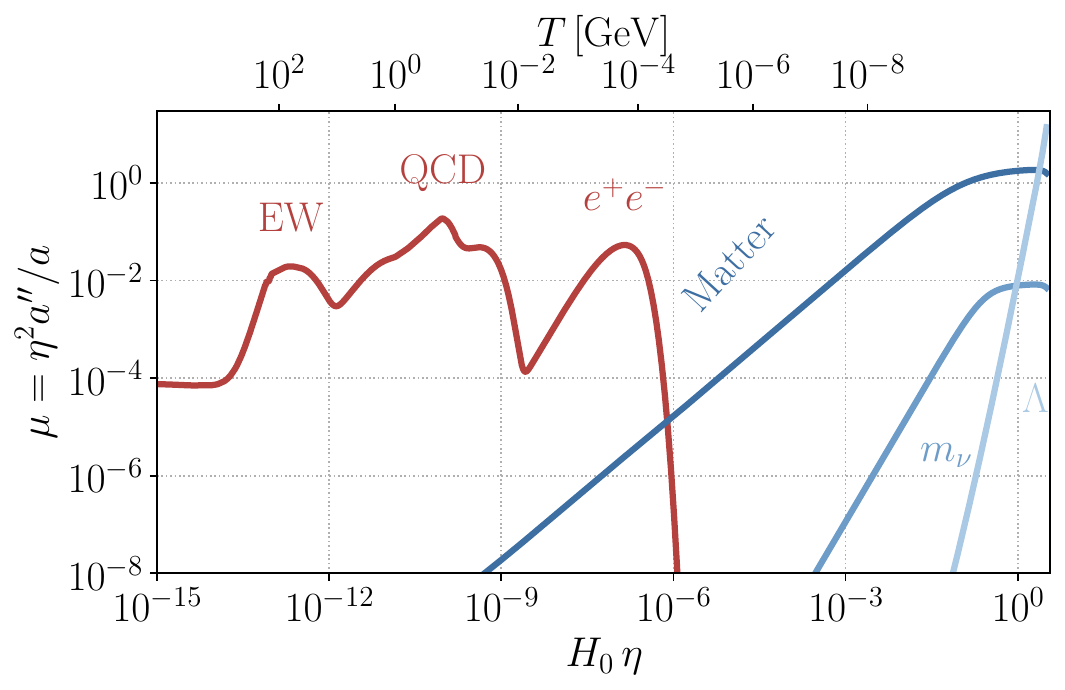}
\caption{The potential $\mu=\eta^2a''/a$ of equation~\eqref{eq:mu_trace} through the whole expansion history: Standard Model plasma (red), cold matter, massive neutrinos, and $\Lambda$ (blue). The top axis gives the photon temperature.}
\label{fig:mu_history}
\end{figure}

We compute $\mu(\eta)$ on $10^5$ logarithmically spaced nodes, from $T=3\times10^5 \,\mathrm{GeV}$ to today. The result is shown in Figure~\ref{fig:mu_history}. During radiation domination $\mu$ is small but not zero: it is $\simeq10^{-4}$ above the electroweak scale with peaks $\mu\simeq0.02$, $0.2$, and $0.05$ at the electroweak crossover, QCD crossover, and $e^+e^-$ annihilation, respectively. In between these events, the plasma is nearly conformal, and after the $e^+e^-$ annihilation peak $\mu$ drops down to $\sim10^{-5}$. It then grows during the transition to the matter era toward the value $\mu=2$ and reaches the maximum $\mu\simeq15$ today because of the cosmological constant $\Lambda$.

\begin{table*}[t]
\caption{Epochs marked in Figure~\ref{fig:full_history_u_tau}: conformal time $H_0\eta_e$ from the background of Figure~\ref{fig:mu_history}, and retarded time $u_e=2(\eta_e-\eta')$ of the corresponding line for a source at $H_0\eta'=10^{-15}$.}
\label{tab:epochs}
\begin{ruledtabular}
\begin{tabular}{rlccc}
 & epoch & $T$ or $z$ & $H_0\eta_e$ & $H_0u_e$ \\
\hline
1 & EW crossover & $T\simeq160 \,\mathrm{GeV}$ & $8.216\times10^{-14}$ & $1.623\times10^{-13}$ \\
2 & QCD crossover & $T\simeq155 \,\mathrm{MeV}$ & $1.137\times10^{-10}$ & $2.273\times10^{-10}$ \\
3 & $\nu$ decoupling & $T\simeq1.5 \,\mathrm{MeV}$ & $1.287\times10^{-8}$ & $2.575\times10^{-8}$ \\
4 & $e^+e^-$ annihilation & $T\simeq0.17 \,\mathrm{MeV}$ & $1.281\times10^{-7}$ & $2.563\times10^{-7}$ \\
5 & BBN & $T\simeq0.07 \,\mathrm{MeV}$ & $3.557\times10^{-7}$ & $7.115\times10^{-7}$ \\
6 & matter-radiation equality & $z\simeq3400$ & $2.537\times10^{-2}$ & $5.073\times10^{-2}$ \\
7 & recombination & $z\simeq1090$ & $6.302\times10^{-2}$ & $1.260\times10^{-1}$ \\
\end{tabular}
\end{ruledtabular}
\end{table*}

\vspace{5 mm}

\subsection{Scattering in the early universe}
\noindent
For a source in the plasma era, $m^2=\mu/\eta^2$ falls steeply after the source. Multiple scatterings in the Born series~\eqref{eq:born_series} are therefore significantly suppressed, and the tail is dominated by single scattering~\eqref{eq:B_born} right after the source time. On the light cone the window starts at the source $\eta'$, while inside the cone the epochs before $\eta'+u/2$ no longer contribute, and, at fixed observation time, $B$ falls from the light cone toward the worldline. This is the opposite of the matter-era source of Section~\ref{sec:res_lcdm}, where $B$ grows toward the worldline due to the large curvature potential $\mu \geq 2$. In the matter era multiple scatterings contribute more and enhance the tail toward the worldline, correcting significantly the approximate prediction from single scattering~\eqref{eq:B_born}. 

\begin{figure}[t]
\centering
\includegraphics[width=\linewidth]{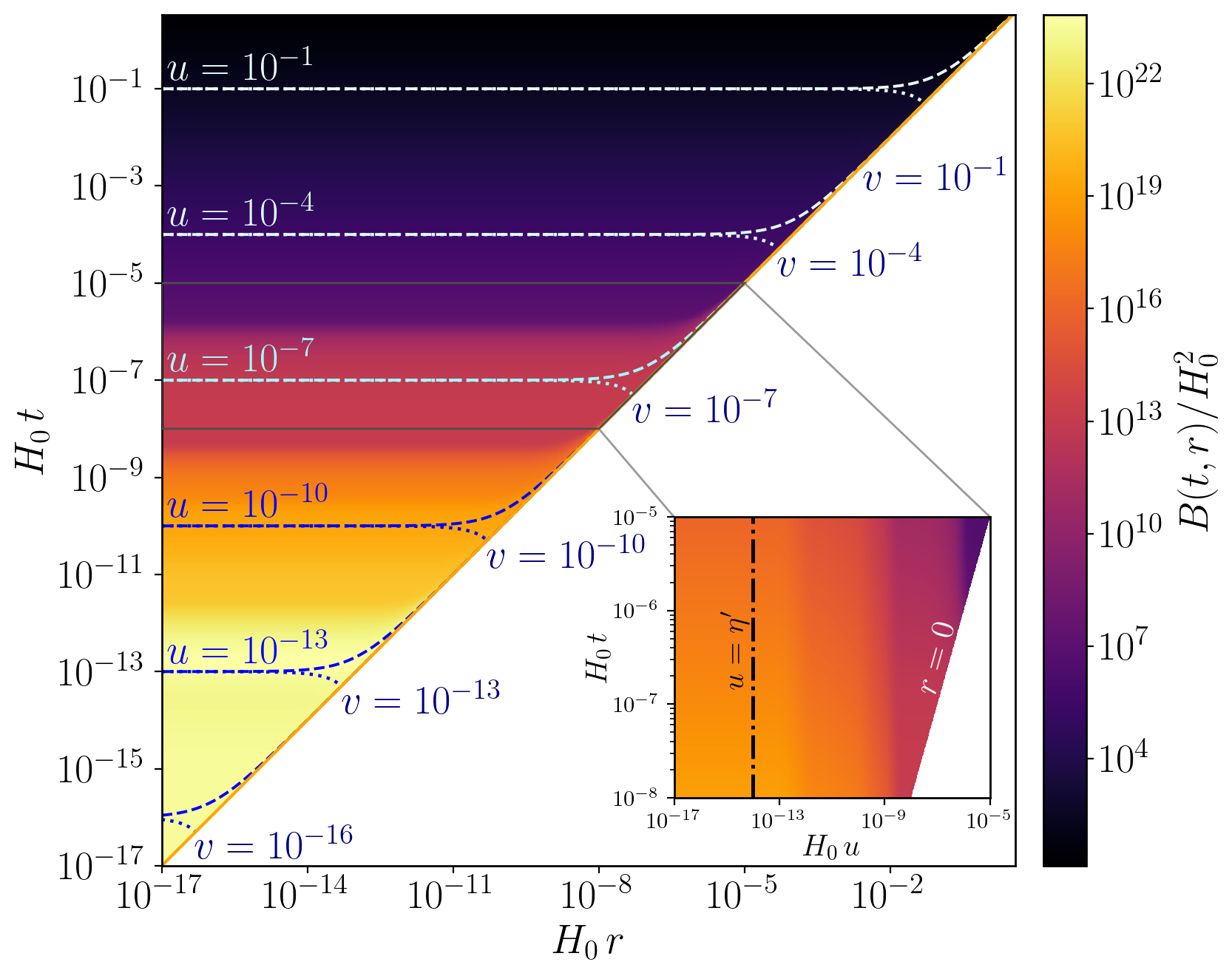}
\caption{Tail $B$, in units of $H_0^2$, of a source at $H_0\eta'=10^{-14}$ over the Goursat wedge, from the source to today, in logarithmic coordinates $(r,{t})$, ${t}=\eta-\eta'$. The light cone $r={t}$ is the diagonal; dashed and dotted lines are the null lines $u={t}-r$ and $v={t}+r$ of the indicated values, and the dash-dotted line marks ${t}=\eta'$. The inset unfolds the strip $10^{-7}\leq H_0{t}\leq10^{-5}$ in the distance from the light cone, the retarded time $u={t}-r$; the gray line is the worldline $u={t}$ ($r=0$), and the dash-dotted line marks $u=\eta'$.}
\label{fig:full_history_wedge}
\end{figure}

\begin{figure}[t]
\centering
\includegraphics[width=\linewidth]{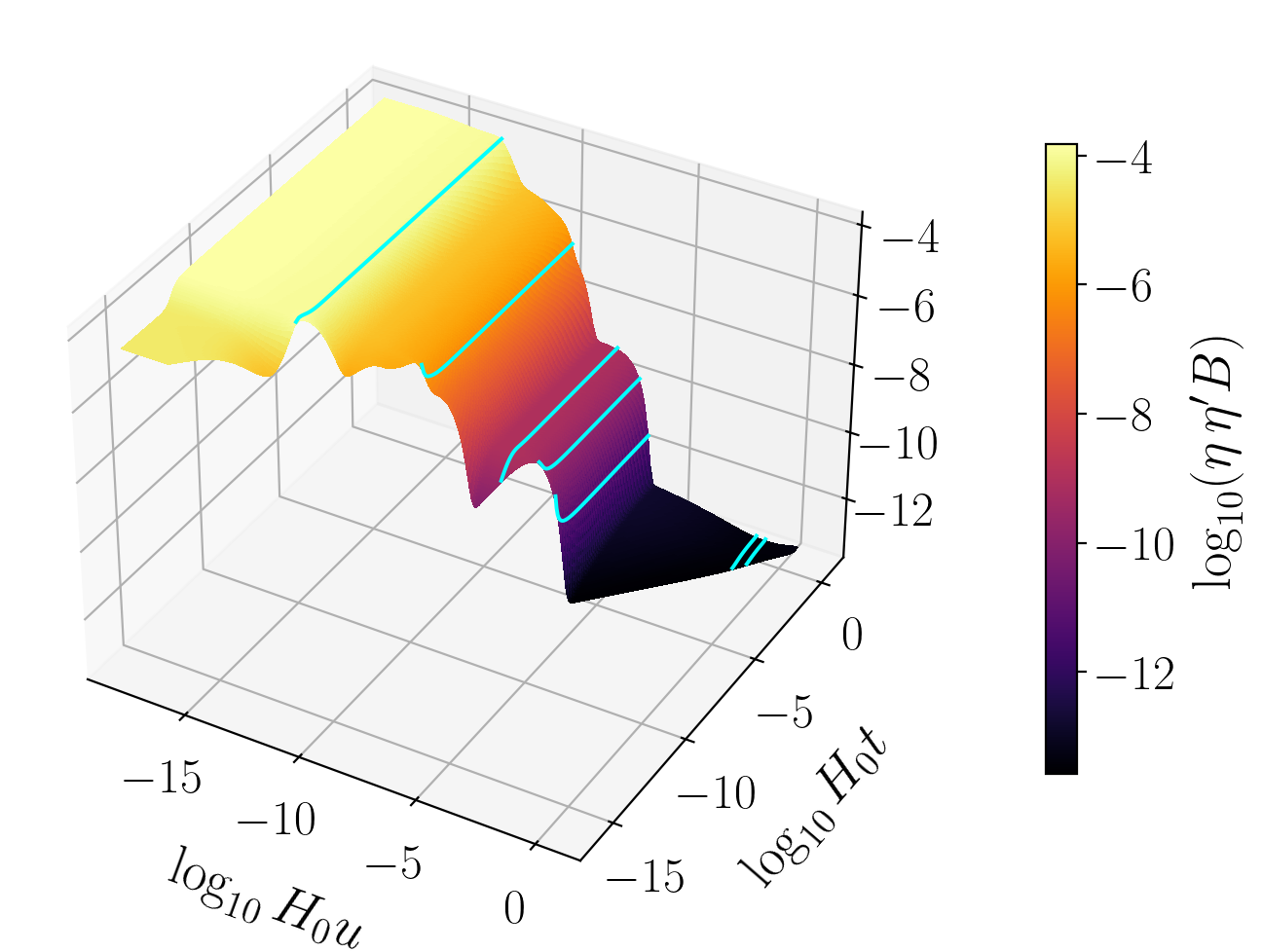}
\caption{Tail $\eta\eta'B$ of a source at $H_0\eta'=10^{-15}$ as a function of the retarded time $u={t}-r$ and of ${t}=\eta-\eta'$ ($u\leq{t}$; $u={t}$ is the worldline, and $u\to0$ is the light cone). The cyan lines mark $u_e=2(\eta_e-\eta')$ for the epochs $\eta_e$ of Table~\ref{tab:epochs}, from left to right.}
\label{fig:full_history_u_tau}
\end{figure}

Figure~\ref{fig:full_history_wedge} shows $B$ for a source at $H_0\eta'=10^{-14}$ ($T\simeq1.3 \,\mathrm{TeV}$), from the source to today. The scale is dominated by the excursion of $1/(\eta\eta')$, which makes $B$ decrease by about twenty orders of magnitude from the source to today, and the imprint of the potential is in the combination $\eta\eta'B$. Near the source, ${t}\lesssim\eta'$, the window~\eqref{eq:B_born} shrinks to $\eta'$, and $\eta\eta'B\simeq\mu(\eta')/2$ is uniform. At later times the tail is set by the distance from the light cone, the retarded time $u$, which fixes the earliest scattering time $\eta'+u/2$. Near the light cone, $u\ll{t}$, the integral in~\eqref{eq:B_born} is dominated by its lower end; the upper end of the window can be sent to infinity, and the tail depends only on $u$:
\begin{equation}
\eta\eta'B\simeq \frac{\eta'}{2}\int_{\eta'+u/2}^{\infty}\frac{\mu(\eta'')}{\eta''^2}\,d\eta''\ ,
\label{eq:B_of_u}
\end{equation}
where we used $\eta\eta'/(2r)\simeq\eta'/2$ for $u\ll t$ and $t\gg\eta'$. On the light cone the window starts at the source, and $\eta\eta'B$ remains at $\simeq 1.2\times10^{-3}$ at all times, set by the epochs right after the source (about two decades in $\eta$). 
For $u\gtrsim\eta'$ these epochs drop out of the window, and the tail falls toward the interior: for $\eta\gg\eta'$ the light cone and the interior of the wedge differ by more than nine orders of magnitude. On the logarithmic axes of the main panel of Figure~\ref{fig:full_history_wedge} this excursion is compressed onto the diagonal, since $u\ll{t}$ means $r$ close to ${t}$, and it is unfolded in the inset by plotting in $u$, the distance from the light cone. As in Figure~\ref{fig:full_history_u_tau}, which shows $\eta\eta'B$ of a source at $H_0\eta'=10^{-15}$ over the whole wedge in the variables $(u,{t})$, the bands, ridges, and dips run parallel to the ${t}$ axis. The features of $\mu$ sit at the retarded times $u_e=2(\eta_e-\eta')$ of the epochs of Table~\ref{tab:epochs}, marked by the cyan lines in Figure~\ref{fig:full_history_u_tau}. Once generated, they are carried outward unchanged along the null lines $u={\rm const}$, at all later times.  
Near the worldline, $r\ll{t}$, the profile is flat in $r$ and $u\simeq{t}$, so that the tail follows the potential at the intermediate time $(\eta+\eta')/2=\eta'+u/2$ (recall equation~\eqref{eq:B_worldline}), and the same features appear in the main panel of Figure~\ref{fig:full_history_wedge} as horizontal bands at ${t}\simeq u_e$. The tail of an early source is therefore a record of the subsequent thermal history, ordered in retarded time, while the signal is almost entirely contained in the light cone.

\begin{figure}[t]
\centering
\includegraphics[width=\linewidth]{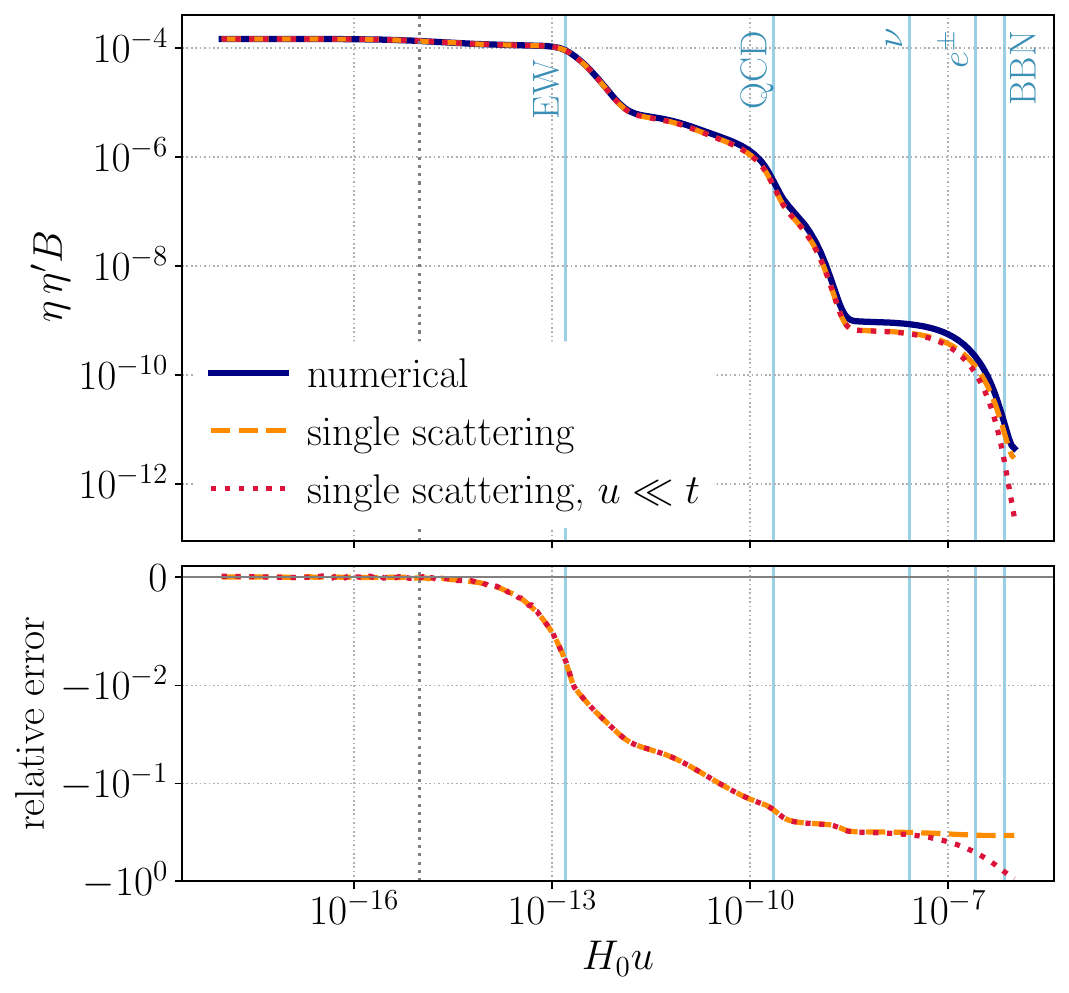}
\caption{Single scattering versus the numerical solution on the section $H_0t=10^{-6}$ of the wedge of a source at $H_0\eta'=10^{-15}$, as a function of the retarded time $u$, from the light cone (left) to the worldline $u=t$ (right). Top: $\eta\eta'B$ from the Goursat solution (solid), from the single-scattering formula~\eqref{eq:B_born} (dashed), and from its limit $u\ll t$, equation~\eqref{eq:B_of_u}, with the upper end of the window at $\eta_0$ (dash-dotted). Bottom: relative error of the two approximations. The dotted line marks $u=\eta'$, the vertical lines the epochs $u_e=2(\eta_e-\eta')$ of Table~\ref{tab:epochs}.}
\label{fig:single_scatter}
\end{figure}

For $u\gtrsim10^{-6}H_0^{-1}$, after $e^+e^-$ annihilation, the fall of $\eta\eta'B$ slows down during matter-radiation equality and recombination. This can only be partially explained by the single-scattering approximation developed in this section. In fact, once the plasma is conformal, only matter contributes to the curvature interaction. During the transition from radiation to matter domination, $a\propto(\eta/\eta_1)^2+2\eta/\eta_1$ with $\eta_1=\eta_{\rm eq}/(\sqrt2-1)\simeq0.061\,H_0^{-1}$, so that
$m^2= 2/[\eta(\eta+2\eta_1)]$, and,   
for $\eta'\ll u\ll{t}$,
\begin{equation}
\eta\eta'B\simeq\frac{\eta'}{2\eta_1}\,\log\Big(1+\frac{4\eta_1}{u}\Big)\ ,
\label{eq:B_floor}
\end{equation}
which depends on $u$ only logarithmically, whereas in the plasma era, where $\mu$ is roughly constant, $B\propto1/u$. The flatter profile of the exact tail can only be described by multiple scatterings, since $\mu\sim 2$ in the matter era. 

Figure~\ref{fig:single_scatter} compares the single-scattering formula~\eqref{eq:B_born} and its limit~\eqref{eq:B_of_u} with the numerical solution. Single scattering is exact on the light cone and reproduces all the features of the profile at the epochs of Table~\ref{tab:epochs}, but inside the cone it underestimates the tail by $\sim10\%$ after the electroweak epoch and by $25$--$35\%$ in the plasma era. The difference is due to double and higher-order scatterings, which are suppressed by $\mu$ but enhanced by the large range of scales. Equation~\eqref{eq:B_of_u} coincides with~\eqref{eq:B_born} for $u\ll t$ and departs from it near the worldline and once the plasma becomes conformal.

\section{Light cone and tail across scales}
\label{sec:fourier}

\noindent
For spectral applications ({\emph{e.g.}}, the gravitational-wave energy spectrum from a population of mergers), one needs the retarded Green's function and its derivative in Fourier space, $\widetilde G(\eta,\eta';k)\equiv\int d^3\bfr\,G(\eta,\bfx;\eta',\bfx')\,e^{-i\bfk\cdot\bfr}$. As done in coordinate space, we split the propagator into light-cone and tail pieces, $\widetilde G_{\rm ret}=\widetilde G_{\LC}+\widetilde G_{\TL}$. The light-cone piece is analytic,
\begin{equation}
    \widetilde G_{\LC}(\eta,\eta';k) = \frac{\sin\!\big(k(\eta-\eta')\big)}{k}\ ,
    \label{eq:GLC_fourier}
\end{equation}
with derivative 
\begin{align}
\partial_{\eta_0}\widetilde G_{\LC}(\eta_0,\eta;k)=\cos\!\big(k(\eta_0-\eta)\big)\ .
\label{eq:GLC_and_deriv}
\end{align}
The tail instead contributes the term
\begin{equation}
    \widetilde G_{\TL}(\eta,\eta';k) = \frac1k\int_0^{{t}}dr\,r\sin(kr)\,B(\eta,\eta',r) \ ,
    \label{eq:GTL_fourier}
\end{equation}
which, for each pair $(\eta,\eta')$ and each wavenumber $k$, reduces to a one-dimensional quadrature of the already-computed radial profile $B(\eta,\eta',r)$ over the causal interval $0\leq r\leq{t} = \eta-\eta^\prime$. Its time derivative is
\begin{align}
    \partial_\eta\widetilde G_{\TL} = \frac1k\int_0^{{t}}dr\,r\sin(kr)\,\partial_\eta B(\eta,\eta',r)  \nonumber \\
    + \frac{{t}\sin(k{t})}{k}\,B(\eta,\eta',{t}) \big|_{u=0} \ .
    \label{eq:dGTL_fourier}
\end{align}
The boundary term is evaluated exactly on the light cone, where $B$ is already known in closed form from \eqref{eq:B_lightcone}. The bulk contribution instead requires new ingredients for evaluating the time derivative $\partial_\eta B$. In Section~\ref{sec:numerics_deriv} we showed that the time derivative can be efficiently computed by adding a companion field to the numerical integration of the Goursat problem. 

As a useful analytic cross-check, we also record the ``constant-tail'' approximation, obtained by neglecting the $r$-dependence of $B$ over the (typically narrow) causal interval, {\emph{i.e.}}, by setting $B(\eta,\eta',r) \simeq B(\eta,\eta')\big|_{u=0}$ and evaluating it on the light cone,
\begin{align}
\widetilde G_{\TL}(\eta,\eta';k)
&=
B(\eta,\eta')\,
\frac{\sin k{t}-k{t}\cos k{t}}{k^3}
\label{eq:GTL_and_deriv_1}
\\
\partial_{\eta}\widetilde G_{\TL}(\eta,\eta';k)
&=
\big[\partial_{\eta}B(\eta,\eta')\big]\,
\frac{\sin k{t}-k{t}\cos k{t}}{k^3}
\nonumber\\
&\quad
+B(\eta,\eta')\,\frac{{t}\sin k{t}}{k} \ .
\label{eq:GTL_and_deriv_2}
\end{align}
Using the $\Lambda \rm CDM$ fit~\eqref{eq:B_tp_fit} and taking $\eta_0$ to be the present time, we obtain
\begin{align}
B(\eta_0,\eta')&\approx\frac{1}{\eta_0\eta'}+\frac{1}{(\eta_m-\eta_0)(\eta_m-\eta')}-0.095\,H_0^2 
\nonumber\\
\partial_{\eta_0}B(\eta_0,\eta')&\approx -\frac{1}{\eta_0^2\eta'}+\frac{1}{(\eta_m-\eta_0)^2(\eta_m-\eta')}\ .
\end{align}
These expressions are fully analytic and provide a useful benchmark for the numerical quadrature \eqref{eq:GTL_fourier}. 
Here we compare the tail propagator obtained from the Goursat solution with this constant-tail approximation, combined with the two-pole fit~\eqref{eq:B_tp_fit}, as used for the memory waveform of a burst in Ref.~\cite{Jokela:2022rhk}.

\begin{figure}[t]
\centering
\includegraphics[width=\linewidth]{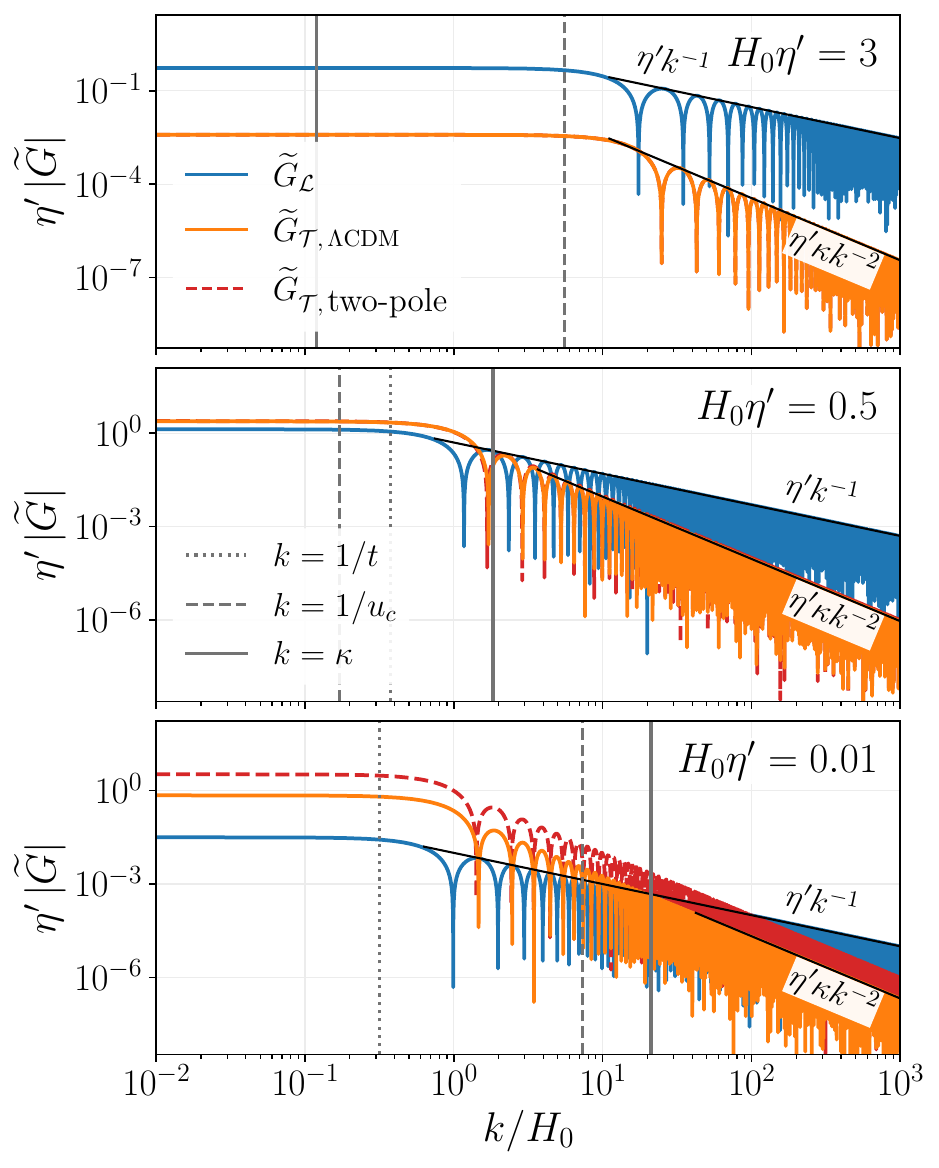}
\caption{Light-cone and tail propagators in Fourier space today, $H_0\eta=H_0\eta_0=3.180$, for sources at $H_0\eta'=3$, $0.5$, and $10^{-2}$ (top to bottom), computed with the interaction potential of Section~\ref{sec:full_history}. Solid blue: light cone~\eqref{eq:GLC_fourier}; solid orange: tail from the Goursat solution in exact $\Lambda$CDM~\eqref{eq:GTL_fourier}; dashed red: two-pole constant-tail approximation~\eqref{eq:GTL_and_deriv_1}, with $H_0\eta_m=4.330$. Black solid lines: the asymptotic envelopes $1/k$ of the light cone and $\kappa/k^2$ of the tail~\eqref{eq:GTL_asymptotic}. Vertical gray lines mark the scales $k{t}=1$ (dotted), $k{u_c}=1$ (dashed), and $k=\kappa$ (solid).}
\label{fig:G_fourier}
\end{figure}

Figure~\ref{fig:G_fourier} shows the light-cone and tail propagators today, $\eta=\eta_0$, for sources at different times. The tail is computed from the Goursat solution with the curvature interaction of Section~\ref{sec:full_history} (Planck parameters and Standard Model plasma, $H_0\eta_0=3.180$), and the two-pole approximation, defined by equations~\eqref{eq:GTL_and_deriv_1} and~\eqref{eq:B_tp_fit}, with $\eta_0$ and the conformal future infinity $H_0\eta_m=4.330$ of the same background.

The profile of the Green's function in momentum space is determined by the discontinuity at the light cone.
The light-cone propagator is the transform of an infinitely thin shell, $\delta(u)/r$. In the radial transform~\eqref{eq:GTL_fourier}, the $\delta$-function picks the integrand at $r={t}$, giving $\widetilde G_{\LC}=\sin(k{t})/k$, so that $|\widetilde G_{\LC}|\simeq{t}$ for $k{t}\lesssim1$ and falls as $1/k$ for $k{t}\gtrsim1$. The tail is instead the transform of the whole ball $r\leq{t}$, over which $g=rB$ is finite inside and jumps to zero at the light-cone front. The oscillations of $\sin(kr)$ cancel over the interior, so that the transform of the tail at $k{t}\gtrsim1$ is set by the profile of $g=rB$ just behind the light-cone front.
In terms of the retarded time $u={t}-r$, equation~\eqref{eq:GTL_fourier} reads
\begin{equation}
\widetilde G_{\TL}=\frac1k\int_0^{t}du\,\sin\!\big(k({t}-u)\big)\,g(u,2{t}-u)\ .
\label{eq:GTL_u}
\end{equation}
The contributions from $u\gg 1/k$ cancel due to the oscillation of the sine, while those from the strip $0\leq u\lesssim 1/k$ behind the front add coherently. Neglecting incoherent oscillations, the envelope of the tail is therefore the average of $g$ over the strip $0\leq u\leq1/k$ behind the light-cone front, namely,
\begin{equation}
k^2\,|\widetilde G_{\TL}|\simeq  k \int_0^{1/k}du\,g(u,2{t}-u)\ ,
\label{eq:GTL_envelope}
\end{equation}
and its local slope is
\begin{equation}
\frac{d\log|\widetilde G_{\TL}|}{d\log k}\simeq-2+n\ ,\qquad n\equiv-\frac{d\log g}{d\log u}\bigg|_{u\simeq1/k}\ .
\label{eq:GTL_slope}
\end{equation}
If $g$ varies behind the front on a scale $u_c$, a mode with $ku_c\gg1$ resolves the profile: over the strip $g$ keeps its light-cone value, $g(u,2{t}-u)\simeq g(0,2{t})$, and equation~\eqref{eq:GTL_envelope} gives $k^2|\widetilde G_{\TL}|\simeq g(0,2{t})$.
Using this value as the amplitude of the envelope, after integration by parts with $g=rB$ odd in $r$, equation~\eqref{eq:GTL_u} reduces to the asymptotic form
\begin{equation}
\widetilde G_{\TL}\simeq-\frac{\kappa}{k^2}\cos k{t}\ ,\qquad \kappa \equiv g(0,2t)=\frac12\int_{\eta'}^{\eta}m^2(\eta'')\,d\eta''\ .
\label{eq:GTL_asymptotic}
\end{equation}
The slope $k^{-2}$ of the tail comes from the jump of $G_{\TL}$ at the light-cone front, the step $\theta(u)$, and its amplitude $\kappa$ is the size of the jump, {\emph{i.e.}}, the total curvature accumulated by the front, which sets the light-cone data.
The light-cone term $G_{\LC}$ is instead a spike $\delta(u)/r$, the derivative of the step.

The propagators thus depend on three scales:
\begin{enumerate}
\renewcommand{\labelenumi}{\roman{enumi}.}
\item the radius ${t}$ of the light cone: for $k{t}\lesssim1$ both propagators are flat, while for $k{t}\gtrsim1$ they oscillate with phase $k{t}$;
\item the width $u_c$ of the region behind the front over which $g$ departs from its light-cone value $\kappa$: for $ku_c\gg1$ the mode resolves the front, and the tail takes the asymptotic form~\eqref{eq:GTL_asymptotic}, while for $1/{t}\lesssim k\lesssim1/u_c$ its envelope follows the profile of $g$ over the strip $u\lesssim1/k$, equation~\eqref{eq:GTL_envelope};
\item the size of the jump at the light cone, $\kappa$, accounting for the curvature accumulated by the front: the envelope $\kappa/k^2$ of the tail meets the envelope $1/k$ of the light cone at $k\simeq\kappa$, so that the tail dominates the propagator for $1/{t}\lesssim k\lesssim\kappa$, and the light cone for $k\gtrsim\kappa$. If $\kappa{t}\lesssim1$, the tail is below the light cone at all $k$.
\end{enumerate}
The width $u_c$ can be read off the slope of $g$ across the front, which is known in closed form from the light-cone data~\eqref{eq:dr_g_lc}. On the section of time $\eta$, $\partial_u=-\partial_r$, so that just behind the front $g\simeq\kappa\,(1-u/u_c)$, with
\begin{equation}
\frac1{u_c}\equiv\frac{\partial_rg|_{r={t}}}{\kappa}=\frac{m^2(\eta)+m^2(\eta')}{4\kappa}-\frac{\kappa}{2}\ .
\label{eq:uc}
\end{equation}
If $B$ does not depend on $r$, as in pure matter or $\Lambda$ domination, $\partial_rg=B=\kappa/{t}$ and $u_c={t}$: $g$ falls linearly from $\kappa$ on the light cone to zero on the worldline, the tail has no scale shorter than the radius ${t}$, and the asymptotic form~\eqref{eq:GTL_asymptotic} holds as soon as $k{t}\gtrsim1$.
The three panels of Figure~\ref{fig:G_fourier} illustrate the hierarchy of the scales $1/{t}$, $1/u_c$, and $\kappa$ for sources at different times as detected today, $H_0\eta_0 = 3.180$. 

For $H_0\eta'=3$, $\mu(\eta')\simeq10$, in the $\Lambda$ era, $B$ is nearly independent of $r$ and $u_c\simeq{t}$. The source has scattered off the background curvature only for a small fraction of a Hubble time, $\kappa{t}\simeq0.02$, and the tail is below the light-cone propagator at all $k$. 

For $H_0\eta'=0.5$, $\mu(\eta')\simeq1.6$, near the end of the matter era, the tail is generated mostly by the scattering right after the source, so that $\kappa\simeq1/\eta'$, while $u_c\simeq2{t}$. The asymptotic form~\eqref{eq:GTL_asymptotic} holds for all $k{t}\gtrsim1$, and the tail exceeds the light cone for $k\lesssim\kappa$. In both cases, the two-pole approximation reproduces well the exact $\Lambda$CDM value.

For $H_0\eta'=10^{-2}$, $\mu(\eta')=0.15$, in the radiation era, $\kappa\simeq0.2/\eta'$ and $u_c\simeq14\,\eta'$, and the three scales are well separated, $1/{t}\ll1/u_c<\kappa$: the tail dominates for all $k\lesssim\kappa$, its envelope follows the profile of $g$ behind the front for $k\lesssim1/u_c$, and it takes the asymptotic form only for $k\gg1/u_c$. The two-pole interaction, which reduces to matter domination at early times, misses the radiation era, where $m^2$ is suppressed, and overestimates the tail by a factor $\simeq 5$ at all $k$. Similarly, the two-pole approximation overestimates the tail for sources before matter-radiation equality.

\subsection{Tail as a phase delay inside the horizon}
\label{sec:phase}
\noindent
Equation~\eqref{eq:GTL_asymptotic} is the leading term of an expansion of the propagator in powers of $1/k$, which follows from integrating equation~\eqref{eq:GTL_fourier} by parts repeatedly in the limit $ku_c\gg1$. Each integration by parts gains a factor $1/k$ and moves one derivative onto $g$, so that the coefficient of $k^{-n-1}$ is set by the $(n-1)$-th derivative of $g$ across the light-cone front, while the worldline never contributes, since $g$ is odd in $r$. 
The leading tail, at order $k^{-2}$, is thus set by the light-cone value of $g$, and the next order, $k^{-3}$, by its radial derivative on the light cone, which is known in closed form~\eqref{eq:dr_g_lc} from the light-cone data. Integrating by parts once more, the retarded propagator is
\begin{multline}
\widetilde G_{\rm ret}=\frac{\sin k{t}}{k}-\frac{\kappa\cos k{t}}{k^2}\\
+\left[\frac{m^2(\eta)+m^2(\eta')}{4}-\frac{\kappa^2}{2}\right]\frac{\sin k{t}}{k^3}+\mathcal O(k^{-4})\ .
\label{eq:G_expansion}
\end{multline}
The first term is the light cone, the second the leading tail~\eqref{eq:GTL_asymptotic}, and the third is the next-to-leading order correction to the tail, set by the slope of $g$ across the light-cone front. The term $[m^2(\eta)+m^2(\eta')]/4$ is local, fixed by the background at the two ends of the path, while the term $-\kappa^2/2$ is accumulated along the whole light cone.

The accumulated term has a notable consequence for the power of the propagator. 
For an incoherent superposition of sources, such as a stochastic background, the strain power involves $|\widetilde G_{\rm ret}|^2$ and the energy density involves $|\partial_\eta\widetilde G_{\rm ret}|^2+k^2|\widetilde G_{\rm ret}|^2$, both averaged over the phase $k{t}$, which for $k{t}\gg1$ varies rapidly from source to source. Squaring equation~\eqref{eq:G_expansion}, with $\langle\sin^2k{t}\rangle=\langle\cos^2k{t}\rangle=1/2$ and $\langle\sin k{t}\cos k{t}\rangle=0$, the interference of the light cone with the leading tail averages out, while the square of the leading tail is canceled exactly by the interference of the light cone with the accumulated term, leaving
\begin{multline}
\langle |\widetilde G_{\rm ret}|^2 \rangle
= \frac{1}{2k^2}\left[1+\frac{m^2(\eta)+m^2(\eta')}{2k^2}\right]+\mathcal O(k^{-5})\ .
\label{eq:G_power}
\end{multline}
The same cancelation holds for the time derivative,
\begin{equation}
\langle|\partial_\eta\widetilde G_{\rm ret}|^2\rangle=\frac12\left[1+\frac{m^2(\eta')-m^2(\eta)}{2k^2}\right]+\mathcal O(k^{-3})\ ,
\end{equation}
so that in the energy density, $\frac12\big(\langle |\partial_\eta\widetilde G_{\rm ret}|^2 \rangle +k^2\langle |\widetilde G_{\rm ret}|^2 \rangle\big)=\frac12\big[1+m^2(\eta')/2k^2\big]$, and only the curvature at the source survives.
As a result, the tail accumulated along the path, which dominates the propagator for $k\lesssim \kappa$, drops out of the averaged power at large $k$, and only the local curvature at the source and at the observer survives.

The cancelation also sets the limits of the constant-tail approximation~\eqref{eq:GTL_and_deriv_1}, $\widetilde G_{\TL}\simeq-\kappa\cos k{t}/k^2+\kappa\sin k{t}/tk^3$. This captures the leading tail, set by the light-cone value alone, but replaces the coefficient of $\sin k{t}/k^3$ in equation~\eqref{eq:G_expansion} by $\kappa/{t}$. The constant tail reproduces the leading tail, but not the averaged power at order $k^{-4}$.

One can derive the same expansion by solving the gravitational-wave equation~\eqref{eq:chi_wave} directly in momentum space, in the limit $k\eta'\gg1$.
We perform this exercise in Appendix~\ref{app:born_k}, which also gives further insight into the picture of scattering off the curvature of Section~\ref{sec:scattering}. In Fourier space the Born series~\eqref{eq:born_series} becomes a series of time integrals, in which each scattering either keeps or reverses the direction of propagation of the wave. The forward scatterings keep the phase of the incident wave and add up coherently along the whole ray, leading exactly to equation~\eqref{eq:GTL_asymptotic}. 
The backward scatterings reverse the direction of the wave, so that their phase depends on the scattering time. Waves reversed at different times interfere destructively and contribute only at the two ends of the path, at relative order $m^2/k^2$. At all orders, the forward scatterings exponentiate into a phase, and for $k\gg\Hell$ along the path the propagator takes the form
\begin{equation}
\widetilde G(\eta,\eta';k)\simeq\frac{\sin\!\big(k{t}-\delta\big)}{\sqrt{\omega(\eta)\,\omega(\eta')}}\ ,\qquad\omega=\sqrt{k^2-m^2}\ ,
\label{eq:G_WKB}
\end{equation}
with $\delta \equiv \kappa/k$.
Inside the horizon the tail is therefore a phase delay $\delta$ of the light-cone wave, accumulated by forward scattering along the path, while the backward scattering at the two ends corrects its amplitude through the factors $\sqrt{\omega(\eta)}$ and $\sqrt{\omega(\eta')}$. Expanding equation~\eqref{eq:G_WKB} in $\delta$ and in $m^2/k^2$, gives back equation~\eqref{eq:G_expansion} (details in Appendix~\ref{app:born_k}), with the additional phase $\delta$ reproducing the second term, equation~\eqref{eq:GTL_asymptotic}, and the amplitude correction $\sqrt{\omega(\eta)\omega(\eta')}$ yielding the third term. In this form the cancelation of the tail is manifest: a phase delay does not change the amplitude of the wave, so it drops out of the power averaged over the phase, $\langle\sin^2(k{t}-\delta)\rangle=1/2$, and only the amplitude, set by the local curvature at the two ends, survives. In position space the same cancelation is carried by the slope of the tail across the light-cone front, equation~\eqref{eq:dr_g_lc}.

\section{Beyond homogeneity}
\label{sec:inhomogeneities}

\noindent
The decomposition~\eqref{eq:G_decomp} into a light-cone and a tail term, and the characteristic problem for the tail, hold on any globally hyperbolic spacetime~\cite{Friedlander:2010eqa,Poisson:2011nh}. Homogeneity and isotropy constrain $B$ to depend on $(\eta,\eta',r)$ only and are needed to reduce the problem to the $(1+1)$-dimensional Goursat problem in $(u,v)$. In a perturbed FLRW universe, $ds^2=a^2[-(1+2\Phi)d\eta^2+(1-2\Psi)d\bfx^2]$, and each component of $\chi_{ij}=ah_{ij}$ obeys the wave equation of the perturbed metric. The operator of equation~\eqref{eq:green_eq} becomes $\mathcal L+\delta\mathcal L$, where $\delta\mathcal L$ is the expansion of the differential operator to first order in $\Phi$ and $\Psi$ that follows from linearizing $\partial_\mu(\sqrt{-g}\,g^{\mu\nu}\partial_\nu h)$ with $h=\chi/a$. Since $\mathcal L\,\bar G=4\pi\delta^{(4)}$, with $\bar G$ the FLRW Green's function~\eqref{eq:G_decomp}, the Green's function of the perturbed universe is
\begin{align}
G(x,x')&=\bar G(x,x')-\frac{1}{4\pi}\int d^4y\,\bar G(x,y)\,\delta\mathcal L_y\,\bar G(y,x') \nonumber \\
& \qquad\qquad\qquad\qquad\qquad + \mathcal O(\Phi^2,\Psi^2)\ .
\end{align}
The product of the two light-cone terms of $\bar G$ deforms the light cone (time delay and lensing), while the terms containing $B$ correct the tail and require the full position-space solution, including the radial profile of $B$. In particular, the light-cone value of the tail, which in FLRW is the integral of $m^2=a''/a$ along the ray, equation~\eqref{eq:B_lightcone}, acquires an integral of $\Phi$, $\Psi$, and their derivatives along the ray, in analogy with the integrated Sachs--Wolfe effect. Perturbations on scales larger than the separation between source and field point act instead as a different local FLRW history, and the Goursat solution applies with the local $m^2(\eta)$. Methods beyond the $(1+1)$-dimensional Goursat characteristic formulation are needed instead to investigate inhomogeneities on smaller scales, for which the problem remains $(3+1)$-dimensional.

\section{Discussion}
\label{sec:discussion}

\noindent
We formulated the tail $B(\eta,\eta',r)$ of the retarded Green's function of gravitational waves in a spatially flat FLRW background as the solution of a Goursat problem, a $(1+1)$-dimensional characteristic problem for $g(\eta,r)=rB(\eta,r)$, with data fixed in closed form on the light cone by a single integral of $m^2=a''/a$, equation~\eqref{eq:B_lightcone}, and by regularity on the worldline. We performed the numerical computation of the tail with a marching scheme in double-null coordinates. Given the expansion history $a(\eta)$, this formulation applies to any homogeneous and isotropic cosmology, and resolves the radial profile of $B$ throughout the interior of the light cone directly in position space, without Fourier transforms and without fitting the interaction.  In $\Lambda$CDM, we verified this construction against a double series expansion around the corner $r\sim t\sim 0$, which follows from the Born series and, equivalently, from the Riemann function of the Goursat problem. We found agreement down to the discretization error of the marching, $\simeq2\times10^{-8}$, within the radius of convergence of the series (Figure~\ref{fig:series_convergence}). On a non-uniform grid, the same scheme follows a single source over the whole expansion history, from the electroweak epoch to today. 

By expanding the Green's function in powers of $m^2$, we make the scattering picture of the tail~\cite{Faraoni:1991xe,Faraoni:1999us} quantitative, which offers a physical intuition for the tail as the sum of multiple scatterings off the curved background. At first order, corresponding to single scattering, $B$ is half the average of $m^2$ over the scattering times $\eta'+u/2\leq\eta''\leq\eta'+v/2$ allowed by the causal geometry, expressed in equation~\eqref{eq:B_born}. This is exact on the light cone, where the window covers the whole ray and equation~\eqref{eq:B_born} reduces to~\eqref{eq:B_lightcone}. Inside the cone, the retarded time $u$ sets the earliest scattering time, $\eta'+u/2$, so that the tail at retarded time $u$ samples the expansion history only after it and is dominated by that time when $m^2$ falls steeply, as in the radiation era. More generally, the Green's function is a Born series of light-cone propagators, each order carrying one more power of $m^2$, which in this picture corresponds to one additional scattering. The Goursat problem sums all its orders at once: in $\Lambda$CDM, for a source at $H_0\eta'=1$, three scatterings reproduce the exact tail to better than $1\%$ (Figure~\ref{fig:scattering_orders}). Multiple scatterings are suppressed by $\mu$ but enhanced by the logarithm of the range of scales, so that their contribution corrects single scattering by $25$--$35\%$ in the plasma era, where $\mu\sim10^{-2}$ (Figure~\ref{fig:single_scatter}), and becomes of order unity in the matter era, where $\mu\simeq2$. We stress that the scattering picture of broken null-paths is an interpretation, not a consequence of the linearized Einstein equations. Solving these equations as the Goursat problem (Section~\ref{sec:goursat}) establishes only that the curvature gives the Green's function a component connecting timelike-separated points.

\emph{Goursat solution and two-pole interaction.} The two-pole model of Refs.~\cite{Jokela:2022rhk,Jokela:2023suz}, defined in equation~\eqref{eq:m2_twopoles}, reproduces the exact $\Lambda$CDM tail to within $1.5\%$ on the light cone and to within $4\%$ everywhere inside the cone, for a source at $H_0\eta'=1$. The largest difference, near the worldline at late times, comes from the excess of the interaction in the transition between matter and $\Lambda$ domination (Section~\ref{sec:res_lcdm}). It fails for sources before matter-radiation equality, whose tail is set by the trace of the Standard Model plasma (Sections~\ref{sec:full_history} and~\ref{sec:fourier}).

\emph{Early sources as probes of the expansion history.} Since the interaction with the curved background is measured by departures from conformality, $\mu\propto a^2(\rho-3p)$, we investigated the tail of an early source as a probe of the trace of the cosmic fluid, {\emph{i.e.}}, of its equation of state. The electroweak and QCD crossovers, the annihilation of $e^\pm$, and the onset of matter domination appear as features at fixed $u$, carried unchanged to later times (Figure~\ref{fig:full_history_u_tau}). Any departure from the Standard Model history that changes the trace, such as additional species becoming non-relativistic, an early matter-dominated era, or a first-order phase transition, would leave its imprint in the tail. 
Whether these features are observable depends on how they enter the memory and is not settled here (see below).

\emph{From the tail to the memory.} This work establishes the tail $B$ with high numerical accuracy of order $10^{-7}$ (Figure~\ref{fig:convergence_goursat}). 
The calculation of the memory requires two further steps. The first is the transverse-traceless projection that enters the strain. It is exact and algebraic in momentum space, where the tail acts identically on the two helicities, and non-local in position space, where the radiation-like projection along the direction of propagation is exact only for the light-cone part of the signal far from the source. The second is the convolution with the source. For a stochastic background of bursts, our estimates (see Section~\ref{sec:phase}) indicate that the phase accumulated by the tail cancels in the gravitational-wave power spectrum, leaving a local correction of relative order $(a''/a)/k^2$. The precise calculation of the stochastic gravitational-wave memory background, as well as the memory waveform of a single burst, is left  for the companion paper. The expansion of Section~\ref{sec:phase} also explains why extending the light-cone value of $B$ as a constant in $r$, as assumed in Ref.~\cite{Jokela:2022rhk}, captures the leading tail, set by the light-cone value alone, but misses the next coefficient, set by the radial slope of the tail behind the front, which controls the residual power after the cancelation of the accumulated tail~\eqref{eq:G_power}.

\emph{Beyond homogeneity.} The reduction to a $(1+1)$-dimensional Goursat problem relies on homogeneity and isotropy. In a perturbed universe the Green's function follows from the FLRW one at first order in the metric perturbations, and the light-cone value of the tail acquires an integral of the potentials along the ray, analogous to the integrated Sachs--Wolfe effect (Section~\ref{sec:inhomogeneities}). In the future, we aim to investigate whether the tail can thereby probe primordial inhomogeneities, encoding information that is absent from the light-cone part of the signal.

\begin{acknowledgments}
\noindent
T.~B. was supported in part by the NSF grant PHY-2409173. L.~G. was supported by Jenny and Antti Wihuri Foundation. N.~J. was supported in part by the Research Council of Finland through grant no. 354533 and the Centre of Excellence in Neutron-Star Physics (project 374062). 

\end{acknowledgments}

\appendix

\section{Double-series expansion from the Riemann method}
\label{app:riemann}

\noindent
Here we give an independent, all-orders derivation of the corner series of Section~\ref{sec:series}, based on the Riemann function of the Goursat problem \eqref{eq:bulk_uv}. Besides serving as a cross-check of \eqref{eq:a0j}--\eqref{eq:closure}, this construction provides a representation of $g$ valid at any point of the wedge, not only within the finite disk of convergence of the Taylor series \eqref{eq:Ansatz}.

\subsection{Analyticity of the tail and the corner series from the Neumann series}
\label{app:analyticity}
\noindent
The Ansatz~\eqref{eq:Ansatz} assumes that $g$ is given by a convergent double power series near the corner. We prove here that this is the case and derive the recursion~\eqref{eq:interior}, the closure~\eqref{eq:closure}, and the number of scattering orders contributing to each coefficient using the Neumann series~\eqref{eq:neumann_1}--\eqref{eq:neumann_n}, which we recall here for clarity:
\begin{align}
g_1(u,v)&=g(0,v)-g(0,u) \label{app:neumann_1}\\
g_{n+1}(u,v)&=\frac14\int_0^u\!\!\int_0^v m^2(\eta'')\,g_n(u',v')\,du'dv'\ .\label{app:neumann_n}
\end{align}

Let us assume that the interaction potential $m^2$ is analytic at the source time $\eta^\prime$. Let $\rho$ be the radius of convergence of the Taylor series around $\eta^\prime$, so that 
\begin{equation}
	m^2(\eta'+t)=\sum_pc_pt^p
\end{equation}
for $|t|<\rho$. We work on the real quadrant
\begin{equation}
    Q=\big\{(u,v):\ u,v\geq0,\ u+v<2\rho\big\}\ ,
    \label{eq:quadrant}
\end{equation}
such that, for every $(u,v)\in Q$, the rectangle $[0,u]\times[0,v]$ is contained in $Q$. On the wedge, $0\leq u\leq v$, the condition $u+v<2\rho$ reads $t<\rho$ for any $r\leq t$.

The light-cone data~\eqref{eq:g0v} are obtained by integrating the series for $m^2$ term by term and can then be expanded as 
\begin{equation}
	g(0,v) = F(v)=\sum_{j\geq1}a_{0j}v^j \ ,
\end{equation}
with $a_{0j}$ given by~\eqref{a_lc}. The first-order term $g_1=F(v)-F(u)$ of the Neumann series~\eqref{app:neumann_1} therefore has the power series $\sum_{i,j}a^{(1)}_{ij}u^iv^j$ with only light-cone coefficients, $a^{(1)}_{0j}=a_{0j}$ and $a^{(1)}_{j0}=-a_{0j}$. One can then proceed with the recursion~\eqref{app:neumann_n} and derive the next terms. 
Inserting 
\begin{equation}
	g_n=\sum_{i,j}a^{(n)}_{ij}u^iv^j 
\end{equation}
and the series for $m^2$ into equation~\eqref{eq:neumann_n}, the Cauchy product and the term-by-term integration give, for $n\geq1$ and $\alpha,\beta\geq0$,
\begin{align}
    a^{(n+1)}_{\alpha+1,\beta+1}&=\frac{1}{4(\alpha+1)(\beta+1)}\sum_{i=0}^{\alpha}\sum_{j=0}^{\beta}a^{(n)}_{ij}\,c_{\alpha-i,\beta-j} \label{eq:order_recursion}\\
     a^{(n+1)}_{0j}&=a^{(n+1)}_{j0}=0\ ,
\end{align}
with
\begin{equation}
	c_{\alpha-i,\beta-j}=\frac{c_{\alpha+\beta-i-j}}{2^{\alpha+\beta-i-j}}\binom{\alpha+\beta-i-j}{\alpha-i} \ .
\end{equation} 
One can show that the triple series $\sum_n\sum_{i,j}|a^{(n)}_{ij}|\,u^iv^j$ converges uniformly on every closed triangle $u,v\geq0$, $u+v\leq2\rho_0$, with $\rho_0<\rho$. Hence each $g_n$ equals its power series, and the sum over $n$ can be rearranged into the Ansatz~\eqref{eq:Ansatz}, $g=\sum_n g_n = \sum_{i,j}a_{ij}u^iv^j$, with
\begin{equation}
    a_{ij}=\sum_{n\geq1}a^{(n)}_{ij}\ ,
    \label{eq:a_sum_orders}
\end{equation}
which converges absolutely at every point of the wedge with $t<\rho$.

Summing equation~\eqref{eq:order_recursion} over $n\geq1$ and using equation~\eqref{eq:a_sum_orders} together with $a^{(1)}_{\alpha+1,\beta+1}=0$ gives exactly the recursion~\eqref{eq:interior}. The corner series is the Neumann series regrouped by total degree. By induction on $n$, equation~\eqref{eq:order_recursion} also gives
\begin{equation}
    a^{(n)}_{ij}\neq0\quad\Longrightarrow\quad\min(i,j)\geq n-1\ .
    \label{eq:depth}
\end{equation}
This holds for $n=1$. If it holds for $n$, then $a^{(n+1)}_{\alpha+1,\beta+1}$ involves only the $a^{(n)}_{ij}$ with $i\leq\alpha$ and $j\leq\beta$, which vanish unless $\min(\alpha+1,\beta+1)\geq n$. Hence $g_n=O(u^{n-1}v^{n-1})$, and the sum~\eqref{eq:a_sum_orders} is finite, 
\begin{equation}
a_{ij}=\sum_{n=1}^{\min(i,j)+1}a^{(n)}_{ij} \ .
\end{equation}
Each coefficient is then completed at the scattering order $\min(i,j)+1$, {\emph{i.e.}}, at the recursion depth $\min(i,j)-1$ of equation~\eqref{eq:interior}.

Since $m^2\big(\eta'+\tfrac{u+v}{2}\big)$ is symmetric under $u\leftrightarrow v$, exchanging $u'$ and $v'$ in equation~\eqref{eq:neumann_n} shows that $g_{n+1}$ is antisymmetric if $g_n$ is. As $g_1$ is antisymmetric, every order is antisymmetric: $a^{(n)}_{ij}=-a^{(n)}_{ji}$. Then 
\begin{equation}
	\sum_{i+j=N}a_{ij}=0
\end{equation}
for every $N$, which is the closure~\eqref{eq:closure}: in this construction the column $a_{j0}=-a_{0j}$ is fixed by $g_1$, and the worldline condition~\eqref{eq:worldline} holds order by order.

\subsubsection*{The Riemann function and the representation formula}

\noindent

Equivalently, one could derive the same corner expansion from the Riemann function of the Goursat problem~\cite{Esposito:2001ry}. For the general linear hyperbolic operator $L[g]\equiv g_{,uv}+a(u,v)g_{,u}+b(u,v)g_{,v}+c(u,v)g$, the Riemann function $R(u,v;\bar u,\bar v)$ is defined as the solution of the adjoint equation $L^\dagger[R]=0$ that equals unity on the two characteristics through the field point $(\bar u,\bar v)$. For equation \eqref{eq:bulk_uv}, $a=b=0$ and $c=-\tfrac14m^2(\eta)$, and the operator is self-adjoint, $L^\dagger=L$. Hence $R$ solves the same bulk equation as $g$, that is,
\begin{equation}
    R_{,uv} = \frac14m^2(\eta)R, \quad R(\bar u,v;\bar u,\bar v)=R(u,\bar v;\bar u,\bar v)=1 \ .
    \label{eq:Rbulk}
\end{equation}
We write the formal solution in the Volterra form
\begin{equation}
    R(u,v;\bar u,\bar v) = 1+\frac14\int_{\bar u}^u\!\!\int_{\bar v}^v m^2(\eta)R(u',v';\bar u,\bar v)\,du'dv' \ ,
    \label{eq:volterra}
\end{equation}
where $\eta = \eta(u',v')$.
Applying Green's theorem to the vector field $(R_{,u}g,Rg_{,v})$ on the rectangle $[0,\bar u]\times[0,\bar v]$, using that $Rg_{,uv}-gR_{,uv}=0$ identically by \eqref{eq:bulk_uv} and \eqref{eq:Rbulk}, and evaluating the four sides using $R=1$ on the two characteristics through $(\bar u,\bar v)$, one obtains the Riemann representation formula
\begin{align}
    g(\bar u,\bar v) = &  R(0,0;\bar u,\bar v)g(0,0)  \nonumber\\
    & +\int_0^{\bar u}\!R(u,0;\bar u,\bar v)g_{,u}(u,0)\,du  \nonumber\\
    & + \int_0^{\bar v}\!R(0,v;\bar u,\bar v)g_{,v}(0,v)\,dv \ .
    \label{eq:repformula0}
\end{align}
For our problem, $g(0,0)=0$, and $F(v)\equiv g(0,v)=\tfrac12\int_0^{v/2}m^2(\eta'+t')dt'$ gives the light-cone data \eqref{eq:g0v}. The derivative is then $g_{,v}(0,v)=F'(v)$ and, by antisymmetry~\eqref{eq:antisym}, $g_{,u}(u,0)=-F'(u)$. Integrating \eqref{eq:repformula0} by parts and using the differentiated Volterra identities $R_{,u}(u,0;\bar u,\bar v)=-\tfrac14\int_0^{\bar v}m^2\big(\eta(u,q)\big)R(u,q;\bar u,\bar v)\,dq$ (and similarly for $R_{,v}$), the two boundary integrals combine into a single double integral over the rectangle,
\begin{widetext}
\begin{equation}
    g(\bar u,\bar v) = F(\bar v)-F(\bar u) + \frac14\int_0^{\bar u}\!\!\int_0^{\bar v}\big[F(q)-F(p)\big]\,m^2\big(\eta(p,q)\big)\,R(p,q;\bar u,\bar v)\,dq\,dp.
    \label{eq:repformula_combined}
\end{equation}
\end{widetext}

The Volterra equation~\eqref{eq:volterra} is solved by Neumann iteration: $R^{(0)}=1$, $R^{(\ell+1)}=1+\tfrac14\int\!\!\int m^2R^{(\ell)}$. We write $R=\sum_{\ell\geq0}R_\ell$, where $R_\ell=R^{(\ell)}-R^{(\ell-1)}$, with $R^{(-1)}\equiv0$. Each $R_\ell$ contains $\ell$ nested integrations of $m^2$ and is of order $(m^2)^\ell$. Inserting this series into equation~\eqref{eq:repformula_combined}, we can evaluate the integrals term by term and match powers $\bar u^i\bar v^j$ to obtain the corner coefficients. The boundary term $F(\bar v)-F(\bar u)$ gives the light-cone coefficients, $a_{0j}$ and $a_{j0}=-a_{0j}$, and the double integral gives the interior ones, order by order in $\ell$.

Equation~\eqref{eq:repformula_combined}, together with the Neumann expansion of the Riemann function $R$, has an immediate connection to the Neumann series for $g$~\eqref{eq:neumann_n}.
The boundary term is $F(\bar v)-F(\bar u)=g_1(\bar u,\bar v)$, while $F(q)-F(p)=g_1(p,q)$ appears in the integrand of the double integral. Since $g_n$ is of order $(m^2)^n$, collecting the powers of $m^2$ in equation~\eqref{eq:repformula_combined} gives
\begin{equation}
    g_{\ell+2}(\bar u,\bar v)=\frac14\int_0^{\bar u}\!\!\int_0^{\bar v}g_1(p,q)\,m^2(\eta)\,R_\ell(p,q;\bar u,\bar v)\,dq\,dp\ .
    \label{eq:g_from_R}
\end{equation}
The $\ell$-th Neumann term of $R$ is therefore the scattering order $g_{\ell+2}$, and its Taylor coefficients are the $a^{(\ell+2)}_{ij}$ of equation~\eqref{eq:order_recursion}. In particular, $R_0=1$ gives $g_2$, whose coefficients are built from the light-cone coefficients alone, {\emph{i.e.}}, from the terms with $m=0$ or $n=0$ in the sum of equation~\eqref{eq:interior}. By equation~\eqref{eq:depth}, $R_\ell$ contributes only to the coefficients with $\min(i,j)\geq\ell+1$, so that $a_{ij}$ is completed at the Neumann order $\ell=\min(i,j)-1$ of $R$, which is the scattering order $\min(i,j)+1$ of $g$. Summing over $\ell$ recovers equation~\eqref{eq:a_sum_orders}, and hence the recursion~\eqref{eq:interior}.

\begin{figure}[t]
\centering
\begin{tikzpicture}[every node/.style={font=\small}]
  \definecolor{edgeCol}{RGB}{178,24,43}
  \definecolor{zeroCol}{RGB}{242,242,242}
  \definecolor{depthZero}{RGB}{198,219,239}
  \definecolor{depthOne}{RGB}{107,174,214}
  \definecolor{depthTwo}{RGB}{49,112,176}
  \definecolor{depthThree}{RGB}{8,48,107}
  \def\S{0.40}
  \def\N{9}   

  \foreach \m in {0,...,\N} {
    \pgfmathtruncatemacro{\nmax}{\N-\m}
    \foreach \n in {0,...,\nmax} {
      \pgfmathsetmacro{\x}{\S*(\m-\n)}
      \pgfmathsetmacro{\y}{\S*(\m+\n)}
      \coordinate (N\m-\n) at (\x,\y);
    }
  }

  \foreach \m in {1,...,8} {
    \pgfmathtruncatemacro{\nmax}{\N-\m}
    \draw[thin, dashed, black!30] (N\m-0) -- (N\m-\nmax);
  }
  \foreach \n in {1,...,8} {
    \pgfmathtruncatemacro{\mmax}{\N-\n}
    \draw[thin, dashed, black!30] (N0-\n) -- (N\mmax-\n);
  }

  \draw[thick] (N0-\N) -- (N0-0) -- (N\N-0);

  \foreach \m in {0,...,\N} {
    \pgfmathtruncatemacro{\nmax}{\N-\m}
    \foreach \n in {0,...,\nmax} {
      \pgfmathtruncatemacro{\ismn}{\m==\n}
      \pgfmathtruncatemacro{\isedge}{(\m==0)||(\n==0)}
      \ifnum\ismn=1
        \node[circle, fill=zeroCol, draw=black!70, line width=0.3pt, inner sep=2.1pt] at (N\m-\n) {};
      \else
        \ifnum\isedge=1
          \node[circle, fill=edgeCol, draw=black!70, line width=0.3pt, inner sep=2.1pt] at (N\m-\n) {};
        \else
          \pgfmathtruncatemacro{\depthmn}{min(\m,\n)-1}
          \ifcase\depthmn
            \node[circle, fill=depthZero, draw=black!70, line width=0.3pt, inner sep=2.1pt] at (N\m-\n) {};
          \or
            \node[circle, fill=depthOne, draw=black!70, line width=0.3pt, inner sep=2.1pt] at (N\m-\n) {};
          \or
            \node[circle, fill=depthTwo, draw=black!70, line width=0.3pt, inner sep=2.1pt] at (N\m-\n) {};
          \or
            \node[circle, fill=depthThree, draw=black!70, line width=0.3pt, inner sep=2.1pt] at (N\m-\n) {};
          \fi
        \fi
      \fi
    }
  }

  \draw[black, line width=0.9pt, dash pattern=on 2.5pt off 1.5pt]
    ($\S*(-6,9)$) -- ($\S*(0,3)$) -- ($\S*(6,9)$);
  \node[font=\scriptsize, anchor=north] at ($\S*(0,3)$) {$g_3$};

  \node[below=3pt] at (N0-0) {\footnotesize $a_{0,0}=0$};
  \node[edgeCol, anchor=south] at ($($(N0-0)!0.5!(N0-\N)$)+(-135:0.62)$) {\footnotesize $a_{0,j}$};
  \node[edgeCol, anchor=south] at ($($(N0-0)!0.5!(N\N-0)$)+(-45:0.62)$) {\footnotesize $a_{j,0}$};
  \node[above=3pt, font=\scriptsize] at (N4-5) {$a_{4,5}$};
  \node[above=3pt, font=\scriptsize] at (N5-4) {$a_{5,4}$};
  \node[above=3pt, font=\scriptsize] at (N0-\N) {$a_{0,\N}$};
  \node[above=3pt, font=\scriptsize] at (N\N-0) {$a_{\N,0}$};
  \draw[->, black!60] ($(N\N-0)+(0.38,-\S*\N)$) -- ($(N\N-0)+(0.38,0)$) node[midway, rotate=90, anchor=north, font=\scriptsize] {total degree $m+n$};

  \begin{scope}[shift={(-1.55,-1.15)}, every node/.style={font=\scriptsize, anchor=west}]
    \node[circle,fill=edgeCol,draw=black!70,line width=0.3pt,inner sep=2.1pt] at (0,0) {};
    \node at (0.2,0) {$g_1$: light-cone data $F(\bar v){-}F(\bar u)$};
    \node[circle,fill=zeroCol,draw=black!70,line width=0.3pt,inner sep=2.1pt] at (0,-0.35) {};
    \node at (0.2,-0.35) {$a_{n,n}=0$ by antisymmetry};
    \node[circle,fill=depthZero,draw=black!70,line width=0.3pt,inner sep=2.1pt] at (0,-0.7) {};
    \node at (0.2,-0.7) {$g_{2}$, $R_{0}$: $a_{1,n}$, $a_{n,1}$};
    \node[circle,fill=depthOne,draw=black!70,line width=0.3pt,inner sep=2.1pt] at (0,-1.05) {};
    \node at (0.2,-1.05) {$g_{3}$, $R_{1}$: $a_{2,n}$, $a_{n,2}$};
    \node[circle,fill=depthTwo,draw=black!70,line width=0.3pt,inner sep=2.1pt] at (0,-1.4) {};
    \node at (0.2,-1.4) {$g_{4}$, $R_{2}$: $a_{3,n}$, $a_{n,3}$};
    \node[circle,fill=depthThree,draw=black!70,line width=0.3pt,inner sep=2.1pt] at (0,-1.75) {};
    \node at (0.2,-1.75) {$g_{5}$, $R_{3}$: $a_{4,n}$, $a_{n,4}$};
    \draw[black, line width=0.9pt, dash pattern=on 2.5pt off 1.5pt] (-0.1,-2.1) -- (0.1,-2.1);
    \node at (0.2,-2.1) {support of $g_3$: $a_{m,n}$ with $m,n\geq2$};
  \end{scope}
\end{tikzpicture}
\caption{The double-indexed array of Taylor coefficients $a_{m,n}$ of the Ansatz \eqref{eq:Ansatz}, up to total degree $m+n\leq9$, with the total degree increasing upward: each horizontal row is one order of the series, built from the rows below it by \eqref{eq:interior}. The corner $a_{0,0}$ and the diagonal $a_{n,n}$ vanish by the antisymmetry \eqref{eq:antisym} (light gray). The two edges leaving $a_{0,0}$ are the light-cone coefficients $a_{0,j}$ and $a_{j,0}$ (red), fixed directly by the boundary term $F(\bar v)-F(\bar u)$ in the representation formula \eqref{eq:repformula_combined}. The scattering order $g_k$ of the field contributes only to the coefficients with $\min(m,n)\geq k-1$, equation~\eqref{eq:depth}, {\emph{i.e.}}, to its own band and to the bands inside it (dashed: the support of $g_3$). Every coefficient is colored by the highest scattering order that contributes to it, $k=\min(m,n)+1$, from $g_1$ on the light-cone edges to $g_5$. By equation~\eqref{eq:g_from_R}, this corresponds to order $\ell=k-2$ in the Neumann series of the Riemann function and to recursion depth $\min(m,n)-1$ in equation~\eqref{eq:interior}.}
\label{fig:coeff_pyramid}
\end{figure}
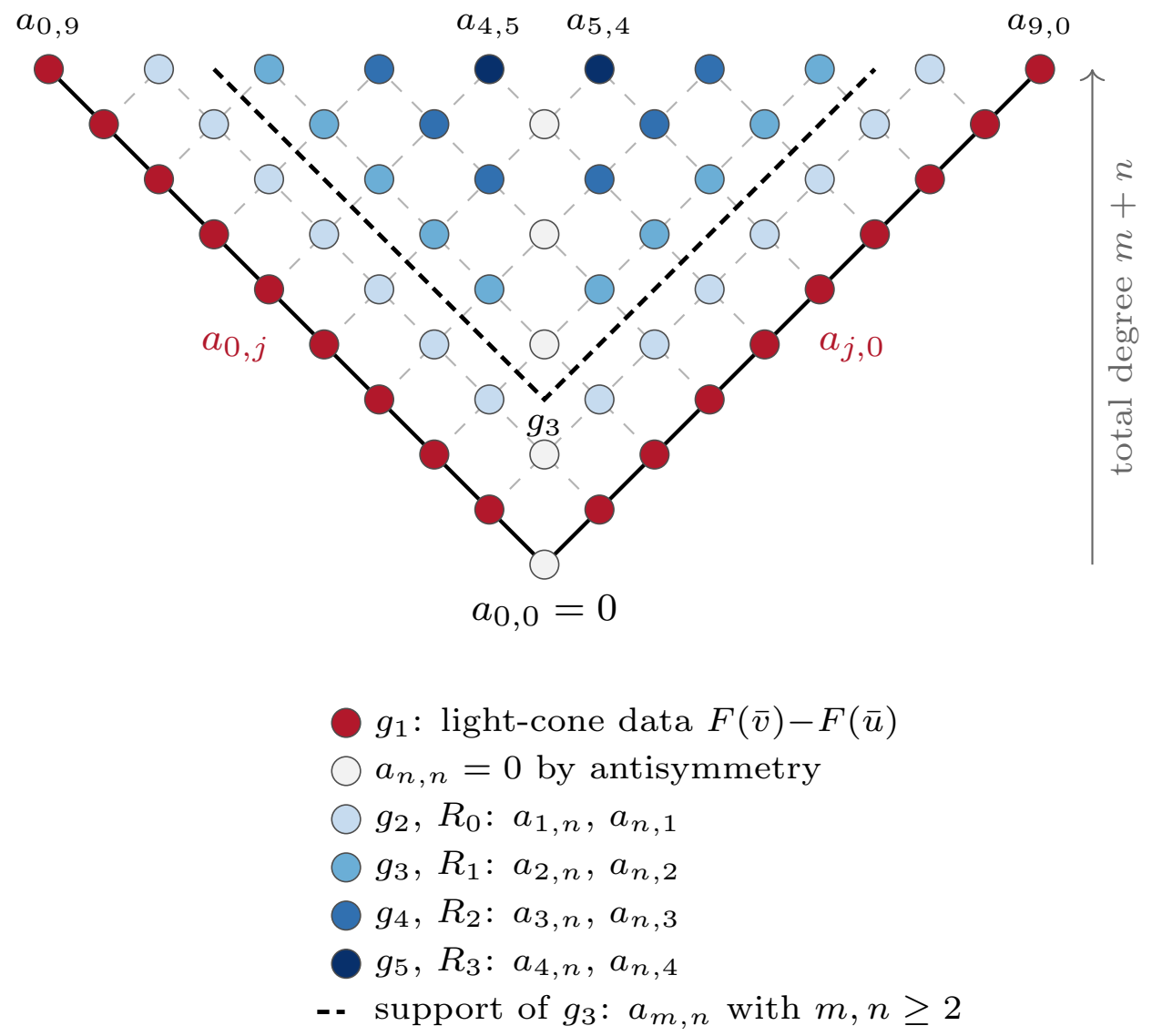

Figure~\ref{fig:coeff_pyramid} displays the array of coefficients $a_{m,n}$ up to total degree $m+n\leq9$. The two edges starting from $a_{0,0}$ collect the coefficients fixed directly by the light-cone data through $F(\bar v)-F(\bar u)$ in \eqref{eq:repformula_combined}. Every other coefficient is colored by the Neumann order of \eqref{eq:volterra} required to reproduce it, equivalently by its recursion depth in \eqref{eq:interior}, or by the highest scattering order of $g$ that contributes to it. Since depth$(m,n)=\min(m,n)-1$ for $m\neq n$, the depth is constant along each pair of diagonals parallel to the two edges: $a_{1,2}$ (and its mirror $a_{2,1}$) sit on the innermost band and are reproduced already at Neumann order $0$, {\emph{i.e.}}, by the scattering order $g_2$; $a_{2,3}$ sits one band further in and requires order $1$ ($g_3$), and so on. Each band extends to all total degrees: $a_{4,n}$ and $a_{n,4}$ are of order $3$ for every $n>4$. Viewed from the field expansion, the scattering order $g_k$ contributes to the band of $g_k$ and to all the bands inside it (dashed outline for $g_3$), so that each coefficient is the finite sum of the orders $g_1,\dots,g_{\min(m,n)+1}$, equation~\eqref{eq:a_sum_orders}.

\section{\texorpdfstring{\NoCaseChange{Large-$k$ limit of the Born series}}{Large-k limit of the Born series}}
\label{app:born_k}

\noindent
In this appendix we derive the large-$k$ form~\eqref{eq:G_WKB} of the propagator from the multiple-scattering expansion of Section~\ref{sec:scattering}, and show that its leading contribution is indeed equation~\eqref{eq:GTL_asymptotic}. We write ${t}=\eta-\eta'$ throughout.

With the normalization~\eqref{eq:GLC_fourier}, the spatial Fourier transform of the wave equation~\eqref{eq:green_eq} is the ordinary differential equation
\begin{equation}
\big[\partial_\eta^2+k^2-m^2(\eta)\big]\,\widetilde G(\eta,\eta';k)=\delta(\eta-\eta')\ ,
\label{eq:app_ode}
\end{equation}
with $\widetilde G=0$ for $\eta<\eta'$. The light-cone propagator $\widetilde G_{\LC}(\eta,\eta')=\theta({t})\sin(k{t})/k$ is the retarded Green's function of $\partial_\eta^2+k^2$. Treating $m^2\widetilde G$ as a source, equation~\eqref{eq:app_ode} is equivalent to the Volterra equation
\begin{equation}
\widetilde G(\eta,\eta')=\widetilde G_{\LC}(\eta,\eta')+\int_{\eta'}^{\eta}d\eta''\,\widetilde G_{\LC}(\eta,\eta'')\,m^2(\eta'')\,\widetilde G(\eta'',\eta')\ ,
\label{eq:app_volterra}
\end{equation}
which is the Fourier transform of the integral equation~\eqref{eq:G_integral}. The convolution over the position $\bfx''$ of the scattering event becomes a product, and only the integral over its time $\eta''$ is left. Iterating equation~\eqref{eq:app_volterra} gives the Born series $\widetilde G=\sum_{n\geq0}\widetilde G^{(n)}$, with $\widetilde G^{(0)}=\widetilde G_{\LC}$ and
\begin{equation}
\widetilde G^{(n)}(\eta,\eta')=\int_{\mathcal T_n} \prod_{j=1}^n d\eta_j \,m^2(\eta_j)\, \prod_{\ell = 0}^n \widetilde G_{\LC}(\eta_{\ell +1},\eta_{\ell}) \ ,
\label{eq:app_born_n}
\end{equation}
where $\eta_0 = \eta^\prime$, $\eta_{n+1} = \eta$, and $\mathcal T_n$ is the ordered domain $\eta'<\eta_1<\dots<\eta_n<\eta$. The $n$-th term represents the wave scattered $n$ times by the potential $m^2$, at the times $\eta_1,\dots,\eta_n$, and is the Fourier transform of the $n$-th term of equation~\eqref{eq:born_series}. Since $|\widetilde G_{\LC}|\leq{t}$ and the volume of $\mathcal T_n$ weighted by $m^2$ is $(\int_{\eta'}^{\eta}m^2)^n/n!$, the series converges for every $k$.
For $n=1$, 
\begin{align}
\widetilde G^{(1)}&=\frac{1}{2k^2}\int_{\eta'}^{\eta}d\eta''\,m^2(\eta'') \left[ \cos\Phi(\eta'')- \cos k{t} \right] \ ,
\label{eq:app_G1}
\end{align}
with $\Phi(\eta'') =  k(\eta+\eta'-2\eta'')$, and $t = \eta-\eta^\prime$.
The two terms correspond to the two ways in which a wave can scatter off the background curvature, backward or forward. 

\emph{Forward scattering:} since the phase $k{t}$ does not depend on the scattering time $\eta''$, the scattered wave $\sim \cos(kt)$ travels in the same direction and in phase with the incident one, so that the scatterings along the whole path add up coherently. The forward scattering alone gives
\begin{equation}
\widetilde G^{(1)}_{\rm fw}=-\frac{\kappa}{k^2}\cos k{t}\ ,\qquad \kappa=\frac12\int_{\eta'}^{\eta}m^2(\eta'')\,d\eta''\ ,
\label{eq:app_fw1}
\end{equation}
which is equation~\eqref{eq:GTL_asymptotic}, where $\kappa=g(0,2{t})={t}\,B|_{u=0}$ is the accumulated curvature along the light-cone front~\eqref{eq:B_lightcone}. In position space the light-cone value is produced by the scattering along the ray itself (Section~\ref{sec:scattering}). In Fourier space the same scatterings are the coherent forward modes. 

\emph{Backward scattering:} the phase $\Phi(\eta'') =  k(\eta+\eta'-2\eta'')$ in the first term of equation~\eqref{eq:app_G1} reflects scatterings away from the light-cone trajectory and, as it depends on the integrated time $\eta^{\prime\prime}$, results in incoherent contributions from all different paths. Integrating by parts, we find
\begin{align}
\widetilde G^{(1)}_{\rm bw}&=\frac{1}{2k^2}\int_{\eta'}^{\eta}d\eta''\,m^2(\eta'')\,\cos \Phi(\eta'')
\nonumber\\
&=\frac{m^2(\eta)+m^2(\eta')}{4k^3}\,\sin k{t} \nonumber\\
& \quad + \frac{1}{4k^3}\int_{\eta'}^{\eta}d\eta''\,\frac{dm^2}{d\eta''}\,\sin \Phi(\eta'')\ .
\label{eq:app_bw1}
\end{align}
The second term is suppressed by both the integration over the incoherent phase $\Phi$ with zero average and the time derivative of the curvature potential, which evolves on cosmological scales and is approximately constant for sub-horizon modes $k\eta^\prime \gg 1$. We therefore keep only the local term of the backward scattering, at the source and at the field point.

Summing forward and backward scatterings up to the term $n=1$ of the Born series~\eqref{eq:app_born_n}, we obtain
\begin{equation}
\widetilde G=\frac{\sin k{t}}{k}\left[1+\frac{m^2(\eta)+m^2(\eta')}{4k^2}\right]-\frac{\kappa}{k^2}\cos k{t}  \ , 
\label{eq:app_first}
\end{equation}
up to corrections of order $\mathcal{O}(\kappa^2/k^3)$ and $\mathcal{O}(k^{-4}{\rm{d}} m^2/\rm{d}\eta)$.

\subsection{Scattering at all orders}
\noindent
For $\eta_{\ell+1}>\eta_{\ell}$ we write each light-cone leg as a sum of two waves,
\begin{equation}
\widetilde G_{\LC} \left(\eta_{\ell+1},\eta_{\ell}\right)=\sum_{\epsilon_{\ell}=\pm}\frac{\epsilon_{\ell}\,e^{i\epsilon_{\ell} k\left(\eta_{\ell+1}-\eta_{\ell}\right)}}{2ik}\ ,
\label{eq:app_legs}
\end{equation}
with $\epsilon_{\ell}=\pm1$ labeling the positive- and negative-frequency parts. In the Born series~\eqref{eq:app_born_n} the times are ordered, $\eta_{\ell+1}>\eta_\ell$ on $\mathcal T_n$, so that $\theta=1$ on every leg and we omit it in the following. 
In coordinate space, after multiplying by the plane wave $e^{i\mathbf k\cdot\mathbf x}$ of the inverse Fourier transform, the two terms of equation~\eqref{eq:app_legs} represent waves traveling in opposite directions along $\hat{\mathbf k}$. 
All the legs of equation~\eqref{eq:app_born_n} carry the same momentum $\mathbf k$, since the potential depends only on time~\footnote{For a potential depending on position, the product in equation~\eqref{eq:app_volterra} would become a convolution in momentum, and the legs would carry different momenta.} and the wave equation~\eqref{eq:app_ode} does not couple different Fourier modes.
Inserting equation~\eqref{eq:app_legs} for every leg, each term of $\widetilde G^{(n)}$ carries one sign $\epsilon_\ell$ per leg. We call the scattering at $\eta_j$ \emph{forward} if the legs before and after it have the same sign, $\epsilon_{j-1}=\epsilon_j$, so that the wave keeps its direction of propagation, and \emph{backward} if $\epsilon_{j-1}=-\epsilon_j$, so that it reverses it, exchanging positive and negative frequencies at fixed $\mathbf k$.
At first order, this is the decomposition of equation~\eqref{eq:app_G1}. For the legs $\eta'\to\eta''\to\eta$ with $\epsilon_0=\epsilon_1$, the phase is $\epsilon k{t}$, independent of $\eta''$, and gives the term $-\cos k{t}$. For $\epsilon_0=-\epsilon_1$, the phase is $\epsilon_1\Phi(\eta'')$, and gives the term $\cos\Phi$.

Inserting contribution~\eqref{eq:app_legs} for the $n+1$ legs of equation~\eqref{eq:app_born_n}, with $\eta_0\equiv\eta'$, $\eta_{n+1}\equiv\eta$, we write
\begin{align}
\widetilde G^{(n)} &=\sum_{\epsilon_0,\dots,\epsilon_n=\pm}\frac{\epsilon_0\cdots\epsilon_n}{(2ik)^{n+1}}\int_{\mathcal T_n}\prod_{j=1}^{n}d\eta_j\,m^2(\eta_j)\, \nonumber\\
& \qquad\qquad \times  \exp\Big[ik\sum_{\ell=0}^{n}\epsilon_{\ell}(\eta_{\ell+1}-\eta_{\ell})\Big]\ .
\label{eq:app_signs}
\end{align}
Collecting the coefficient of each time, the exponent is
\begin{equation}
ik\sum_{\ell=0}^{n}\epsilon_\ell(\eta_{\ell+1}-\eta_\ell)=ik\Big[\epsilon_n\eta-\epsilon_0\eta'+\sum_{\ell=1}^{n}(\epsilon_{\ell-1}-\epsilon_\ell)\,\eta_\ell\Big]\ ,
\label{eq:app_exponent}
\end{equation}
so that the scattering time $\eta_\ell$ enters the phase only through the difference $\epsilon_{\ell-1}-\epsilon_\ell$. If the wave keeps its direction at $\eta_j$ (forward scattering, $\epsilon_{j-1}=\epsilon_j$), this term vanishes. If it reverses direction (backward scattering), the phase is $\pm2ik\eta_j$. This means that, for the terms with all signs equal, $\epsilon_j=\epsilon$, the exponent is $i\epsilon k{t}$ for every choice of the scattering times, and the time integral is the integral of a symmetric function over the ordered domain. Since the $n!$ orderings of $\eta_1,\dots,\eta_n$ tile the cube $[\eta',\eta]^n$,
\begin{equation}
\int_{\mathcal T_n}\prod_{j=1}^{n}d\eta_j\,m^2(\eta_j)=\frac{1}{n!}\left(\int_{\eta'}^{\eta}m^2\,d\eta''\right)^{n}=\frac{(2\kappa)^n}{n!}\ .
\label{eq:app_ordered}
\end{equation}
Therefore, the forward scattering of order $n$ is 
\begin{align}
\widetilde G^{(n)}_{\rm fw} & =\sum_{\epsilon=\pm}\frac{\epsilon^{n+1}e^{i\epsilon k{t}}}{(2ik)^{n+1}}\,\frac{(2\kappa)^n}{n!} \nonumber \\
 & =\sum_{\epsilon=\pm}\frac{\epsilon\,e^{i\epsilon k{t}}}{2ik}\,\frac{1}{n!}\left(-\frac{i\epsilon \kappa}{k}\right)^{n}\ ,
\label{eq:app_fwn}
\end{align}
which for $n=1$ reproduces equation~\eqref{eq:app_fw1}. The sum over $n$ is the exponential series,
\begin{equation}
\sum_{n\geq0}\widetilde G^{(n)}_{\rm fw}=\sum_{\epsilon=\pm}\frac{\epsilon\,e^{i\epsilon k{t}}}{2ik}\,e^{-i\epsilon \kappa/k}=\frac{\sin\!\big(k{t}-\delta\big)}{k}\ ,
\label{eq:app_fwsum}
\end{equation}
with $\delta \equiv \kappa/k$.
The forward scatterings of all orders sum exactly to the light-cone wave delayed in phase by $\delta$.

The terms in the exponential function of equation~\eqref{eq:app_signs} with at least one reversal contain a factor $e^{\pm2ik\eta_j}$ for each reversal. Integrating by parts in $\eta_j$ over the interval $[\eta_{j-1},\eta_{j+1}]$ allowed by the ordering, each reversal costs a factor of order $m^2/k^2$ and leaves boundary terms where $\eta_j$ meets its neighbors.
Rather than summing these terms one by one, we look for an approximate solution of equation~\eqref{eq:app_ode} in the limit of small and slowly varying curvature potential $m^2$, that is, $|m^2|\ll k^2$ and $|dm^2/d\eta|\ll k\,|m^2|$. In other words, we consider the limit $k\gg\Hell$ along the path. Two independent solutions of the homogeneous equation are then
\begin{equation}
\psi_\pm(\eta)=\frac{1}{\sqrt{\omega(\eta)}}\exp\Big(\pm i\int^{\eta}\omega\,d\eta''\Big)\ ,
\end{equation}
with effective frequency $\omega(\eta)=\sqrt{k^2-m^2(\eta)}$ and the Wronskian $\mathcal{W}(\eta') = \psi_+\psi_-'-\psi_+'\psi_-=-2i$. The retarded Green's function, with boundary conditions $\widetilde G=0$ and $\partial_\eta\widetilde G=1$ at $\eta=\eta'$, is the linear combination of homogeneous solutions, $\widetilde G(\eta,\eta';k)=\left[ \psi_+(\eta')\psi_-(\eta)-\psi_+(\eta)\psi_-(\eta')\right]/ \mathcal{W}(\eta')$, yielding the approximate solution
\begin{equation}
\widetilde G(\eta,\eta';k)=\frac{\sin\Phi(\eta,\eta')}{\sqrt{\omega(\eta)\,\omega(\eta')}}\ ,\qquad \Phi(\eta,\eta')=\int_{\eta'}^{\eta}\omega\,d\eta''\ .
\label{eq:app_wkb}
\end{equation}
Expanding $\omega=k-m^2/2k-m^4/8k^3+\dots$, we find
\begin{align}
\Phi &=k{t}-\delta-\frac{1}{8k^3}\int_{\eta'}^{\eta}m^4\,d\eta''+\ldots \\
\frac{1}{\sqrt{\omega(\eta)\,\omega(\eta')}} &=\frac1k\left[1+\frac{m^2(\eta)+m^2(\eta')}{4k^2}+\ldots\right] \ .
\label{eq:app_wkb_exp}
\end{align}
The phase contains the resummed forward scattering~\eqref{eq:app_fwsum}, and the amplitude the backward scattering at the two ends, equation~\eqref{eq:app_bw1}. To first order in $m^2$, the WKB approximation~\eqref{eq:app_wkb} recovers equation~\eqref{eq:app_first}, proving the expression~\eqref{eq:G_WKB} used in Section~\ref{sec:fourier}.

\begin{figure*}[t]
\centering
\includegraphics[width=\textwidth]{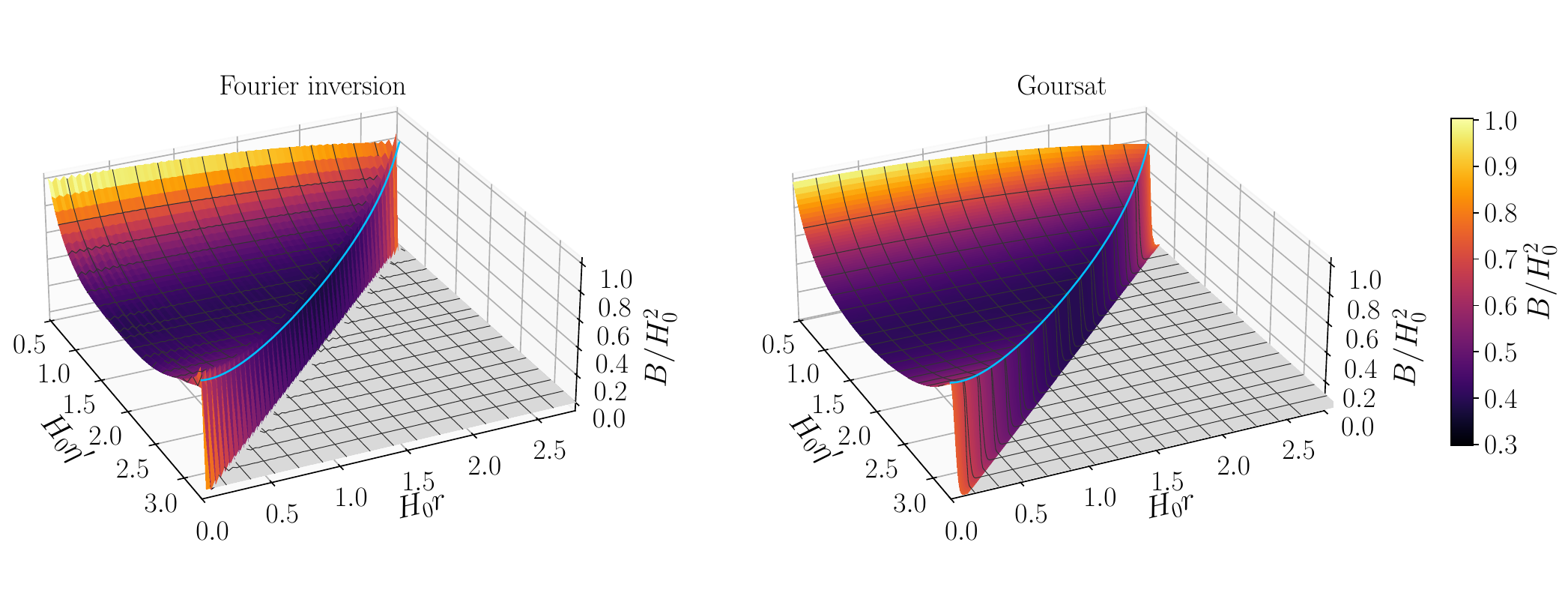}
\caption{Comparison of tail solution obtained through Fourier inversion of Ref.~\cite{Jokela:2022rhk} (left panel) and direct Goursat integration of Section~\ref{sec:goursat} (right panel) for the exact $\Lambda$CDM interaction. The observation time is fixed at $H_0\eta=3.305$, with cutoff $K=100\,H_0$. The image reproduces Figure~4 of Ref.~\cite{Jokela:2022rhk}. A fine grid of steps $0.06\,H_0^{-1}$ in $\eta'$ and $0.03\,H_0^{-1}$ in $r$ resolves the oscillations. The light-blue line is the exact light-cone value~\eqref{eq:B_lightcone}, and the gray plane is the exterior of the light cone, $r>\eta-\eta'$, where $B=0$.}
\label{fig:jks_extended_3D}
\end{figure*}

\section{Comparison to other methods}
\label{app:literature}

\subsection*{Solving the tail in Fourier space}

\begin{figure}[t]
\centering
\includegraphics[width=\linewidth]{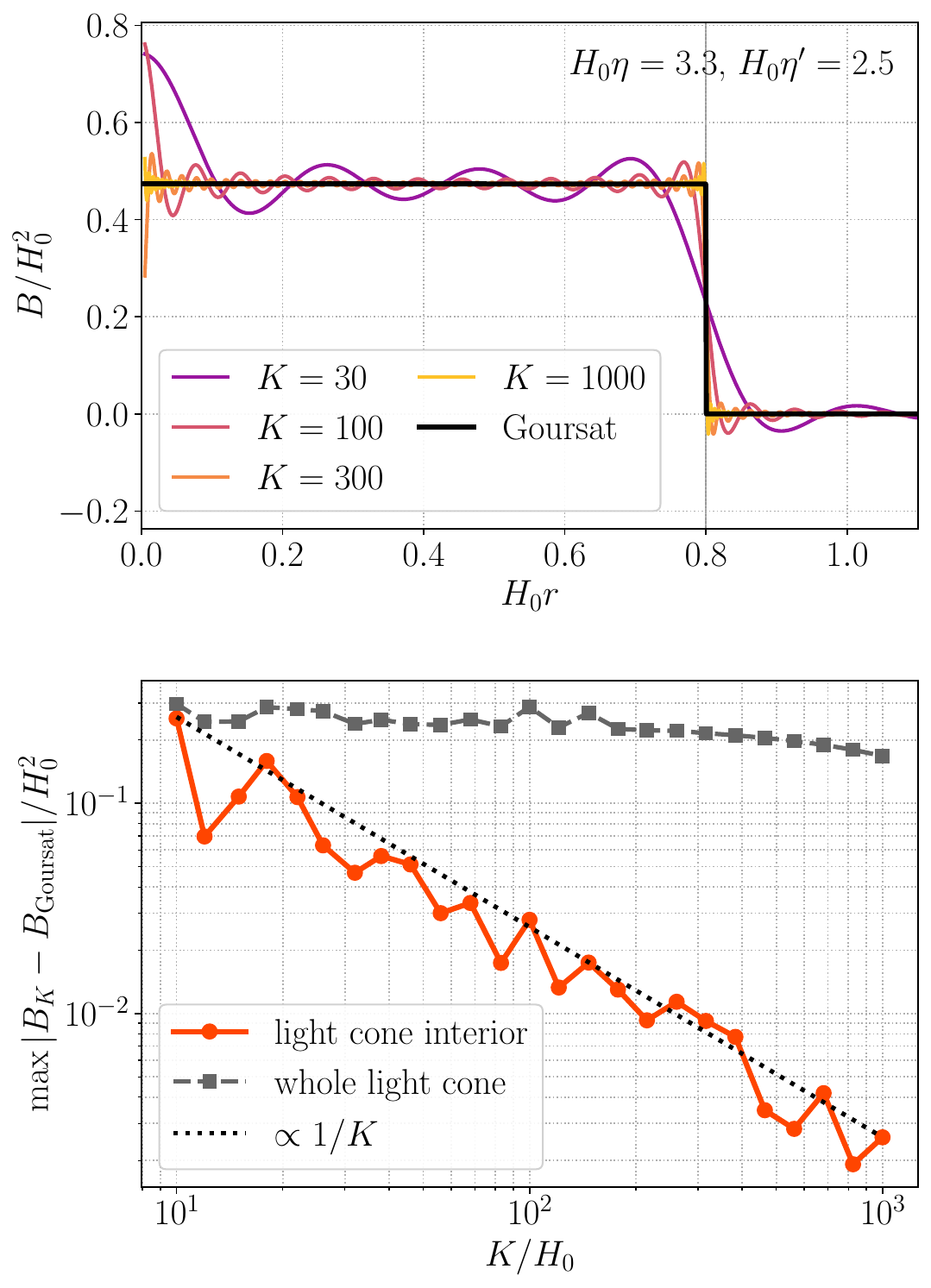}
\caption{Convergence of the Fourier method of Ref.~\cite{Jokela:2022rhk} with the cutoff $K$, at $H_0\eta=3.3$ for a source at $H_0\eta'=2.5$ (${t}=0.8\,H_0^{-1}$). Top: $B_K$, equation~\eqref{eq:jks_inverse}, for different cutoff values $K/H_0$ against the Goursat solution (thick line); the vertical line marks the light cone, $r={t}$. Bottom: largest deviation $|B_K-B_{\rm Goursat}|$ as a function of $K$, in the interior of the cone, $0.1<H_0r<H_0{t}-0.1$ (circles), and over the whole cone, $0<r<{t}$ (squares). The interior deviation decreases as $1/K$ (dotted line), equation~\eqref{eq:jks_ripples}, while the deviation over the whole cone remains approximately constant.}
\label{fig:jks_extended_ripples}
\end{figure}

\noindent
Ref.~\cite{Jokela:2022rhk} computed the $\Lambda$CDM tail at the fixed observation time $H_0\eta=3.3$ in Fourier space (Figure~4 of Ref.~\cite{Jokela:2022rhk}). Subtracting the light-cone mode~\eqref{eq:GLC_fourier} from the mode equation, the tail mode~\eqref{eq:GTL_fourier} of each source time $\eta'$ solves the ordinary differential equation
\begin{equation}
\partial_\eta^2\widetilde G_{\TL}+\left(k^2-m^2\right)\widetilde G_{\TL}=m^2\,\frac{\sin k{t}}{k}\ ,
\label{eq:jks_mode}
\end{equation}
with $\widetilde G_{\TL}=\partial_\eta\widetilde G_{\TL}=0$ at $\eta=\eta'$, and the tail follows from the inverse transform of equation~\eqref{eq:GTL_fourier}, truncated at a wavenumber $K$,
\begin{equation}
B_K(\eta,\eta',r)=\frac{2}{\pi r}\int_0^K dk\,k\sin(kr)\,\widetilde G_{\TL}(\eta,\eta';k)\ .
\label{eq:jks_inverse}
\end{equation}
For a comparison with the novel Goursat solutions, we independently implement the same method (fixed-step Runge--Kutta for the modes, $\Delta k=0.01\,H_0$, and $K=100\,H_0$). 
The result of the integration is shown in Figure~\ref{fig:jks_extended_3D} and reproduces the plot of Figure~4 of Ref.~\cite{Jokela:2022rhk} well. A more detailed comparison is given in Figure~\ref{fig:jks_extended_ripples}.

The most evident difference between $B_K$ and the Goursat solution is in the ripples of the truncated function~\eqref{eq:jks_inverse}, which arise from the cutoff $K$.
The tail $B\,\theta({t}-r)$ jumps from the light-cone value $B_{\rm lc}\equiv B(\eta,\eta',{t})$~\eqref{eq:B_lightcone} to zero across the light cone. Its transform therefore falls only as $\widetilde G_{\TL}\simeq-{t}\,B_{\rm lc}\cos(k{t})/k^2$ at large $k$ [cf.\ equation~\eqref{eq:GTL_and_deriv_1}], so that the integrand of equation~\eqref{eq:jks_inverse} decreases only as $1/k$. Truncating the integral at $K$ leaves, for $K|{t}-r|\gg1$,
\begin{equation}
B_K-B\simeq\frac{{t}\,B_{\rm lc}}{\pi Kr}\left[\frac{\cos K({t}+r)}{{t}+r}-\frac{\cos K({t}-r)}{{t}-r}\right]\ ,
\label{eq:jks_ripples}
\end{equation}
ripples of period $2\pi/K$ in $r$, whose amplitude decreases as $1/K$ and grows toward the light cone and the worldline. 
Away from the worldline and from the light cone, $0.1\leq H_0r\leq H_0{t}-0.1$, $B_K$ agrees with the Goursat solution to $\sim 0.03\,H_0^2$ with $K=100\,H_0$. 
At the two edges the truncated integral does not converge uniformly, and the amplitude of the ripples stays approximately constant. These are the bumps and tilted walls noted in Figure~4 of Ref.~\cite{Jokela:2022rhk}.

The Goursat solution (right panel of Figure~\ref{fig:jks_extended_3D}) does not suffer from these numerical artifacts, since the jump at the light cone enters exactly as the boundary data~\eqref{eq:B_lightcone}, the tail vanishes outside the cone by construction, and $B$ is regular on the worldline. The error of the Goursat solution is controlled by the grid spacing $h$ only, and decreases as $h^2$, remaining below $\sim 2\times10^{-7}$ over the whole wedge for $\Ho h=0.002$ (Figure~\ref{fig:convergence_goursat}). By contrast, the error of the Fourier method decreases only as $1/K$ inside the cone and not at all at its edges. The Goursat march is also cheaper: for each source it gives $B$ at all $r$ and $\eta$ in a single pass, while the Fourier method requires one ordinary differential equation for each of $\sim K/\Delta k$ wavenumbers, and one oscillatory integral for each point $(\eta',r)$.

\subsection*{Tail solution as a recursive Volterra series}

\noindent
In position space, tails have recently been included in the linearized field of localized sources~\cite{Compere:2026xwy}.
In that work, the Green's function is integrated over the two directions transverse to $\bfx-\bfx'$, which gives the retarded Green's function of a $(1+1)$-dimensional problem, $\frac12\theta(t-|r|)J(u,v)$, with
\begin{equation}
J(u,v)=1+\int_r^{t}d\tilde r\,\tilde r\,B(\eta,\eta',\tilde r)\ .
\label{eq:J_def}
\end{equation}
The characteristic function $J$ solves the same equation as $g$, $4J_{,uv}=m^2J$, with unit data on both characteristics, $J(0,v)=J(u,0)=1$. Ref.~\cite{Compere:2026xwy} provides an analytic representation for $J$ by the Volterra series
\begin{equation}
J=\sum_{n\geq0}J_n\ ,\quad J_{n+1}(u,v)=\frac14\int_0^u\!\!du'\!\int_0^v\!\!dv'\,m^2\,J_n(u',v')\ ,
\label{eq:J_volterra}
\end{equation}
with $J_0=1$
[equations~(3.9)--(3.11) of Ref.~\cite{Compere:2026xwy}, with their characteristic coordinates $U=u/2$, $V=v/2$]. The tail follows from equation~\eqref{eq:J_def} by a radial derivative at fixed $t$,
\begin{equation}
g=rB=-\partial_rJ=-(\partial_v-\partial_u)J\ .
\label{eq:g_from_J}
\end{equation}
This formulation is therefore equivalent, term by term, to our Born series~\eqref{eq:born_series}: $-\partial_rJ_1=\frac12\int_{u/2}^{v/2}m^2(\eta'+t')\,dt'$ is the single-scattering term~\eqref{eq:B_born}, and $-\partial_rJ_n$ is the $n$-th scattering order $g_n$ of the Neumann series~\eqref{eq:neumann_n}.  However, the construction of Ref.~\cite{Compere:2026xwy}, stated for $J$ with trivial data, requires the numerical differentiation of a smooth function to extract the tail, thus propagating an additional numerical error.

For $\Lambda$CDM with radiation, Ref.~\cite{Compere:2026xwy} gives the scale factor in closed form,
\begin{equation}
a(\eta)=\frac{-H_0\sqrt{\Omega_r}\,\wp'(\eta)+\frac12H_0^2\Omega_m\,\wp(\eta)}{2\wp(\eta)^2-\frac12H_0^4\Omega_\Lambda\Omega_r}\ ,
\label{eq:a_weierstrass}
\end{equation}
with $\wp$ the Weierstrass function of invariants $g_2=H_0^4\Omega_\Lambda\Omega_r$ and $g_3=-H_0^6\Omega_\Lambda\Omega_m^2/16$, and $m^2=a''/a=H_0^2\big(\Omega_m/(2a)+2\Omega_\Lambda a^2\big)$. Inserted into the series~\eqref{eq:J_volterra}, this gives an analytic representation of the Green's function. Below, we evaluate this representation numerically and compare it with the Goursat solution.

We compare the two constructions in $\Lambda$CDM with radiation, $\Om=0.3$, $\Omega_r=9.1\times 10^{-5}$, $\OL=1-\Om-\Omega_r$, for a source in the radiation era, $H_0\eta'=0.01$, observed at $H_0\eta=0.1$, after matter-radiation equality ($H_0\eta_{\rm eq}\simeq0.026$). Our solutions use $a(\eta)$ from the numerical inversion of $H_0\eta(a)=\int_0^a db\,(\Omega_r+\Om b+\OL b^4)^{-1/2}$, and $g$ from the marching~\eqref{eq:marching}. The representation of Ref.~\cite{Compere:2026xwy}, evaluated numerically here, takes $a(\eta)$ from equation~\eqref{eq:a_weierstrass}, with $\wp$ evaluated through the Jacobi function ${\rm cn}$. The iteration~\eqref{eq:J_volterra} is computed by a two-dimensional trapezoidal rule, and $B$ is obtained from equation~\eqref{eq:g_from_J} by second-order finite differences.  The two Born series are computed on the same uniform double-null grid of step $H_0h=5\times10^{-4}$. 

\begin{figure}[t]
\centering
\includegraphics[width=\linewidth]{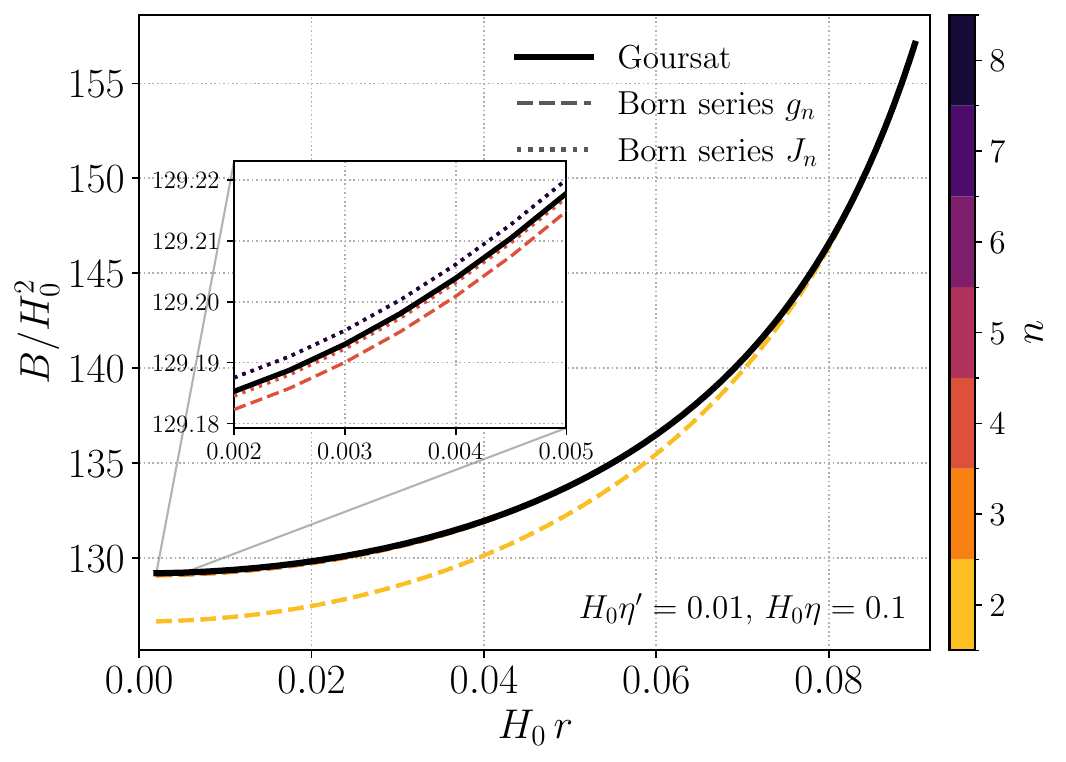}
\caption{Born series of the tail in $\Lambda$CDM with radiation ($\Om=0.3$, $\Omega_r=9.1\times 10^{-5}$), on the section $H_0\eta=0.1$ of a source at $H_0\eta'=0.01$. Goursat solution (solid black), partial sums $g_1+\dots+g_n$ of the Neumann series~\eqref{eq:neumann_n} (dashed, colored by the order $n$), and partial sums $J_0+\dots+J_n$ of the Volterra series~\eqref{eq:J_volterra} used in Ref.~\cite{Compere:2026xwy} (dotted, same colors). Grid step $H_0h=5\times10^{-4}$.}
\label{fig:goursat_vs_compare}
\end{figure}

Since the two constructions $g_n$~\eqref{eq:neumann_n} and $J_n$~\eqref{eq:J_volterra}  are the same Born series, we compare them order by order. Figure~\ref{fig:goursat_vs_compare} shows them together with the Goursat solution. At second order, the two series overlap and visibly underestimate the tail near the worldline, and from $n\simeq4$, the partial sums can no longer be distinguished from the Goursat solution on the scale of the main panel. The inset, near the worldline, resolves the residual differences, with the partial sum $g_n$ converging to the exact Goursat solution at $n\simeq 3$, while the partial sums $J_n$ stop slightly above the exact Goursat curve. 

\begin{figure}[t]
\centering
\includegraphics[width=\linewidth]{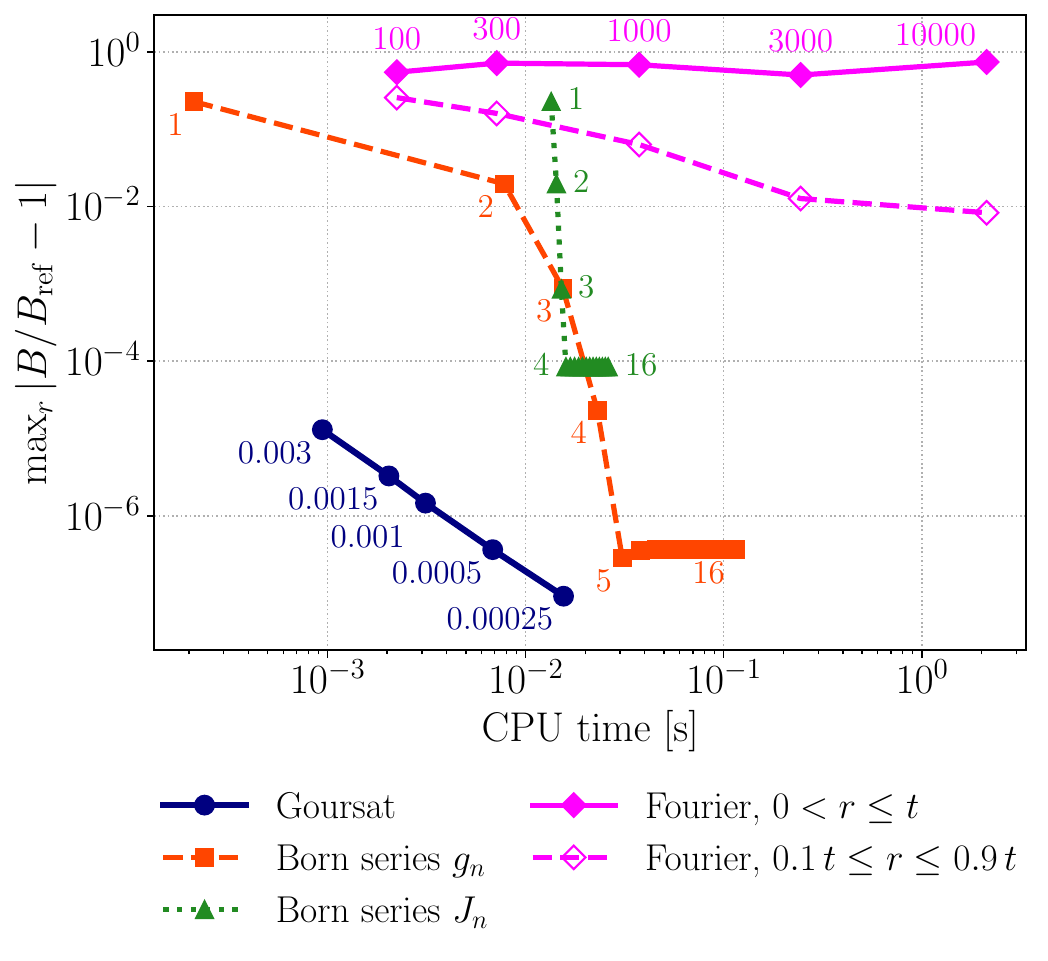}
\caption{Accuracy versus computational cost of the methods for the tail, on the section of Figure~\ref{fig:goursat_vs_compare}. Vertical axis: largest relative deviation from the reference solution (Richardson extrapolation of the Goursat solution at $H_0h=2.5\times10^{-5}$) over the section, including the light cone. Horizontal axis: CPU time. Navy circles: direct Goursat marching, labeled by the grid step $H_0h$. Orange squares and green triangles: partial sums $g_1+\dots+g_n$ and $J_0+\dots+J_n$ of the two Born series at $H_0h=5\times10^{-4}$, labeled by the order $n$, with cumulative times. Magenta diamonds: Fourier method of Ref.~\cite{Jokela:2022rhk}, equations~\eqref{eq:jks_mode} and~\eqref{eq:jks_inverse}, labeled by the cutoff $K/H_0$.}
\label{fig:CPU_time}
\end{figure}

The accuracy of the two series is more evident in Figure~\ref{fig:CPU_time}, which compares accuracy and computational cost of the direct Goursat marching at several grid steps, of the two Born series, and of the Fourier method of Ref.~\cite{Jokela:2022rhk}, equations~\eqref{eq:jks_mode} and~\eqref{eq:jks_inverse}, at several cutoffs $K$. As a reference, we take the Richardson extrapolation $(4B_h-B_{2h})/3$ of the Goursat solution at $H_0 h=2.5\times 10^{-5}$. Figure~\ref{fig:CPU_time} shows that the two Born series have the same truncation error up to $n\simeq3$. At higher orders, however, the partial sum $g_n$ reaches the accuracy of the direct marching on the same grid~\footnote{The floor that appears in the Born series $g_n$ is the discretization error of the marching itself, of order $h^2$.} from $n\simeq5$, while the error of the series $J_n$ saturates more than two orders of magnitude higher, from $n\simeq4$. This floor is the error of the numerical derivative~\eqref{eq:g_from_J}, which is largest near the light cone, where it requires one-sided differences.

The times recorded in Figure~\ref{fig:CPU_time} are CPU times of our Python implementations. Each order of the Born series $g_n$ beyond the first requires one march over the whole wedge, so that the series reaches the accuracy of the direct solution only at several times the computational cost. The representation of Ref.~\cite{Compere:2026xwy} spends most of its time before the first order evaluating the potential on the two-dimensional grid, and little for each further order, but its error stops at the floor of the numerical derivative. The series of Ref.~\cite{Compere:2026xwy} has to be summed term by term, whereas the marching~\eqref{eq:marching} sums all the orders at once, at the cost of a single order. A null-grid evolution of $J$, also outlined in Ref.~\cite{Compere:2026xwy}, would likewise sum all the orders, but would not remove the numerical derivative. The error of the Fourier method over the whole section does not decrease with $K$, because of the oscillations near the light cone and the worldline~\eqref{eq:jks_ripples}, while in the interior it decreases roughly as $1/K$. At the largest cutoff it is still at the level of $10^{-2}$ in the interior, for a CPU time longer than that of all the other methods. At equal cost, the direct marching is more accurate than the Fourier method by several orders of magnitude.

\bibliography{refs_B}

\end{document}